\documentclass[aps,amsmath,amssymb,nofootinbib]{revtex4} 
\usepackage{graphicx} 
\usepackage{amsmath}
\usepackage{amsfonts,amsbsy}
\usepackage{amssymb}
\allowdisplaybreaks

\usepackage{xcolor}
\definecolor{lcolor}{rgb}{0.5,0,0}
\definecolor{citcolor}{rgb}{0,0.3,0.0}
\usepackage[breaklinks,colorlinks,urlcolor=blue,citecolor=citcolor,linkcolor=lcolor]{hyperref}

\usepackage{mciteplus}

\def\gsim{ \,\, \vcenter{\hbox{$\buildrel{\displaystyle >}\over\sim$}}
 \,\,}

\def\be{\begin{equation}}
\def\ee{\end{equation}}
\def\bea{\begin{eqnarray}}
\def\eea{\end{eqnarray}}

\newcommand{\dd}{{\rm d}}
\newcommand{\nn}{\nonumber}

\begin{document}

\title{\bf Sub-femtometer scale color charge fluctuations
  in a proton made of three quarks and a gluon}

\author{Adrian Dumitru}
\email{adrian.dumitru@baruch.cuny.edu}
\affiliation{Department of Natural Sciences, Baruch College, CUNY,
17 Lexington Avenue, New York, NY 10010, USA}
\affiliation{The Graduate School and University Center, The City University
  of New York, 365 Fifth Avenue, New York, NY 10016, USA}

\author{Risto Paatelainen} \email{risto.sakari.paatelainen@cern.ch}
\affiliation{Helsinki Institute of Physics, FI-00014 University of
  Helsinki, Finland}
\affiliation{Theoretical  Physics  Department,  CERN,  CH-1211  Geneve  23,
  Switzerland}

\begin{abstract}
  The light-front wave function of a proton composed of three quarks
  and a perturbative gluon is computed.  This is then used to derive
  expressions for the color charge density correlator
  $\langle\rho^a(\vec q_1)\, \rho^b(\vec q_2)\rangle$ at ${\cal
    O}(g^4)$ due to the emission of a gluon by one of the quarks in
  light-cone gauge. The correlator exhibits the soft and collinear singularities. Albeit, we employ exact gluon emission and absorption
  vertices, and hence the gluon is not required to carry very small
  light-cone momentum, or to be collinear to the emitting quark.

  We verify that the correlator satisfies the Ward identity and that
  it is independent of the renormalization scale, i.e.\ that
  ultraviolet divergences cancel. Our expressions provide
  $x$-dependent initial conditions for Balitsky-Kovchegov evolution of
  the $C$-even part of the dipole scattering matrix to higher
  energies. That is, we determine the first non-trivial moment of the
  color charge fluctuations which act as sources for soft color fields
  in the proton with wavelengths greater than approximately $1/x \sim
  10 - 100$.
\end{abstract}

\maketitle

\tableofcontents

\newcommand{\Ptp}{{\mathbf{P}'}}
\newcommand{\Pt}{{\mathbf{P}}}
\newcommand{\qtp}{{\mathbf{q}'}}
\newcommand{\ktp}{{\mathbf{k}'}}
\newcommand{\ktpp}{{\mathbf{k}''}}
\newcommand{\ptp}{{\mathbf{p}'}}
\newcommand{\ptpp}{{\mathbf{p}''}}
\newcommand{\xtp}{{\mathbf{x}'}}
\newcommand{\ytp}{{\mathbf{y}'}}
\newcommand{\ztp}{{\mathbf{z}'}}
\newcommand{\rtp}{{\mathbf{r}'}}
\newcommand{\rt}{{\mathbf{r}}}
\newcommand{\xt}{{\mathbf{x}}}
\newcommand{\bt}{{\mathbf{b}}}
\newcommand{\yt}{{\mathbf{y}}}
\newcommand{\zt}{{\mathbf{z}}}
\newcommand{\pt}{{\mathbf{p}}}
\newcommand{\qt}{{\mathbf{q}}}
\newcommand{\kt}{{\mathbf{k}}}
\newcommand{\ktpzero}{{\mathbf{k}_{0'}}}
\newcommand{\nt}{{\mathbf{n}}}
\newcommand{\mt}{{\mathbf{m}}}
\newcommand{\Rt}{{\mathbf{R}}}
\newcommand{\ot}{\mathbf{0}}
\newcommand{\itt}{\mathbf{i}}
\newcommand{\jt}{\mathbf{j}}
\newcommand{\hht}{\mathbf{h}}
\newcommand{\Kt}{\mathbf{K}}
\newcommand{\Lt}{\mathbf{L}}
\newcommand{\at}{\mathbf{a}}
\newcommand{\ptt}{p_T} 
\newcommand{\ktt}{k_T} 
\newcommand{\qtt}{q_T} 
\newcommand{\lt}{\mathbf{l}}
\newcommand{\nabt}{\boldsymbol{\nabla}}
\newcommand{\epst}{\boldsymbol{\varepsilon}}
\newcommand{\At}{\mathbf{A}}
\newcommand{\Deltat}{{\boldsymbol{\Delta}}}
\newcommand{\Deltatt}{{\Delta_T}}
\newcommand{\kvec}{{\vec{k}}}
\newcommand{\pvec}{{\vec{p}}}
\newcommand{\lvec}{{\vec{l}}}
\newcommand{\ppvec}{{{\vec{p}}{\, '}}}
\newcommand{\pppvec}{{{\vec{p}}{\, ''}}}
\newcommand{\kpvec}{{{\vec{k}}{\, '}}}
\newcommand{\kpzerovec}{{{\vec{k}_{0'}}}}
\newcommand{\kppzerovec}{{{\vec{k}_{0''}}}}
\newcommand{\kponevec}{{{\vec{k}_{1'}}}}
\newcommand{\kpponevec}{{{\vec{k}_{1''}}}}
\newcommand{\kppvec}{{{\vec{k}}{\, ''}}}
\newcommand{\qvec}{{\vec{q}}}
\newcommand{\qpvec}{{{\vec{q}}{\, '}}}
\newcommand{\xvec}{{\vec{x}}}
\newcommand{\yvec}{{\vec{y}}}
\newcommand{\nvec}{{\vec{n}}}
\newcommand{\mvec}{{\vec{m}}}
\newcommand{\Kvec}{{\vec{K}}}
\newcommand{\Pvec}{{\vec{P}}}
\newcommand{\hvec}{{\vec{h}}}

\newcommand{\lo}{{\textnormal{LO}}}
\newcommand{\nlo}{{\textnormal{NLO}}}

\newcommand{\epsl}{{\varepsilon\!\!\!/}}
\newcommand{\pasl}{{p\!\!\!/_{a}}}
\newcommand{\pbsl}{{p\!\!\!/_{b}}}
\newcommand{\psl}{{p\!\!\!/}}
\newcommand{\ppsl}{{p\!\!\!/}'}
\newcommand{\ksl}{{k\!\!\!/}}


\newcommand{\dk}{{\widetilde{\mathrm{d} k}}}
\newcommand{\dkp}{{\widetilde{\mathrm{d} k}{'}}}
\newcommand{\dkpzero}{{\widetilde{\mathrm{d} k_{0'}}}}
\newcommand{\dkppzero}{{\widetilde{\mathrm{d} k_{0''}}}}
\newcommand{\dkpone}{{\widetilde{\mathrm{d} k_{1'}}}}
\newcommand{\dkpp}{{\widetilde{\mathrm{d} k}{''}}}
\newcommand{\ddp}{{\widetilde{\mathrm{d} p}}}
\newcommand{\dpp}{{\widetilde{\mathrm{d} p}{'}}}
\newcommand{\dppp}{{\widetilde{\mathrm{d} p}{''}}}
\newcommand{\dqp}{{\widetilde{\mathrm{d} q}{'}}}
\newcommand{\dq}{{\widetilde{\mathrm{d} q}}}

\newcommand{\dPp}{{\widetilde{\mathrm{d} P}{'}}}

\newcommand{\qemit}{V}
\newcommand{\qemitas}{\mathcal{V}}
\newcommand{\qbemit}{\overline{V}}
\newcommand{\qbemitas}{\overline{\mathcal{V}}}
\newcommand{\paircr}{A}
\newcommand{\paircras}{\mathcal{A}}
\newcommand{\annih}{\overline{A}}
\newcommand{\annihas}{\overline{\mathcal{A}}}

\newcommand{\ud}{\, \mathrm{d}}
\newcommand{\uc}{{\mathrm{c}}}
\newcommand{\ul}{{\mathrm{L}}}
\newcommand{\intd}{\int \!}
\newcommand{\tr}{\, \mathrm{tr} \, }
\newcommand{\R}{\mathrm{Re}}
\newcommand{\nc}{{N_\mathrm{c}}}
\newcommand{\nf}{{N_\mathrm{F}}}
\newcommand{\half}{\frac{1}{2}}
\newcommand{\hc}{\mathrm{\ h.c.\ }}
\newcommand{\nosum}[1]{\textrm{ (no sum over } #1 )}
\newcommand{\na}{\, :\!}
\newcommand{\nb}{\!: \,}
\newcommand{\cf}{C_\mathrm{F}}
\newcommand{\ca}{C_\mathrm{A}}
\newcommand{\df}{d_\mathrm{F}}
\newcommand{\da}{d_\mathrm{A}}
\newcommand{\nr}[1]{(\ref{#1})}
\newcommand{\dadj}{D_{\mathrm{adj}}}
\newcommand{\ra}{R_A}
\newcommand{\rp}{R_p}

\newcommand{\mev}{\ \textrm{MeV}}
\newcommand{\tev}{\ \textrm{TeV}}
\newcommand{\gev}{\ \textrm{GeV}}
\newcommand{\fm}{\ \textrm{fm}}
\newcommand{\mb}{\ \textrm{mb}}
\newcommand{\ls}{\Lambda_\mathrm{s}}
\newcommand{\qs}{Q_\mathrm{s}}
\newcommand{\qsprime}{Q_\mathrm{s}'}
\newcommand{\qsprimep}{Q_{\mathrm{s}p}'}
\newcommand{\qsprimea}{Q_\mathrm{sA}'}
\newcommand{\qsa}{Q_{\mathrm{s}A}}
\newcommand{\qsp}{Q_{\mathrm{s}p}}
\newcommand{\qsadj}{\widetilde{Q}_{\mathrm{s}}}
\newcommand{\qsoadj}{\widetilde{Q}_{\mathrm{s0}}}
\newcommand{\qso}{Q_\mathrm{s0}}

\newcommand{\lqcd}{\Lambda_{\mathrm{QCD}}}
\newcommand{\as}{\alpha_{\mathrm{s}}}

\newcommand{\subA}{A}
\newcommand{\subB}{B}
\newcommand{\B}{A_{\subB}}

\newcommand{\fig}{Fig.~}
\newcommand{\figs}{Figs.~}
\newcommand{\eq}{eq.~}
\newcommand{\se}{Sec.~}
\newcommand{\eqs}{eqs.~}

\newcommand{\npart}{{N_\mathrm{part}}}
\newcommand{\nch}{{N_\mathrm{ch}}}

\newcommand{\xbj}{{x}}
\newcommand{\sigmaa}{{ \sigma^A_\textrm{dip} }}
\newcommand{\sigmap}{{ \sigma^\textrm{p}_\textrm{dip} }}
\newcommand{\sigmadip}{{ \sigma_\textrm{dip} }}
\newcommand{\dsigmap}{{\frac{\ud \sigma^\textrm{p}_\textrm{dip}}{\ud^2 \bt}}}
\newcommand{\dsigmaa}{{\frac{\ud \sigma^A_\textrm{dip}}{\ud^2 \bt}}}
\newcommand{\dsigmab}{{\frac{\ud \sigma^B_\textrm{dip}}{\ud^2 \bt}}}
\newcommand{\dsigma}{{\frac{\ud \sigma_\textrm{dip}}{\ud^2 \bt}}}
\newcommand{\dsigmaadj}{{\frac{\ud \tilde{\sigma}_\textrm{dip}}{\ud^2 \bt}}}
\newcommand{\xpom}{{x_\mathbb{P}}}

\newcommand{\llsim}{{\underset{\sim}{\ll}}}
\newcommand{\ggsim}{{\underset{\sim}{\gg}}}

\newcounter{diag}
\setcounter{diag}{0}

\newcommand{\namediag}[1]{\refstepcounter{diag} \thediag \label{#1}}
\renewcommand{\thediag}{(\alph{diag})}

\newcommand{\ps}{\textnormal{PS}}
\newcommand{\epsmsbar}{\varepsilon_{\overline{\textnormal{MS}}}}

\section{Introduction} \label{sec:Intro}

The purpose of this paper is to derive expressions for the (light-cone
gauge) color charge correlator $\langle\rho^a(\vec q_1)\, \rho^b(\vec
q_2)\rangle$ in a proton boosted to large momentum $P^+$, on the
$x^+=0$ light front.  The charge operator $\rho^a$ corresponds to the
plus component of the color current due to valence quarks, and a
perturbative gluon which is not required to carry small momentum
fraction. It sums color charges with light-cone (L.C.) momentum fractions
greater than a cutoff $x$, which collectively generate the color field
from which the projectile scatters eikonally. The kinematic region of
interest here corresponds to moderately small L.C.\ momentum fractions $x
\sim 0.01 - 0.1$.\\

Color charge correlations in impact parameter space are obtained via
2d Fourier transform of the charge correlator,
\be
\frac{1}{2}\, g^2\, \delta^{ab}\, G_2(\vec q_1, \vec q_2) \equiv
\langle\rho^a(\vec q_1)\, \rho^b(\vec
q_2)\rangle~~~~,~~~~
G_2(\vec b, \vec q_1- \vec q_2) =
\int\limits_{\vec K}\, e^{-i \vec b \cdot \vec K}\,
G_2\left( \vec q_1,\vec q_2\right)~.
\ee
Here\footnote{In general, we use the light-cone coordinates $x^{\mu} =
  (x^{+},x^{-},\vec x)$, where arrow notation denotes two-dimensional
  transverse vectors, $\vec x = (x_1,x_2) = (x^i)$, $\ud x_1 \ud x_2 =
  \ud^2x$ and $\xvec^2 = \vert \xvec\vert^2$.}, we assumed that $\vec
P=0$ for the incoming proton; $\vec K \equiv -\vec q_1- \vec q_2$
denotes its transverse momentum after scattering via two gluon
exchange. Also, $\int_{\vec K} \equiv \int \frac{\dd^2
  K}{(2\pi)^2}$. The brackets $\langle \cdots\rangle$ denote the
expectation value of a given operator over all possible superpositions
of quark and gluon states in the incoming and scattered protons,
respectively. The precise definition is given in
eq.~(\ref{eq:<K|..|P>}) below.

Ref.~\cite{Dumitru:2020fdh} showed that $G_2(\vec b,
\vec q_1- \vec q_2)$ exhibits non-trivial behavior as a function of
impact parameter $\vec b$ and relative transverse momentum $\vec q_1-
\vec q_2$ of the probes (and their relative angle), changing from
``repulsion'' at small $b$ and $\vec q_1- \vec q_2$ to ``attraction''
at large $b$, $\vec q_1- \vec q_2$. Their analysis restricted to the
valence quark state of the proton; here, we derive the corrections to
the color charge correlator due to the emission of a gluon by one of
the quarks. The exclusive $\gamma^{(*)} + p \to J/\Psi + p$ cross
section, for example, is determined by the impact parameter $\vec b$
and dipole size $\vec r$ dependence of the dipole-proton scattering
amplitude~\cite{Munier:2001nr,Kowalski:2003hm, Kowalski:2006hc,
  Armesto:2014sma, Dumitru:2019qec, Mantysaari:2019jhh,
  Mantysaari:2018zdd, Mantysaari:2017dwh, Mantysaari:2020axf} which,
in turn, is related to $G_2(\vec q_1, \vec q_2)$ (see below and
refs.~\cite{Dumitru:2018vpr,Dumitru:2019qec}).\\

In covariant gauge, $\langle\rho^a(\vec q_1)\, \rho^b(\vec
q_2)\rangle$ determines the $S$-matrix for scattering of a quark -
antiquark dipole from the color fields in the target proton (in the
``dilute'' limit $1-{\cal S} \ll {\cal S}$).  The $S$-matrix for
eikonal scattering can be expressed as
(e.g.\ ref.~\cite{Mueller:2001fv})
\be \label{eq:S_dipole_b}
   {\cal S} (\vec r,\vec b) =
   \frac{1}{N_c}\,{\rm tr} \,\left< U\left(\vec b +
     \frac{\vec r}{2}\right)\,
     U^\dagger\left( \vec b - \frac{\vec r}{2}\right)\right> \,
     =
1 - {\cal T} (\vec r,\vec b)~.
\ee
(Following the standard convention in the small-$x$ literature we
define the scattering amplitude without a factor of $i$).

When integrated over impact parameters $\vec b$, the scattering
amplitude is related to the so-called dipole gluon
distribution~\cite{Dominguez:2011wm}.  $U$ ($U^\dagger$) are
(anti-)path ordered Wilson lines representing the eikonal scattering
of the dipole of size $\vec r$ at impact parameter $\vec b$:
\be \label{eq:WilsonLines}
U(\vec x) = {\cal P} e^{ig \int dx^- A^{+a}(x^-,\vec x)\, t^a}, \quad\quad
U^\dagger(\vec x) = \overline{\cal P} e^{-ig \int dx^- A^{+a}(x^-,\vec x)\, t^a} ~.
\ee
Expanding ${\cal T} (\vec r,\vec b)$ to second order in $gA^+$,
i.e.\ neglecting exchanges of more than two gluons and the resummation
of two gluon exchanges, allows us to write it in terms of correlators
of the field integrated over the longitudinal coordinate $x^-$.
%
%
This field is related to the two-dimensional (2d) color charge density
in covariant gauge via
\be - \nabla_\perp^2 A^{+a}(\vec x) = \rho^a(\vec x)~.
\label{eq:Poisson_A+}
\ee
We refer to ref.~\cite{Altinoluk:2019wyu} for a thorough discussion of
the relation of Wilson line correlators at small $x$ to Wigner
distributions.

The gauge transformation from covariant to light-cone gauge involves
the color charge density
itself~\cite{Iancu:2003xm,Burkardt:2003yg}. Therefore, to quadratic
order in the charge density, the charge correlators in the two gauges
are the same.\\

From eqs.~(\ref{eq:S_dipole_b}, \ref{eq:Poisson_A+})
one obtains the $C$-even two gluon exchange
amplitude~\cite{Dumitru:2018vpr}
\bea
{\cal T}_{gg}(\vec r,\vec b) &=& 
  - \frac{g^4}{2} C_F\,\int\limits_{\vec K, \vec q}
  \frac{e^{-i\vec b \cdot \vec K}}{(\vec q -\frac{1}{2}\vec K)^2
    \, (\vec q+ \frac{1}{2} \vec
  K)^2}\,\left(
\cos\left(\vec r \cdot {\vec q}\right) 
- \cos\left(\frac{{\vec r}\cdot \vec K}{2}\right)\right)
\,
G_2\left({\vec q}-\frac{1}{2}\vec K,-{\vec q}-\frac{1}{2}\vec K\right)
.    
\label{eq:Pomeron}
\eea
Since $G_2(\vec q_1, \vec q_2)$ is symmetric under a simultaneous sign
flip of both arguments it follows that ${\cal T}_{gg}(\vec r,\vec b)$
is real. $G_2$ satisfies a Ward identity and vanishes when
either one of the gluon momenta goes to
zero~\cite{Bartels:1999aw,Ewerz:2001fb}: $G_2\left({\vec
  q}-\frac{1}{2}\vec K,-{\vec q}-\frac{1}{2}\vec K\right) \sim
({\vec q}\pm\frac{1}{2}\vec K)^2$ as ${\vec q}\to\pm \frac{1}{2}\vec
K$.\\

The computation presented here corresponds to explicit ``evolution''
of the three valence quark Fock state of the proton to $x$ of order a
few times $10^{-2}$.~\footnote{The emission of a gluon which is not soft
  or collinear to the valence charges has been considered previously
  by Altinoluk and Kovner
  in ref.~\cite{Altinoluk:2011qy}. However, their focus was on
  single-inclusive particle production in the collision of such a
  proton with a nucleus rather than on color charge correlations in
  the proton. Therefore, they did not require non-forward matrix
  elements.}  The dipole scattering amplitude at yet smaller $x$ can
be obtained by adding additional soft gluons to the
proton~\cite{Mueller:1993rr}.  This is achieved by the
Balitsky-Kovchegov (BK) evolution
equation~\cite{Balitsky:1995ub,Kovchegov:1999yj} which also accounts
for multiple scattering (i.e.\ the resummation of two-gluon exchanges
in covariant gauge) as one approaches the unitarity limit.

Detailed fits of BK evolution with running coupling corrections to the
$\gamma^* - p$ cross section measured at HERA have been performed by
Albacete {\it et al.} in ref.~\cite{Albacete:2009fh,Albacete:2010sy}
(see also refs.~\cite{Lappi:2013zma, Beuf:2020dxl}). Improved recent
analyses employ a collinearly improved NLO BK evolution equation
(refs.~\cite{Iancu:2015joa,Ducloue:2019jmy} and references
therein). However, these fits of small-$x$ QCD evolution to HERA DIS
data for the inclusive cross section typically impose {\it ad-hoc}
initial conditions for the dipole scattering amplitude on the proton,
tuned to obtain the best match of the evolution equation to the
data. Moreover, a change in the initial value of $x$ requires
uncontrolled (by theory) retuning of the initial condition for the
dipole scattering amplitude.

Here, continuing previous
work~\cite{Dumitru:2020fdh,Dumitru:2018vpr,Dumitru:2019qec} which
restricted to the three valence quark Fock state, we attempt to
provide initial conditions based explicitly on the light-front wave
function (LFwf) of the proton. That way one may take advantage of
``proton imaging'' performed at the future electron-ion collider
(EIC)~\cite{Boer:2011fh,Accardi:2012qut,Aschenauer:2017jsk,Proceedings:2020eah}.

Our initial condition for BK evolution is obtained by cutting off the
divergent integral over the {\em plus momentum} $k_g^+$ of the gluon in
the right-moving proton. However, the BK equation in its standard
formulation evolves the wave function of the projectile, and the
evolution ``time'' is then related to the minus component of the
momentum of the emitted gluon~\cite{Beuf:2014uia, Ducloue:2019ezk,
  Boussarie:2020fpb}. Duclou\'e {\em et al.} have
reformulated~\cite{Ducloue:2019ezk} BK evolution at NLO in terms of
the target rapidity (or Bjorken-$x$).  They obtained an evolution
equation which is non-local in rapidity and which depends explicitly
on the initial rapidity (or $x=k_g^+/P^+$).  \emph{This underscores the
importance of a controlled $x$-dependence of the ``initial condition''
for the dipole scattering amplitude which we compute here}.

\section{Three quark Fock state of the proton} \label{sec:LFwf_val-q}
  
The light cone state\footnote{For a detailed presentation of the light
  cone formalism and its application to high energy scattering, see
  ref.~\cite{Brodsky:1997de}.} of an on-shell proton with
four-momentum $P^\mu = (P^+, P^-,\vec{P})$ composed of three
quarks is written as~\cite{Lepage:1980fj}
\bea
|P\rangle &=& \frac{1}{\sqrt{2 N_c}}
\int \left[\dd x_i\right]
\int \left[\dd^2 k_i\right]\,
\Psi_{qqq}(x_1, \vec k_1; x_2, \vec k_2; x_3, \vec k_3)
\sum_{i_1, i_2, i_3}\epsilon_{i_1 i_2 i_3}\,
|p_1,i_1; \, p_2,i_2; \, p_3,i_3\rangle \,
|S\rangle~.
\label{eq:valence-proton}
\eea
Here, $N_c=3$ is the number of colors while $|S\rangle$ is the helicity
wave function of the proton described in sec.\ref{sec:SpinWF}
below. It is normalized to $\langle S | S\rangle=1$. Furthermore, the following compact notation has been introduced:
\be \left[\dd x_i\right] \equiv \frac{\dd x_1 \dd x_2 \dd x_3}{8 x_1 x_2 x_3}
\delta(1-x_1-x_2-x_3),\quad\quad
\left[\dd^2 k_i\right] \equiv
\frac{\dd^2 k_1 \dd^2 k_2 \dd^2 k_3}{(2\pi)^6}\,
\delta(\vec{k}_1+\vec{k}_2+\vec{k}_3)~.
\label{eq:[dxi][dki]}
\ee
The three on-shell quark momenta are specified by their lightcone
momentum components $p_i^+ = x_i P^+$ and their transverse components
$\vec{p}_{i} = x_i \vec{P} + \vec{k}_i$.  The quark colors are
denoted as $i_{1,2,3}$.  $\Psi_{qqq}$ is the probability amplitude for
finding exactly three quarks (and no gluons) with the specified
momenta, colors, and helicities, in the proton. It is symmetric under
exchange of any two of the quarks: $\Psi_{qqq}(x_1, \vec k_1; x_2,
\vec k_2; x_3, \vec k_3) = \Psi_{qqq}(x_2, \vec k_2; x_1, \vec k_1;
x_3, \vec k_3)$ etc.

For simplicity, we will assume that the momentum space wave function
$\Psi_{qqq}$ does not depend on the helicities $h_i=\pm 1$ of the
quarks, i.e.\ that the helicity wave function factorizes from the
color-momentum wave function. It is presented in more detail in the
next section.

We neglect plus momentum transfer so that $\xi = (K^+ -
P^+)/P^+ \to 0$. This approximation is valid at high energies.
The proton state is then normalized according to
\bea
\langle K | P\rangle &=& 16\pi^3 \, P^+ \delta(P^+ - K^+)
\, \delta(\vec{P} - \vec{K}) \label{eq:ProtonNorm1}
~.
\eea
\\

The one-particle quark states introduced above are created by the
action of the quark creation operator on the vacuum $|0\rangle$:
\be
|p,i,\lambda\rangle =   b^\dagger_{i\lambda}(p) |0\rangle~.
\ee
The quark creation and annihilation operators satisfy the
anti-commutation relation
\be
\{ b_{j\sigma}(k) , b^\dagger_{i\lambda}(p) \} = \delta^{ji}_{\sigma
  \lambda}\,\, 16\pi^3\, k^+\delta(k^+-p^+)\, \delta(\vec k - \vec
p)~,\label{anticomm} 
\ee
therefore,
\be
\langle k, j, \sigma\,|\, p,i,\lambda\rangle = \delta^{ji}_{\sigma
  \lambda}\,\, 16\pi^3\, k^+\delta(k^+-p^+)\, \delta(\vec k - \vec
p)~.
\ee

These relations determine the normalization of the valence quark wave
function to be
\be \label{eq:Norm_psi3}
\frac{1}{2}
\int \left[\dd x_i\right]
  \int \left[\dd^2 k_i\right]\,
  |\Psi_{qqq}(x_1, \vec k_1; x_2, \vec k_2; x_3, \vec k_3)|^2 = 1~.
\ee

For later use we also write the commutation relations of the operators
which create or destroy a gluon
\be
[ a_{a\lambda}(k) , a^\dagger_{b\rho}(p) ] =
  \delta^{ab}_{\lambda\rho}\,\, 16\pi^3\, k^+\delta(k^+-p^+)\,
  \delta(\vec k - \vec p)~.\label{eq:acomm}
\ee

\subsection{Helicity wave function} \label{sec:SpinWF}

The flavor structure of the proton plays no role in our analysis, so
we may assume that the first two quarks are always $u$ quarks, and the
third quark is always a $d$ quark. Further, we are interested in
matrix elements of operators which are diagonal in helicity. However,
the $q\to qg$ vertex does involve the quark and gluon helicities and
so we need to properly count states to ensure the correct
normalization. Since we consider an unpolarized proton we assume that
in the three quark Fock state the quarks couple with equal probability
to positive or negative proton helicity,
\be
|S \rangle = 
\frac{1}{\sqrt 2} \left( |p_\uparrow\rangle +
|p_\downarrow\rangle\right)~~~~~~,~~~~~~
\langle S | S \rangle = 1~.
\ee
In Schlumpf's
notation~\cite{Schlumpf:1992vq} the spin wave function of the
$|uud\rangle$ proton with positive helicity is
%
\bea
|p_\uparrow\rangle &\sim& \chi_{\uparrow}^{\rho_1} +
\chi_{\uparrow}^{\rho_2}~, \label{eq:p_spin}\\
\chi_{\uparrow}^{\rho_1} &=& \frac{1}{\sqrt 2}
\left( \left|\uparrow\uparrow\downarrow\right> -
\left|\downarrow\uparrow\uparrow\right>\right)\\
\chi_{\uparrow}^{\rho_2} &=& \frac{1}{\sqrt 2}
\left( \left|\uparrow\uparrow\downarrow\right> -
\left|\uparrow\downarrow\uparrow\right>\right)~.
\eea
For $|p_\downarrow\rangle$ all arrows (quark helicities) are reversed.
Hence, the squared norm of the state (\ref{eq:p_spin}) is
$\chi_{\uparrow}^{\rho_1\, \dagger}\chi_{\uparrow}^{\rho_1} +
\chi_{\uparrow}^{\rho_2\, \dagger}\chi_{\uparrow}^{\rho_2} +
\chi_{\uparrow}^{\rho_1\, \dagger}\chi_{\uparrow}^{\rho_2} +
\chi_{\uparrow}^{\rho_2\, \dagger}\chi_{\uparrow}^{\rho_1} = 3$.
Therefore, we take
\be
|p_\uparrow\rangle = \frac{1}{\sqrt 3}\left(\chi_{\uparrow}^{\rho_1} +
\chi_{\uparrow}^{\rho_2}\right) =\frac{1}{\sqrt 6}\bigl(
2\, \left|\uparrow\uparrow\downarrow\right> -
\left|\downarrow\uparrow\uparrow\right> -
\left|\uparrow\downarrow\uparrow\right>
\bigr)
~.
\ee
Helicity matrix elements of diagonal operators are given by
\bea
\langle S \vert O \vert S\rangle
&=& \frac{1}{2}\left( \left< p_\uparrow\right| O \left| p_\uparrow\right>
+ \left< p_\downarrow\right| O \left| p_\downarrow\right>
\right) = \nn \\
& & \frac{1}{12}\left[
4 \left<\uparrow\uparrow\downarrow\right| O
\left|\uparrow\uparrow\downarrow\right> 
+ \left<\downarrow\uparrow\uparrow\right| O
\left|\downarrow\uparrow\uparrow\right> 
+ \left<\uparrow\downarrow\uparrow\right| O
\left|\uparrow\downarrow\uparrow\right>
+ (\uparrow \leftrightarrow \downarrow)
\right]~. \label{eq:HelicityMatrixElement}
\eea
We will use this expression below to sum the gluon emission vertex
over quark helicities. However, since we are not concerned with
helicity dependent processes we shall symmetrize over permutations of
the three quarks. For example, eq.~(\ref{eq:HelicityMatrixElement})
gives $\langle h_1 h_2\rangle = \frac{1}{3}$ but when we symmetrize
over permutations, $\langle h_1 h_2\rangle_s = (\langle h_1 h_2\rangle
+ \langle h_1 h_3\rangle +\langle h_2 h_3\rangle)/3 = - \frac{1}{3}$.


\section{The three quark plus one gluon Fock state}

\subsection{Quark to quark + gluon splitting}
\label{sec:q->qg}

The light-cone wave function (LCwf) for $q \to q g$ splitting is given in LC perturbation theory by
\be
\label{eq:qqgLCWF}
\psi_{q\to qg}(p;k_q,k_g) = \frac{-g(t^{a})_{ji}}
    {\Delta^{-}_{qg}}\biggl [\bar{u}_{\lambda}(k_q)
      \epsl^{\ast}_{\sigma}(k_g)u_h(p)\biggr ],
\ee
where $p$ denotes the momentum of the incoming quark;
and
\be
\label{eq:qgqLCWF}
\psi_{qg\to q}(k_q,k_g; p) = \frac{-g(t^{a})_{ij}}
    {\Delta^{-}_{q}}\biggl [\bar{u}_{h}(p)
      \epsl_{\sigma}(k_g)u_{\lambda}(k_q)\biggr ],
\ee
where $p$ denotes the momentum of the outgoing quark. The quantities $k_q = p - k_g$ and $k_g$ are the
momenta of the other quark and of the gluon, respectively\footnote{Note that in LCwf only the plus and transverse momentum components are conserved.}. Also, $a$ is the adjoint color index for the gluon and
$i, j$ are the fundamental color indices for the quarks. The
quarks are assumed massless so that their helicity is conserved. Note
that the expression above does not assume that the plus momentum of the daughter quark or gluon is small. Using the on-shell relation $p^- =
\pvec^2/2p^+$, and similar for $k_q^-$ and $k_g^-$, the energy
denominator $\Delta^{-}_{qg}$ is given by
\be
\Delta^{-}_{qg} =  p^- - (k_q^- + k_g^-) = \frac{-p^+}{2k_q^+k_g^+}\biggl (\kvec_g - (k_g^+/p^+)\pvec \biggr )^2 = -\frac{\nvec^2}{2p^+z(1-z)} = -\Delta^{-}_{q}.
\ee
Here, $z = k_g^+/p^+$ with $0 \leq z \leq1$ is the LC momentum
fraction of the gluon and
\be
\nvec = \kvec_g - z\pvec
\ee
is the center-of-mass transverse momentum. If we do account for
a non-zero quark mass in the energy denominator, i.e.\ $p^- = ( \pvec ^2
+ m^2)/2p^+$ and $k^- = (\kvec_q^2 + m^2)/2k_q^+$, then the numerator
in r.h.s.\ of the last expression turns into $\nvec^2 + z^2 m^2$. We
will use this form whenever needed to regularize infrared divergences
but take $m^2\to0$ where possible.

The quark-gluon vertex can be decomposed into its
symmetric and antisymmetric parts as~\cite{Hanninen:2017ddy}
\be
\bar{u}_{\lambda}(k_q)\epsl^{\ast}_{\sigma}(k_g)u_h(p) = \frac{1}{p^+ z(1-z)}\biggl [\left (1 - \frac{z}{2} \right )\delta^{ij} \bar{u}_{\lambda}(k_q)\gamma^{+}u_h(p) - \frac{z}{4}\bar{u}_{\lambda}(k_q)\gamma^{+}[\gamma^i,\gamma^j]u_h(p)\biggr ]n^i\epst^{\ast j}_{\sigma}
\ee
and
\be
\bar{u}_{h}(p)\epsl_{\sigma}(k_g)u_\lambda(k_q) = \frac{1}{p^+ z(1-z)}\biggl [\left (1 - \frac{z}{2} \right )\delta^{ij} \bar{u}_{h}(p)\gamma^{+}u_\lambda(k_q) + \frac{z}{4}\bar{u}_{h}(p)\gamma^{+}[\gamma^i,\gamma^j]u_\lambda(k_q)\biggr ]n^i\epst_{\sigma}.
\ee
where $k_q^+ = p^+(1-z)$ and $k_g^+ = zp^+$. Note that $\delta^{ij}\delta^{ij} = D-2$ and $\gamma^i \gamma^{i} =
2-D$ and that this expression is valid in arbitrary spacetime
dimensions and automatically accounts for the conservation of plus and transverse momentum.

Putting all together yields 
\be
\label{eq:qqgLCWFfull}
\psi_{q\to qg}(p; k_q, k_g) =
2g(t^{a})_{ji}\biggl [\left (1 - \frac{z}{2} \right
  )\delta^{lm} \bar{u}_{\lambda}(k_q)\gamma^{+}u_h(p) -
  \frac{z}{4}\bar{u}_{\lambda}(k_q)\gamma^{+}[\gamma^l,\gamma^m]u_h(p)\biggr
]\frac{n^l\epst^{\ast m}_{\sigma}}{\nvec^2}
\ee
and
\be
\label{eq:qgqLCWFfull}
\psi_{qg\to q}(k_q,k_g;p) = -2gt^{a}_{ij}\biggl
    [\left (1 - \frac{z}{2} \right )\delta^{lm}
      \bar{u}_{\lambda}(p)\gamma^{+}u_h(k_q) +
      \frac{z}{4}\bar{u}_{\lambda}(p)\gamma^{+}[\gamma^l,\gamma^m]u_h(k_q)\biggr
    ]\frac{n^l\epst^{m}_{\sigma}}{\nvec^2}~.  \ee

In $D=4$, the expressions in \eq\nr{eq:qqgLCWFfull} and \eq\nr{eq:qgqLCWFfull}
can be expressed very compactly in the helicity basis by first noting
that~\cite{Hanninen:2017ddy}
\be
\bar{u}_{\lambda}(k_q)\gamma^{+}[\gamma^i,\gamma^j]u_h(p) =
-2ih\epsilon^{ij}\bar{u}_{\lambda}(k_q)\gamma^{+}u_h(p),
\ee
where the remaining matrix element is simple,
$\bar{u}_{\lambda}(k_q)\gamma^{+}u_h(p) =
\sqrt{(2p^+)(2k_q^+)}\delta_{\lambda h} = 2p^+\sqrt{1-z}\,
\delta_{\lambda h}$. Hence, we find that in $D=4$,
eqs.~(\ref{eq:qqgLCWFfull},\ref{eq:qgqLCWFfull}) can be reworked to
\be
\label{eq:qqgLCWFfull4D}
\psi_{q\to qg}(p; k_q, k_g) = 2g(t^{a})_{ji}\,
2p^+ \sqrt{1-z}\biggl [\left (1 - \frac{z}{2} \right )\delta^{lm} +
  \frac{z}{2}i\lambda\epsilon^{lm}\biggr ] \delta_{\lambda h}
\frac{n^l\epst^{\ast m}_{\sigma}}{\nvec^2}
\ee
and
\be
\label{eq:qgqLCWFfull4D}
\psi_{qg\to q}(k_q, k_g; p) = -2g(t^{a})_{ij} 2p^+
\sqrt{1-z}\biggl [\left (1 - \frac{z}{2} \right )\delta^{lm}
  - \frac{z}{2}i\lambda\epsilon^{lm}\biggr
] \delta_{\lambda h} \frac{n^l\epst^{m}_{\sigma}}{\nvec^2}.
\ee

As a check, note that in $D=4$
\be
\frac{1}{N_c}
\sum_{h, \sigma, a, i, j}\vert \psi_{q\to qg}(p; k_q, k_g)\vert^2 =
16g^2(p^+)^2\cf (1-z)\biggl [\left (1 - \frac{z}{2} \right )^2\delta^{lm}
  \delta^{kn} + \frac{z^2}{4}\epsilon^{lm}\epsilon^{kn} \biggr ]
\frac{n^l n^k}{\nvec^4}\sum_{\sigma}\epst^{\ast m}_{\sigma}\epst^{n}_{\sigma},
\ee
where the sum over the helicity states of the gluon yields
$\sum_{\sigma}\epst^{\ast m}_{\sigma}\epst^{n}_{\sigma} =
\delta^{mn}$ and the Fierz identity, $\epsilon^{lm}\epsilon^{kn} =
\delta^{lk}\delta^{mn} - \delta^{ln}\delta^{mk}$, simplify the
Kronecker delta contraction. This gives the following result
\be \frac{1}{N_c} \sum_{h, \sigma, a, i, j}\vert
\psi_{q\to qg}(p; k_q, k_g)\vert^2 = 8g^2\cf
(p^+)^2 (1-z)\biggl [1 + (1-z)^2\biggr ]\frac{1}{\nvec^2}.
\ee
The result is proportional to the splitting function $P_{q \to
  gq}(z) \sim \cf \left (1 + (1-z)^2\right )$ as it should be. Also, in
the $z \to 0$ soft gluon limit, the LCwf is independent of the helicity
of the quark
\be
\label{eq:qqgLCWFzto0}
\psi_{q\to qg}(\pvec; \kvec_q, \kvec_g) = 4g(t^{a})_{ji} \, p^+  \frac{\nvec \cdot \epst^{\ast}_{\sigma}}{\nvec^2}\delta_{\lambda h} .
\ee

For our applications below it will be convenient to take LC momentum
fractions of the daughter partons relative to the proton plus momentum
$P^+$ rather than relative to the parent quark. Hence, in the $q\to
qg$ amplitude, $z$ is then given by $z=x_g/x$ (or $1-z = x_q/x$) where
$p^+ = x P^+$ for the parent quark, $k^+_q = x_q P^+$ for the daughter
quark, and $k^+_g = x_g P^+$ for the gluon. On the other hand, in the
$qg\to q$ amplitude, $z=x_g/(x_g+x_q)$.
\\

\subsection{The quark wavefunction renormalization factor $Z_q(p^+)$ at
  order $g^2$}
\label{sec:Zq}

The full physical incoming one-particle quark state can be written as a simultaneous perturbative and Fock state decomposition in terms of the bare states 
\be
\begin{split}
\label{eq:qpertexp}
\vert q(p, h, i)\rangle = Z^{1/2}_q(p^+) \biggl
(\vert q(p, h, i)\rangle_0 +
\sum_{\lambda,\sigma,j,a}\int \dk_q
\dk_g(2\pi)^{3}&\delta(p^+ - k_q^+ - k_g^+)\delta(\pvec - \kvec_q - \kvec_g)\psi_{q\to
  qg}(p; k_q, k_g)\\
& \times \vert q(k_q, \lambda, j) \,
g(k_g,\sigma,a)\rangle_0 + \dots \biggr )~.
\end{split}
\ee
Here, the LCwf for $q \to q g$ splitting is denoted as $\psi_{q\to qg}$ and the Lorentz invariant measures $\dk_q$ and $\dk_g$ are defined as
\be
\begin{split}
  \int \dk & \equiv \int \frac{\ud k^+}{2k^+}\frac{\ud^{2}k}{(2\pi)^{3}}
  \to (\mu^2)^{2-D/2}\int \frac{\ud k^+}{(2\pi)\, 2k^+}
  \frac{\ud^{D-2}k}{(2\pi)^{D-2}}
~.
\end{split}
\ee
The latter form will be used when we regularize ultraviolet (UV) divergences by integrating over the momenta of all particles in $D$ dimensions. Here, an arbitrary scale $\mu^2$ is introduced so that the transverse integrals preserve their natural dimensions.  The quark wave function
renormalization coefficient $Z_q$ can be calculated from the
normalization requirement
\be
\label{eq:1qnorm}
\langle q(p, h, i)\vert q(p, h, i)\rangle =
        {}_{0}\langle q(p, h, i)\vert q(p, h, i)\rangle_0 =
        2p^+ (2\pi)^{3}\delta^{(3)}(0)~.
\ee

At order $g^2$ for $Z_q(p^+)$ we find 
\be
\label{eq:Zqatg2}
Z_q(p^+) = 1 - \frac{1}{2p^+}\int \dk_g \frac{1}{2k_q^+}
\vert \psi_{q\to qg}(p; k_q,
k_g)\vert^2 \equiv 1 - C_q(p^+),
\ee
where $\vert \psi_{q\to qg}\vert^2$ is summed over the internal gluon
and quark helicities and colors. Substituting \eq\nr{eq:qqgLCWFfull}
into \eq\nr{eq:Zqatg2} leads to
\be
\begin{split}
Z_q(p^+) & = 1 - \frac{1}{2p^+}\int \dk_g \frac{1}{2k_q^+} 4g^2\cf (2p^+)(2k_q^+)\biggl [\left (1 - \frac{z}{2}\right )^2 + \frac{z^2}{4}(D-3) \biggr ] \frac{1}{\nvec^2}\\
& = 1 - 2g^2\cf \int_{0}^{p^+} \frac{\ud k^{+}_g}{(2\pi)2k^{+}_g}(\mu^2)^{2-D/2} \int \frac{\ud^{D-2}k_g}{(2\pi)^{D-2}}\biggl [1 + (1-z)^2 + \frac{z^2}{2}(D-4) \biggr ] \frac{1}{\nvec^2},
\end{split}
\ee
where $k^{+}_g \geq 0$. Changing integration variables from $k^+_g$ to
$z$ and $\kvec_g$ to $\nvec$ gives
\be
\begin{split}
Z_q(p^+)  & = 1 - \frac{g^2\cf}{2\pi} \int_{0}^{1} \frac{\ud z}{z}\biggl [1 + (1-z)^2 + \frac{z^2}{2}(D-4) \biggr ] (\mu^2)^{2-D/2} \int \frac{\ud^{D-2} n}{(2\pi)^{D-2}}\frac{1}{\nvec^2}.
\end{split}
\ee
Finally, regulating the soft IR divergence in $z \to 0$ by a cutoff
$\alpha > 0$ with $\alpha < z < 1$ and the collinear IR divergence
with a quark mass parameter (as discussed in section~\ref{sec:q->qg}),
we arrive at
\be
\begin{split}
  Z_q(p^+)  & = 1 - \frac{g^2\cf}{2\pi} \int_{\alpha}^{1} \frac{\ud z}{z}\biggl [1 + (1-z)^2 + \frac{z^2}{2}(D-4) \biggr ] (\mu^2)^{2-D/2} \int \frac{\ud^{D-2}n}{(2\pi)^{D-2}}\frac{1}{(\nvec^2 + z^2m^2)} \\
& = 1 - \frac{g^2\cf}{8\pi^2} \int_{\alpha}^{1} \frac{\ud z}{z}\biggl [1 + (1-z)^2 + \frac{z^2}{2}(D-4) \biggr ]A_0(\Delta)~.
\label{eq:eq:Zqatg2Dform}
\end{split}
\ee
Here we have introduced the following notation for the UV divergent integral (see the discussion in appendix~\ref{sec:app-psi-psi*})
\be
\begin{split}
A_0(\Delta) & \equiv 4\pi(\mu^2)^{2-D/2} \int \frac{\ud^{D-2}n}{(2\pi)^{D-2}}\frac{1}{(\nvec^2 + \Delta)}\\
& = \Gamma\left (2 - \frac{D}{2}\right )\biggl [
  \frac{\Delta}{4\pi\mu^2}\biggr ]^{D/2 - 2} =
\frac{(4\pi)^{2-D/2}}{(2-D/2)} \Gamma\left (3 - \frac{D}{2}\right ) -
\log\left ( \frac{\Delta }{\mu^2}\right ) + \mathcal{O}(D-4)~,
\end{split}
\ee
where $\Delta \equiv z^2m^2$. In the above expression, we keep the universal constants
together with the $D \to 4$ pole. This corresponds to the MSbar scheme
for UV renormalization. Taking $D = 4-2\varepsilon$ in
\eq\nr{eq:eq:Zqatg2Dform} and expanding in $\varepsilon$, we find
\be
\begin{split}
\label{eq:eq:Zqatg2epsform}
Z_q(p^+) & = 1 + \frac{g^2\cf}{8\pi^2} \, \Theta(1-\alpha)\, \biggl
\{\biggl [\frac{1}{\varepsilon_{\rm MSbar}} + \log\left (
  \frac{\mu^2}{m^2}\right ) \biggr ]\biggl (\frac{3}{2} +
2\log\alpha \biggr ) + 4  - 2\log^2(\alpha) + 
\mathcal{O}(\varepsilon)\biggr \} \\
& \equiv 1 - C_q(p^+)~,
\end{split}
\ee
where $1/\varepsilon_{\rm MSbar} \equiv 1/\varepsilon - \gamma_E +
\log(4\pi)$, the parameter $\gamma_E$ is the Euler–Mascheroni constant and the scaleless cutoff $\alpha = p^+_{\rm min}/p^+$ with
$p^+_{\rm min} > 0$.

In the coming sections, we will also need the following $D$-dimensional integral
\bea \label{eq:Int-psi-psi*}
\frac{(2\pi)^{D-1}}{2p_1^+} \int \frac{\dk_g}{2(p_1^+ - k_g^+)}
\hat\psi_{q\to qg}(p_1; p_1- k_g, k_g)\,
\hat\psi^*_{q\to qg}(p_1-l; p_1- k_g - l_1,
k_g- l + l_1) &=& \nn\\
& & \hspace{-3cm}
2\pi^3 \biggl \{\frac{C_q(p_1^+)}{g^2\cf} + F(\vec l, \vec l_1;
\alpha_1, m^2) \biggr \} ~,
\eea
where two transverse momenta $\vec l$ and $\vec l_1$ are arbitrary.
This integral is done in detail in appendix~\ref{sec:app-psi-psi*}
where we also provide an explicit expression for the finite function
$F(\vec l, \vec l_1; \alpha_1, m^2)$ which includes the contribution
from the collinear DGLAP~\cite{Gribov:1972ri, Gribov:1972rt,
  Altarelli:1977zs, Dokshitzer:1977sg} IR singularity. Lastly, the UV
coefficient $C_q(p_1^+) = 1 - Z_q(p_1^+)$ is related to the quark wave
function renormalization factor which is given in
\eq\nr{eq:eq:Zqatg2epsform}.

\subsection{Proton with a gluon} \label{sec:LFwf_qqqg}

We replace each quark state vector in \eq\nr{eq:valence-proton} by the
perturbative expansion in \eq\nr{eq:qpertexp}. This yields
\be
\label{eq:qqqg-A-proton}
\begin{split}
&\vert q(p_1,h_1,i_1)\, q(p_2,h_2,i_2)\,
  q(p_3,h_3,i_3)\rangle\, \vert S\rangle
   = \\
  &\biggl [\left(1
  - \frac{C_q(p^+_1)}{2}\right)\vert q(p_1,h_1,i_1)\rangle_0
  + \sum_{\lambda,\sigma,j ,a}2g(t^{a})_{j i_1}\int \frac{\dk_g}{2(p_1^+ - k_g^+)}\hat \psi_{q\to qg}(p_1; p_1- k_g, k_g)\\
  &\hspace{5cm} \times
  \vert q(p_1-k_g,\lambda,j )\, g(k_g,\sigma,a)\rangle_0
  + \cdots \biggr ]\\
  \otimes & \biggl [\left(1
  - \frac{C_q(p^+_2)}{2}\right)\vert q(p_2,h_2,i_2)\rangle_0
  + \sum_{\lambda,\sigma,j ,a}2g(t^{a})_{j i_2}\int \frac{\dk_g}{2(p_2^+ - k_g^+)}\hat \psi_{q\to qg}(p_2; p_2-k_g, k_g)\\
  &\hspace{5cm} \times
  \vert q(p_2-k_g,\lambda,j )\, g(k_g,\sigma,a)\rangle_0
  + \cdots   \biggr ]\\
  \otimes & \biggl [\left(1
  - \frac{C_q(p^+_3)}{2}\right)\vert q(p_3,h_3,i_3)\rangle_0
  + \sum_{\lambda,\sigma,j ,a}2g(t^{a})_{j i_3}\int \frac{\dk_g}{2(p_3^+ - k_g^+)}\hat \psi_{q\to qg}(p_3; p_3-k_g, k_g)\\
  &\hspace{5cm} \times
  \vert q(p_3-k_g,\lambda,j )\, g(k_g,\sigma,a)\rangle_0
  + \cdots   \biggr ]\, \vert S\rangle~.
\end{split}
\ee
We have extracted the common factor $2g(t^a)_{j i_i}$ from
$\psi_{q\to qg}$ via definition $\psi_{q\to qg}(p_i; p_i- k_g,
k_g) \equiv 2g(t^a)_{j i_i}\hat \psi_{q\to qg}(p_i;
p_i- k_g, k_g)$. Note that the quark helicities $h_i$ enter
the $q \to qg$ amplitudes $\hat \psi_{q\to qg}$, see
sec.~\ref{sec:q->qg}. Also, we note that $C_q(p^+_i) \sim \mathcal{O}(g^2)$ while
$\psi_{q \to qg} \sim \mathcal{O}(g)$, and that terms of order
$\mathcal{O}(g^3)$ and higher must be dropped. Finally, the integration
over the plus momentum of the gluon extends up to the plus momentum of
the parent quark; for example, $k_g^+ < p_1^+$ in the first line, and
so on.\\

We also need to add to the r.h.s.\ of eq.~(\ref{eq:qqqg-A-proton}) the
${\cal O}(g^2)$ contributions from two-body two-quark states, where
one quark emits a gluon which is then absorbed by a second (distinct)
quark. For example, if the first quark emits and the second quark
absorbs the gluon,
\be
\begin{split}
  & \sum_{\lambda_1,\lambda_2,\sigma,j ,n,a}
  4g^2 (t^a)_{j i_1} (t^a)_{n i_2} \\
  & \int
  \frac{\dk_g}{2(p_1^+ - k_g^+)}
  \hat \psi_{q\to qg}(p_1; p_1-k_g, k_g)\,
  \frac{1}{2(p_2^+ + k_g^+)}
  \hat \psi_{qg\to q}(p_2, k_g; p_2 + k_g)\\
  &
  \vert q(p_1- k_g,\lambda_1,j)\,
  q(p_2+k_g,\lambda_2,n)\rangle_0 \otimes
  \vert q(p_3,h_3,i_3)\rangle_0\, \vert S\rangle
  ~.
\end{split}
\ee
Here, the integration over $k_g^+$ extends up to $\mathrm{min}(p_1^+,
P^+-p_2^+)$. There are analogous contributions corresponding to gluon
emission from quark 2 and absorption by quark 1 as well as from other
pairings. Since we sum over all permutations of emitter and absorber,
to avoid double counting, we should either multiply the above
expression by $\frac{1}{2}$ or else include this factor in the
symmetry factors of the corresponding diagrams. We choose the latter
option.

\subsection{Wave function normalization} \label{sec:LFwf_Norm}

We recompute $\langle K | P\rangle$ to match its normalization to
eq.~(\ref{eq:ProtonNorm1}),
\bea
\langle K | P\rangle &=& 16\pi^3 \, P^+ \delta(P^+ - K^+)
\, \delta(\Pvec - \Kvec)~.
\eea

\begin{figure}[htb]
\centerline{
  \includegraphics[width=0.22\linewidth]{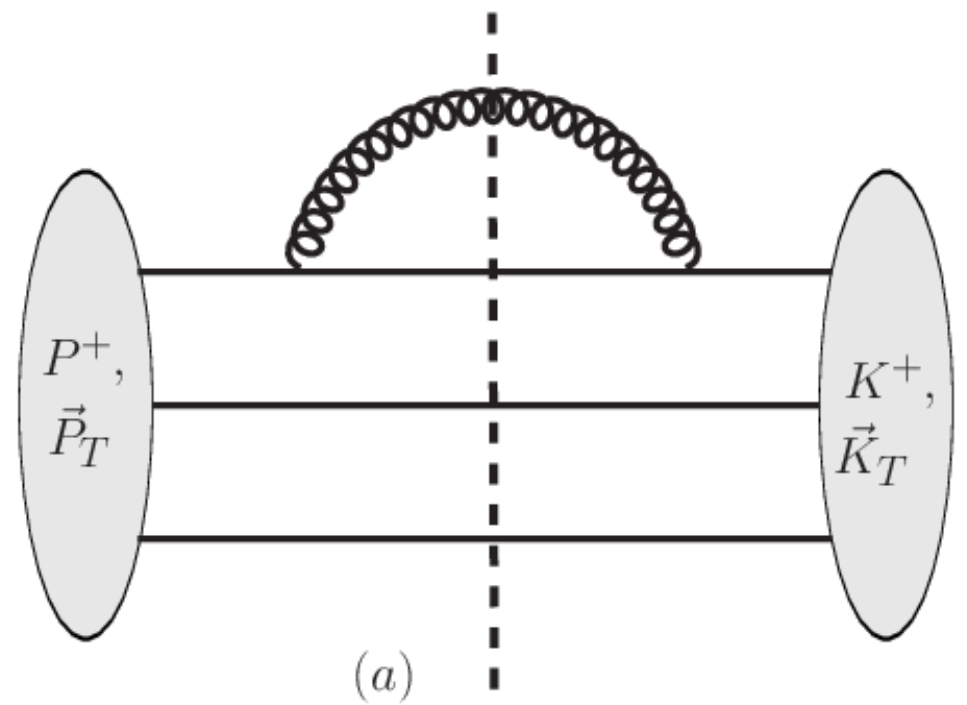}
  \hspace*{1cm}
  \includegraphics[width=0.22\linewidth]{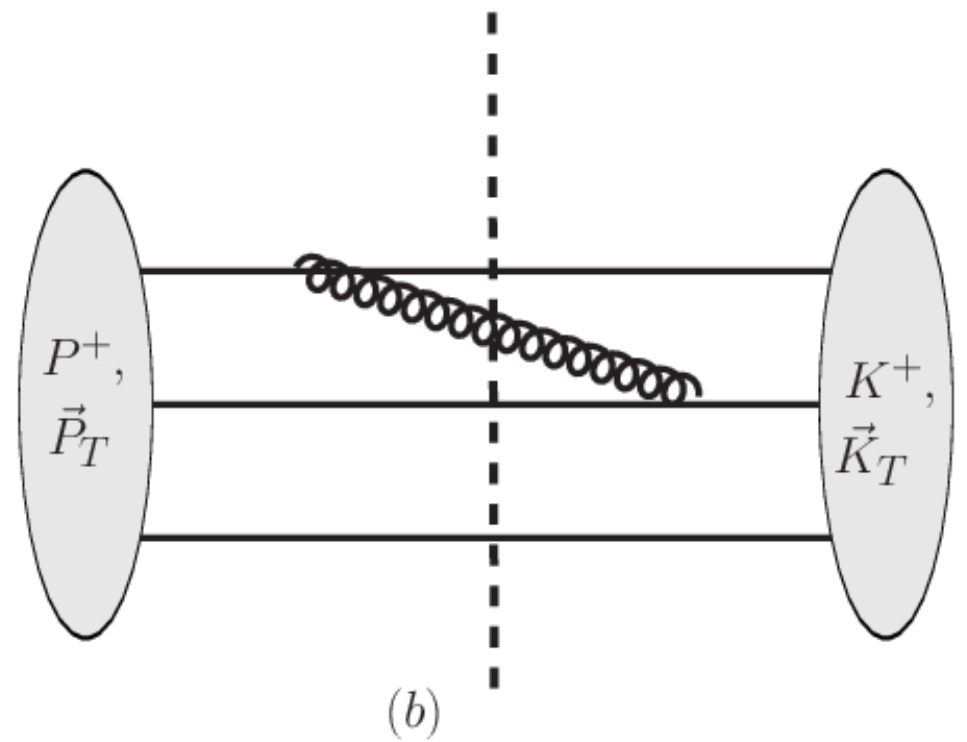}
  \hspace*{1cm}
  \includegraphics[width=0.22\linewidth]{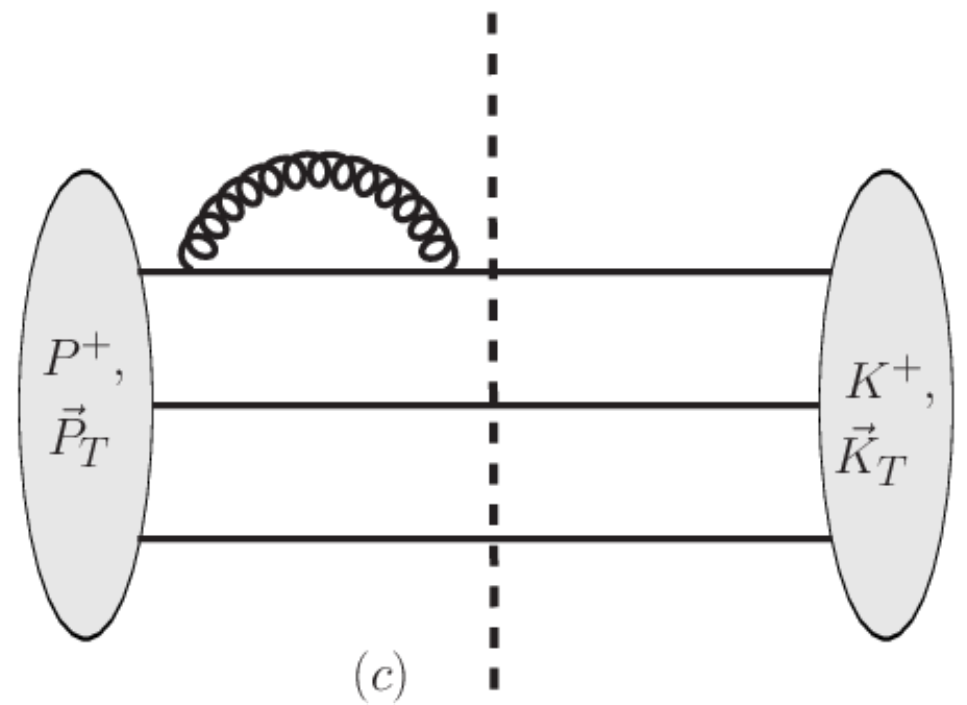}
  \hspace*{1cm}
  \includegraphics[width=0.22\linewidth]{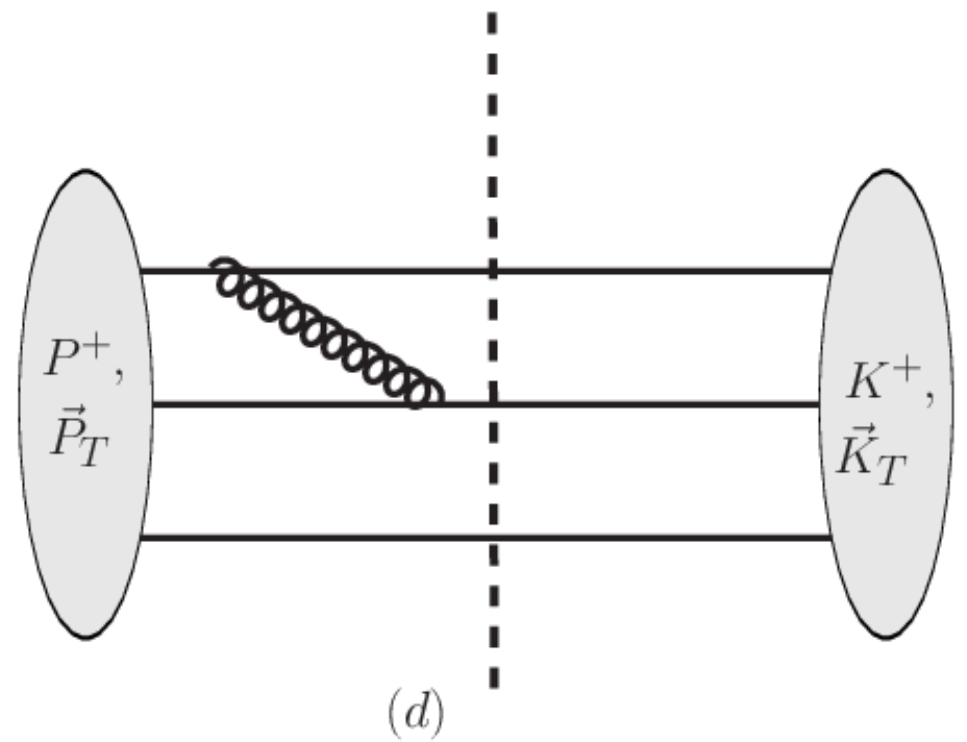}}
\caption{Diagrams for the computation of the proton wave function
  normalization at $\mathcal{O}(g^2)$. In figures the notation $\Kvec = \Kvec_T$ is used. 
}
\label{fig:Norm_g2}
\end{figure}
The $\mathcal{O}(g^2)$ corrections to $\langle K\vert P\rangle$ are depicted in
fig.~\ref{fig:Norm_g2}. For diagram~\ref{fig:Norm_g2}(a) we get
\be
\begin{split}
  \langle K | P\rangle_{\mathrm{fig}.~\ref{fig:Norm_g2}(a)} & =
(2\pi)^3\,
  P^+\delta(P^+ - K^+) \, \delta(\Pvec-\Kvec)
\int [\ud x_i] \int [\ud^2 k_i]
\,\, \frac{1}{2p_1^+}\int\dk_g\frac{1}{2(p_1^+-k_g^+)}
\left| \psi_{q\to qg}(p_1; p_1- k_g, k_g)\right|^2 \\
& \hspace{5cm}\vert \Psi_{qqq}(x_1, \kvec_1; x_2, \kvec_2; x_3, \kvec_3)\vert^2\,
\langle S|S\rangle
  \\
  & = (2\pi)^3\,
  P^+\delta(P^+ - K^+) \, \delta(\Pvec-\Kvec)
\int [\ud x_i] \int [\ud^2 k_i]
\,\, C_q(p_1^+)\, \vert \Psi_{qqq}(x_1, \kvec_1; x_2, \kvec_2; x_3, \kvec_3)\vert^2\,
\langle S|S\rangle~.
\end{split}
\label{eq:<K|P>a}
\ee
This diagram has a symmetry factor of 3 since the gluon may also be emitted
and reabsorbed by quarks 2 or 3.

For diagram~\ref{fig:Norm_g2}(c) we get
\be
\begin{split}
  \langle K | P\rangle_{\mathrm{fig}.~\ref{fig:Norm_g2}(c)} & = - (2\pi)^3\,
  P^+\delta(P^+ - K^+) \, \delta(\Pvec-\Kvec)
\int [\ud x_i] \int [\ud^2 k_i]
\,\, \frac{C_q(p_1^+)}{2}\,
\vert \Psi_{qqq}(x_1, \kvec_1; x_2, \kvec_2; x_3, \kvec_3)\vert^2\,
\langle S|S\rangle~.
\end{split}
\label{eq:<K|P>c}
\ee
This will be multiplied by a symmetry factor of 6. These two UV
divergent contributions cancel.\\

Continuing to diagram~\ref{fig:Norm_g2}(b) we find
\be
\begin{split}
  \langle K | P\rangle_{\mathrm{fig}.~\ref{fig:Norm_g2}(b)} & = - 16\pi^3\,
  P^+\delta(P^+ - K^+) \, \delta(\Pvec-\Kvec)\\
  & \frac{g^2C_F N_c}{12}
  \int [\ud x_i] \int [\ud^2 k_i]
\Psi_{qqq}(x_1, \kvec_1; x_2, \kvec_2; x_3, \kvec_3)
\int \frac{\dk_g}{p_1^+ - k_g^+} \frac{1}{p_2^+ + k_g^+} \\
& \Psi^*_{qqq}(x_1-x_g, \kvec_1-\kvec_g + x_g\Pvec; x_2+x_g,
\kvec_2+\kvec_g-x_g\Pvec; x_3, \kvec_3) \\
&
\langle S\vert\hat \psi_{q\to qg}(p_1; p_1-k_g, k_g)\,
\hat \psi_{q\to qg}^*(p_2 + k_g; p_2, k_g)\,
\vert S\rangle~.
\end{split}
\label{eq:<K|P>b-interm}
\ee
Note that the summations over the polarization of the gluon and the helicities
  of the daughter quarks are not indicated; the helicities $h_1, h_2$
  of the parent quarks which appear in the $q\to q g$ amplitudes are
  those from $\vert S\rangle$. Also, there is an upper limit for the
  integration over $k_g^+$ which is given by min($p_1^+,P^+-p_2^+$).

The integral over $\kvec_g$ converges in the UV because it shifts the
arguments of $ \Psi^*_{qqq}$. Therefore, we can immediately insert the
$D=4$ form of $\psi_{q\to qg}$ from eq.~(\ref{eq:qqgLCWFfull4D}):
\be
\begin{split}
\frac{\hat \psi_{q\to qg}(p_1; p_1-k_g, k_g)\, 
  \hat \psi_{q\to qg}^*(p_2 + k_g; p_2, k_g)}
     {4 p_1^+ (p_2^+ + k_g^+) \sqrt{1-z_1}\sqrt{1-z_2}}
     & =
\left[\left(1-\frac{z_1}{2}\right)\delta^{ij} + \frac{z_1}{2} ih_1\epsilon^{ij}\right]
\left[\left(1-\frac{z_2}{2}\right)\delta^{kl} - \frac{z_2}{2} ih_2\epsilon^{kl}\right]\delta^{jl}\\
& ~~~
\, \frac{(k_g^i-z_1p_1^i)}{(\kvec_g-z_1\pvec_1)^2}
\frac{((1-z_2)k_g^k-z_2 p_2^k)}{((1-z_2)\kvec_g-z_2\pvec_2)^2}\\
& \hspace{-3cm}
= \left[ \left(\left(1-\frac{z_1}{2}\right)\left(1-\frac{z_2}{2}\right)
  +\frac{z_1z_2}{4} h_1 h_2 \right) \delta^{ik} +
  \left(ih_1\frac{z_1}{2}\left(1-\frac{z_2}{2}\right) -
  ih_2\frac{z_2}{2}\left(1-\frac{z_1}{2}\right)\right) \epsilon^{ik}\right]\\
& ~~~~
\frac{(k_g^i-z_1p_1^i)}{(\kvec_g-z_1\pvec_1)^2}
\frac{((1-z_2)k_g^k-z_2p_2^k)}{((1-z_2)\kvec_g-z_2\pvec_2)^2}
~,
\end{split}
\ee
where $z_1=\frac{x_g}{x_1}$ and $z_2=\frac{x_g}{x_2+x_g}$.  We now
have to sandwich this expression between proton helicity states as given
in eq.~(\ref{eq:HelicityMatrixElement}). Note that $h_1=\uparrow$ and
$h_1=\downarrow$ occur an equal number of times (same for $h_2$) so
that terms linear in helicity drop out, while $\langle S\vert h_1
h_2\vert S\rangle = -\frac{1}{3}$ (incl.\ symmetrization over
permutations of quark helicities, c.f.\ sec.~\ref{sec:SpinWF}). Hence,
\be \label{eq:<S|psi-psi*|S>}
\begin{split}
\langle S\vert\hat \psi_{q\to qg}(p_1; p_1-k_g, k_g)\,
\hat \psi_{q\to qg}^*(p_2 + k_g; p_2, k_g)\,
\vert S\rangle &=
4 p_1^+ (p_2^+ + k_g^+) \sqrt{1-z_1}\sqrt{1-z_2}
\left(1-\frac{z_1+z_2}{2}
+\frac{z_1z_2}{6} \right) \\
& ~~~
\frac{(\kvec_g-z_1\pvec_1)}{(\kvec_g-z_1\pvec_1)^2} \cdot
\frac{((1-z_2)\kvec_g-z_2\pvec_2)}{((1-z_2)\kvec_g-z_2\pvec_2)^2}~.
\end{split}
\ee
With this we finally obtain
\be
\begin{split}
  \langle K | P\rangle_{\mathrm{fig}.~\ref{fig:Norm_g2}(b)} & = - 16\pi^3\,
  P^+\delta(P^+ - K^+) \, \delta(\Pvec-\Kvec)\\
  & g^2C_F \int [\ud x_i] \int [\ud^2 k_i]
\Psi_{qqq}(x_1, \kvec_1; x_2, \kvec_2; x_3, \kvec_3)
\int\limits^{\mathrm{min}(x_1,1-x_2)}_x
\frac{\ud x_g}{x_g} \frac{x_1}{x_1 - x_g} \int\frac{\ud^2 k_g}{16\pi^3}\\
& \Psi^*_{qqq}(x_1-x_g, \kvec_1-\kvec_g + x_g\Pvec; x_2+x_g,
\kvec_2 + \kvec_g - x_g\Pvec; x_3, \kvec_3) \\
&
\sqrt{1-z_1}\sqrt{1-z_2}
\left(1-\frac{z_1+z_2}{2}
+\frac{z_1z_2}{6} \right)\,
\frac{(\kvec_g-z_1\pvec_1)}{(\kvec_g-z_1\pvec_1)^2} \cdot
\frac{((1-z_2)\kvec_g-z_2\pvec_2)}{((1-z_2)\kvec_g-z_2\pvec_2)^2}
~.
\end{split}
\label{eq:<K|P>b}
\ee
Recall that $z_1=\frac{x_g}{x_1}$, $z_2=\frac{x_g}{x_2+x_g}$ and
$\pvec_{1,2} = \kvec_{1,2} + x_{1,2} \Pvec$; the shift $\kvec_g \to \kvec_g +
x_g \Pvec$ makes this expression independent of $\Pvec$. Here, $x$ is the
minimal allowed LC momentum fraction of the gluon, i.e.\ in subsequent
sections we will evaluate correlators of color charges with LC momenta
greater than $xP^+$.  In eq.~(\ref{eq:<K|P>b}) the first quark emits
and the second quark absorbs the gluon. By symmetry of the wave
function $\Psi_{qqq}$ under exchange of the quarks, reversing emission
and absorption leads to the same result.  Also, thanks to the fact
that we have averaged over permutations of the helicities of the three
quarks, in all we can simply multiply this diagram by a symmetry
factor of 6 to include the contributions where quarks 1 and 3 or
quarks 2 and 3 exchange the gluon.\\

Lastly,
\be
\begin{split}
  \langle K | P\rangle_{\mathrm{fig}.~\ref{fig:Norm_g2}(d)} & = - 16\pi^3\,
  P^+\delta(P^+ - K^+) \, \delta(\Pvec-\Kvec)\\
  & \frac{g^2C_F N_c}{12}
  \int [\ud x_i] \int [\ud^2 k_i]
\Psi_{qqq}(x_1, \kvec_1; x_2, \kvec_2; x_3, \kvec_3)
\int \frac{\dk_g}{p_1^+ - k_g^+} \frac{1}{p_2^+ + k_g^+} \\
& \Psi^*_{qqq}(x_1-x_g, \kvec_1-\kvec_g + x_g\Pvec; x_2+x_g,
\kvec_2 + \kvec_g - x_g\Pvec; x_3, \kvec_3) \\
&
\langle S\vert\hat \psi_{q\to qg}(p_1; p_1-k_g, k_g)\,
\hat \psi_{qg\to q}(p_2, k_g; p_2 + k_g)\,
\vert S\rangle~.
\end{split}
\label{eq:<K|P>d-interm}
\ee
Here, the product
\be
\begin{split}
\langle S\vert\hat \psi_{q\to qg}(p_1; p_1-k_g, k_g)\,
\hat \psi_{qg\to q}(p_2, k_g ; p_2 + k_g)\,
\vert S\rangle &=
- 4 p_1^+ (p_2^+ + k_g^+) \sqrt{1-z_1}\sqrt{1-z_2}
\left(1-\frac{z_1+z_2}{2}
+\frac{z_1z_2}{6} \right) \\
& ~~~
\frac{(\kvec_g-z_1\pvec_1)}{(\kvec_g-z_1\pvec_1)^2} \cdot
\frac{((1-z_2)\kvec_g-z_2\pvec_2)}{((1-z_2)\kvec_g-z_2\pvec_2)^2}~,
\end{split}
\ee
where $z_1 = k_g^+/p_1^+ = x_g/x_1$ and $z_2 = k_g^+/(p_2^+ + k_g^+) =
x_g/(x_2+x_g)$.  Therefore, we find the result
\be
\begin{split}
  \langle K | P\rangle_{\mathrm{fig}.~\ref{fig:Norm_g2}(d)} & = 16\pi^3\,
  P^+\delta(P^+ - K^+) \, \delta(\Pvec-\Kvec)\\
  & g^2C_F
  \int [\ud x_i] \int [\ud^2 k_i]
\Psi_{qqq}(x_1, \kvec_1; x_2, \kvec_2; x_3, \kvec_3)
\int\limits^{\mathrm{min}(x_1,1-x_2)}_x
\frac{\ud x_g}{x_g} \frac{x_1}{x_1 - x_g}
\int\frac{\ud^2 k_g}{16\pi^3}\\
& \Psi^*_{qqq}(x_1-x_g, \kvec_1 - \kvec_g + x_g\Pvec; x_2+x_g,
\kvec_2 + \kvec_g - x_g\Pvec; x_3, \kvec_3) \\
&
\sqrt{1-z_1}\sqrt{1-z_2}
\left(1-\frac{z_1+z_2}{2}
+\frac{z_1z_2}{6} \right)
\frac{(\kvec_g-z_1\pvec_1)}{(\kvec_g-z_1\pvec_1)^2} \cdot
\frac{((1-z_2)\kvec_g-z_2\pvec_2)}{((1-z_2)\kvec_g-z_2\pvec_2)^2}
~.
\end{split}
\label{eq:<K|P>d}
\ee
The shift $\kvec_g \to \kvec_g + x_g \Pvec$ again
shows that this expression is in fact independent of $\Pvec$. 

The symmetry factor for this diagram is
$\frac{1}{2}\times12=6$. Hence, the diagrams in
fig.~\ref{fig:Norm_g2}(d), eq.~(\ref{eq:<K|P>d}), and in
fig.~\ref{fig:Norm_g2}(b), eq.~(\ref{eq:<K|P>b}), also cancel. The
normalization condition for $\Psi_{qqq}$ therefore remains as written
in eq.~(\ref{eq:Norm_psi3}).

\section{Color charge correlators}

\subsection{Color charge operators}
\label{sec:ChargeOperators}

The color charge operator $\rho^a(\vec x)$ measures the color charge
density at transverse coordinate $\vec x$, integrated over the
longitudinal coordinate $x^-$. Its 2d Fourier transform is
$\rho^a(\vec k)$. The contribution to this Fock space operator due to
quarks is given by the plus component of their color current $g (t^a)_{ij}
\bar\psi_i \gamma^+ \psi_j$~\cite{Dumitru:2018vpr},
\be
\rho^a_\mathrm{qu}(\vec k) = g \sum_{i,j,\sigma} (t^a)_{ij}
\int\frac{\dd x_q \dd^2q}{16\pi^3\, x_q}
 b^\dagger_{i\sigma}(x_q,\vec q)\, b_{j\sigma}(x_q,\vec k+\vec q) ~,
  \label{eq:rho_q} 
\ee
where $t^a$ are the generators of the fundamental representation of
color-SU(3).
\\

The contribution of gluons to the color current in light-cone gauge is
(see, for example, ref.~\cite{Burkardt:2003yg})
\be
J^{+a}_\mathrm{gl}(x) = g f^{abc} A^{ib}(x)\, \partial^+ A^{ic}(x)~,
\ee
with $\partial^+ = \partial_- = \frac{\partial}{\partial x^-}$. This
follows from the quadratic in $A^i$ part of $(D_\nu)^{ab} \,
(F^{\nu+})^b = J^{+a}_{\mathrm{qu}}$ which we shift to the r.h.s. of
the equation. Using $(D_\nu)_{ab} = \delta_{ab}\partial_\nu + ig
A_\nu^c (T^c)_{ab}$ and $F^{-+}_b = 0$ we obtain $J^{+a}_\mathrm{gl}(x)
= ig A^{ic}(T^c)^{ab} F^{i+b}$ with $F^{i+}_b = -\partial_- A^i_b$.
$(T^c)^{ab}=-if^{cab}$ are the generators of the adjoint
representation of color-SU(3).

Next we introduce the standard plane wave
expansion of the bare gluon field in LC gauge on the $x^+=0$ front:
\be
A^{ia}(x^-,\vec x) = \sum_{\lambda} \int \frac{\dd k^+ \dd^2k}{16\pi^3\, k^+}
\left[\epsilon^i_\lambda a_{a\lambda}(k) e^{-i k\cdot x} +
  \epsilon^{*i}_\lambda a^\dagger_{a\lambda}(k) e^{i k\cdot x}
\right] \Bigg|_{x^+=0} ~.
\ee
Integrating $J^+$ over $x^-$ leads to
\be
\rho_\mathrm{gl}^a(\vec x) = 2\pi i g f^{abc} \sum_{\lambda b c}
\int \frac{\dd k^+ \dd^2k}{16\pi^3\, k^+}
\int \frac{\dd \ell^+ \dd^2\ell}{16\pi^3\, \ell^+} \ell^+\delta(k^+-\ell^+)
\left[
a_{b\lambda}(k) \, a^\dagger_{c\lambda}(\ell) e^{i (\vec k - \vec
    \ell)\cdot \vec x} -
  a^\dagger_{b\lambda}(k) \, a_{c\lambda}(\ell) e^{-i (\vec k - \vec
    \ell)\cdot \vec x}
  \right]~.
\ee
Here, $a_{b\lambda}(k) \, a^\dagger_{c\lambda}(\ell)$ can be replaced
  by $a^\dagger_{c\lambda}(\ell)\, a_{b\lambda}(k)$ because the
    commutator is proportional to $\delta_{bc}$, eq.~(\ref{eq:acomm}).
Finally, performing a Fourier transform to transverse momentum space
we find
\be \label{eq:rho_gl}
\rho_\mathrm{gl}^a(\vec k) = g  \sum_{\lambda b c} (T^a)_{bc}
\int \frac{\dd x_g \dd^2q}{16\pi^3\, x_g}
a^\dagger_{b\lambda}(x_g,\vec q) \, a_{c\lambda}(x_g,\vec q + \vec k)~.
\ee
\\

The eikonal currents formally sit at $x=0$ as we have integrated over
$x^-$. They sum up the charge of all particles with LC momenta
$>0$. However, as already discussed previously in
sec.~\ref{sec:LFwf_Norm}, the integral over the L.C.\ momentum
fraction $x_g$ of the gluon diverges at $x_g\to0$, and so we introduce
a cut off $x$ to exclude gluon fluctuations with $x_g < x$. Hence, our
proton does not ``contain'' any gluons below $x$ so that, in effect,
the correlator $\langle \rho^a\, \rho^b\rangle$ which we compute below
excludes contributions from softer gluons.  It would make no
difference in our analysis if our $\rho^a(q)$ sat at non-zero $q^+$,
as long as this $q^+$ is less than $x P^+$.  In other words, we work
in the ``shockwave'' limit where there exists a separation of scales
such that the plus momentum of the gluon fluctuations in the proton
exceeds that of the probes corresponding to the $\rho^a$ charge
operators. In practical applications, one would choose the cutoff $x$
to correspond to the L.C.\ momentum fraction of color charges probed
by the kinematics of the process.

\subsection{Correlator of two color charge operators, $\langle
  \rho^a(\vec q_1)\,\rho^b(\vec q_2) \rangle$}

In this section we compute color charge correlations for two external
probes\footnote{The expectation value of a single charge operator is
  proportional to the trace of a generator of color-SU(3), in either
  the fundamental or adjoint representation.}, $\langle\rho^a(\vec q_1)\,
\rho^b(\vec q_2)\rangle$. Since $\rho^a(\vec q) = \rho^a_{\mathrm{qu}}(\vec q) +
\rho^a_{\mathrm{gl}}(\vec q)$ we have
\be
\langle\rho^a(\vec q_1)\, \rho^b(\vec q_2)\rangle =
\langle\rho^a_{\mathrm{gl}}(\vec q_1)\, \rho^b_{\mathrm{gl}}(\vec q_2)\rangle
+
\langle\rho^a_{\mathrm{gl}}(\vec q_1)\, \rho^b_{\mathrm{qu}}(\vec q_2)\rangle
+
\langle\rho^a_{\mathrm{qu}}(\vec q_1)\, \rho^b_{\mathrm{gl}}(\vec q_2)\rangle
+
\langle\rho^a_{\mathrm{qu}}(\vec q_1)\, \rho^b_{\mathrm{qu}}(\vec q_2)\rangle
~.
\ee
We define expectation
values of products of color charge operators by stripping off the
delta functions for conservation of LC and transverse momentum:
\be
\left< K|\rho^{a_1}(\vec q_1)\cdots \rho^{a_n}(\vec
q_n)|P\right> = 16\pi^3 \, P^+ \delta(P^+ - K^+)\,\, \delta\left(\vec P -\vec
K - \sum_{i=1}^{n} \vec q_i\right)\, \left< \rho^{a_1}(\vec q_1)\cdots
\rho^{a_n}(\vec q_n) \right>.
\label{eq:<K|..|P>}
\ee
It is understood that the color charge correlators correspond to a
transverse momentum of the scattered proton of $\vec K = \vec P -
\sum \vec q_i$ and light-cone momentum $K^+ = P^+$.
We will also abbreviate $\vec q = \sum_{i=1}^{n} \vec q_i$.
\\

\subsubsection{Coupling to gluon, $\langle \rho^a_\mathrm{gl}(\vec q_1)\,
\rho^b_\mathrm{gl}(\vec q_2)\rangle$}

We begin with the diagrams where both external probes couple to the
gluon in the proton. (We amputate the propagators of the external gluons
in the following diagrams to obtain the expectation value of the color
charge correlator.)
\begin{figure}[htb]
\centerline{
  \includegraphics[width=0.35\linewidth]{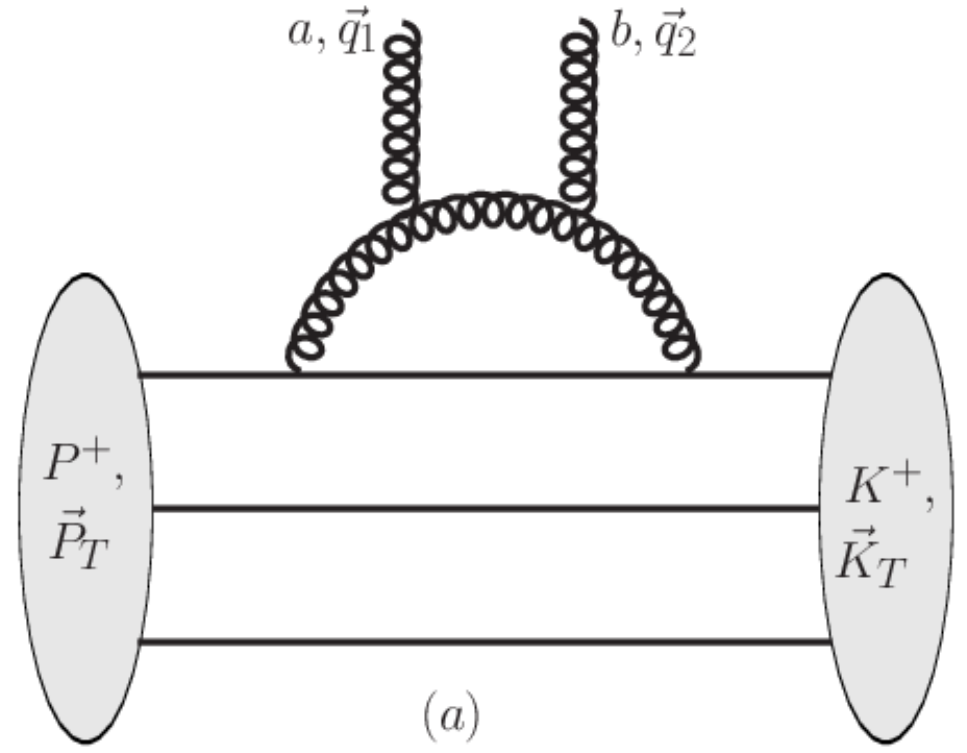}
    \hspace*{.6cm}
  \includegraphics[width=0.35\linewidth]{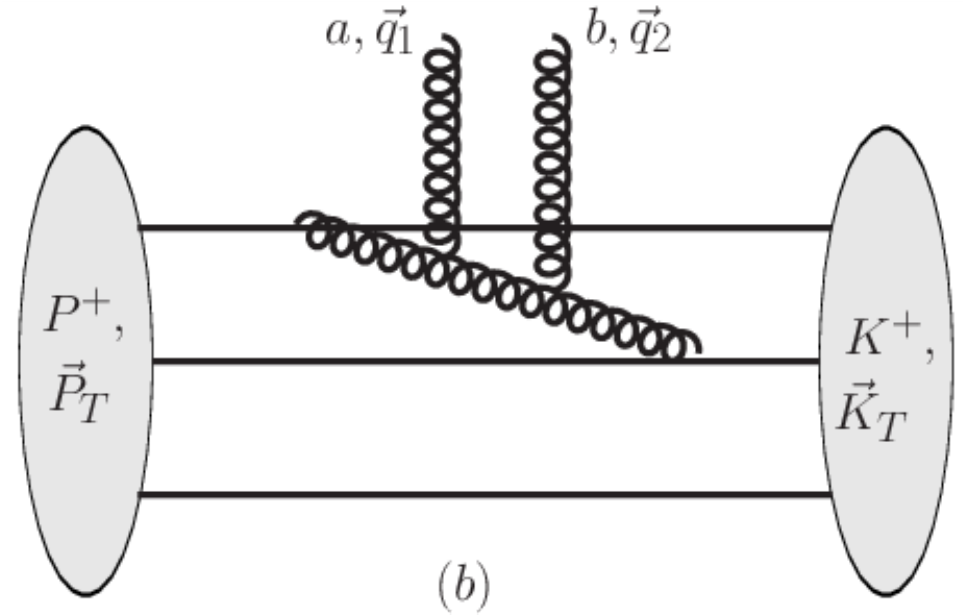}
  }
\caption{Diagrams for $\langle \rho^b_\mathrm{gl}(\vec q_2)\,
  \rho^a_\mathrm{gl}(\vec q_1)\rangle$. The cut is located at the
  insertion of the two color charge operators.}
\label{fig:rho-rho_gl}
\end{figure}

To prepare, we first compute the matrix element of
$\rho^a_\mathrm{gl}(\vec q_1)\, \rho^b_\mathrm{gl}(\vec q_2)$ between
one-gluon states:
\be \label{eq:<g|rho_gl-rho_gl|g>}
\left< \ell, \rho, d \vert\, \rho^a_\mathrm{gl}(\vec q_1)\,
\rho^b_\mathrm{gl}(\vec q_2)\, \vert k_g, \sigma, c\right> = g^2
\left(T^a T^b\right)_{d c}\, \delta_{\rho\sigma}\,
(2\pi)^{D-1}\, 2k_g^+ \delta(k_g^+ -
\ell^+)\, \delta(\vec k_g - \vec \ell -\vec q)~.
\ee
Here we used the color charge operator from eq.~(\ref{eq:rho_gl})
and the commutation relations~(\ref{eq:acomm}).
\\

We now proceed to compute fig.\ref{fig:rho-rho_gl}a which reads
\bea
\left<K \vert \rho^a_\mathrm{gl}(\vec q_1)\,
\rho^b_\mathrm{gl}(\vec q_2) \vert P\right> &=& \frac{4g^2}{6}
\int[\dd x_i] \int[\dd^2 k_i]\, \Psi_{qqq}(x_i,\vec k_i)\,
\epsilon_{i_1 i_2 i_3} \int[\dd x_i'] \int[\dd^2 k_i']\,
\Psi_{qqq}^*(x_i',\vec k_i') \, \epsilon_{i_1' i_2' i_3'}\, (t^c)_{j
  i_1}(t^d)_{i_1' j'} \nn\\ & & \int \frac{\dk_g}{2(p_1^+ - k_g^+)}
\int \frac{\dk_g'}{2(p_1^{+\prime} - k_g^{+\prime})}
\langle S\vert \hat\psi_{q\to qg}(p_1; p_1-k_g, k_g)  \hat\psi_{q\to
  qg}^*(p_1'; p_1'-k_g', k_g') \vert S\rangle   \nn\\
& & \langle p_2',i_2' | p_2,
i_2\rangle\, \langle p_3',i_3' | p_3, i_3\rangle\, \langle
p_1'-k_g',j' | p_1-k_g,j\rangle\, \left< k_g',d \vert\,
\rho^a_\mathrm{gl}(\vec q_1)\, \rho^b_\mathrm{gl}(\vec q_2)\, \vert
k_g, c\right>~.
\eea
Here, $p_i$ are the momenta of the quarks in
$|P\rangle$ (with $p_i^+ = x_i P^+$ and $\vec p_i = \vec k_i + x_i
\vec P$), and $p_i'$ those of the quarks in $\langle K|$ (with
$p_i^{+\prime} = x_i' K^+$ and $\vec p_i' = \vec k_i' + x_i' \vec K$).

If we now evaluate the quark state overlaps and insert the
result~(\ref{eq:<g|rho_gl-rho_gl|g>}) we obtain
\bea
\mathrm{fig.}\, \ref{fig:rho-rho_gl}\mathrm{a} &=&
\frac{2g^4}{3\cdot 16\pi^3} \,\tr T^a T^b
\int \left[\dd x_i\right]
  \int \left[\dd^2 k_i\right]\,
  \Psi_{qqq}(x_1,\vec k_1; x_2,\vec k_2; x_3,\vec k_3)\,
\nn\\
& &     \Psi_{qqq}^*(x_1,\vec k_1-(1-x_1)\vec q;
x_2,\vec k_2 +x_2\vec q;
x_3,\vec k_3 +x_3\vec q) \nn\\
& &
\frac{(2\pi)^{D-1}}{2p_1^+} \int \frac{\dk_g}{2(p_1^+ - k_g^+)}
\langle S| \hat\psi_{q\to qg}(p_1; p_1-k_g, k_g)\,
\hat\psi^*_{q\to qg}(p_1 - q; p_1 - k_g,
k_g - q) |S\rangle~,
     \label{eq:rho-rho_gl-gl-a}
\eea
where $\tr T^a T^b = N_c\, \delta^{ab}$. Also, recall that the plus
component of $\vec q_1, \vec q_2$ is zero. The symmetry factor for
this diagram is 3.  The expression for the integral of $\hat\psi_{q\to
  qg}\, \hat\psi^*_{q\to qg}$ over $k_g$ is given in
eq.~(\ref{eq:Int-psi-psi*}) above.\\

For fig.\ref{fig:rho-rho_gl}b we get
\bea
\left<K \vert \rho^a_\mathrm{gl}(\vec q_1)\,
\rho^b_\mathrm{gl}(\vec q_2) \vert P\right>
 &=&
\frac{4g^2}{6}
\int[\dd x_i] \int[\dd^2 k_i]\, \Psi_{qqq}(x_i,\vec k_i)
\epsilon_{i_1 i_2 i_3} \int[\dd x_i'] \int[\dd^2 k_i']\,
\Psi_{qqq}^*(x_i',\vec k_i') \epsilon_{i_1' i_2' i_3'} \nn\\
& & \int \frac{\dk_g}{2(p_1^+ - k_g^+)}
\frac{\dk_g'}{2(p_2^{+\prime} - k_g^{+\prime})}
\langle S\vert \hat\psi_{q\to qg}(p_1; p_1-k_g, k_g)
\hat\psi_{q\to qg}^*(p_2'; p_2'-k_g', k_g') \vert S\rangle\nn\\
& &
\langle p_3',i_3' | p_3, i_3\rangle\, \langle
p_1',i_1' | p_1-k_g,j_1\rangle\,
\langle p_2'-k_g',j_2 | p_2, i_2\rangle \nn\\
& & 
\left< k_g',d \vert\,
\rho^a_\mathrm{gl}(\vec q_1)\, \rho^b_\mathrm{gl}(\vec q_2)\, \vert
k_g, c\right> \, (t^c)_{j_1 i_1}\, (t^d)_{i_2' j_2} \\
&=&
-\frac{g^4}{6} \mathrm{tr}\, T^a T^b\,
P^+\delta(P^+ - K^+)\, \delta(\vec P - \vec K -\vec q) \nn\\
& &
\int[\dd x_i] \int[\dd^2 k_i]\, \Psi_{qqq}(x_i,\vec k_i)
\int\frac{\dd k_g^+}{k_g^+} \frac{\dd^2 k_g}{2(p_1^+-k_g^+)} \nn\\
& &
\langle S\vert \hat\psi_{q\to qg}(p_1; p_1-k_g, k_g)
\hat\psi_{q\to qg}^*(p_2+k_g-q; p_2, k_g-q) \vert S\rangle
\, \frac{x_2}{x_2+x_g} \frac{1}{2p_2^+} \nn\\
& &
\Psi_{qqq}^*(x_1-x_g,\vec k_1+x_1\vec q - \vec k_g + x_g\vec K;
x_2+x_g,\vec k_2 -(1-x_2)\vec q + \vec k_g - x_g\vec K;
x_3,\vec k_3 +x_3\vec q)~.
\eea
We again evaluate the helicity matrix element of $\hat\psi_{q\to qg}
\hat\psi_{q\to qg}^*$ as in eq.~(\ref{eq:<S|psi-psi*|S>}).  This leads
to the finite result
\bea
\mathrm{fig.}\, \ref{fig:rho-rho_gl}\mathrm{b}
& = &
-\frac{g^4}{6\cdot 16\pi^3} \,\tr T^a T^b
\int \left[\dd x_i\right]
  \int \left[\dd^2 k_i\right]\,
  \Psi_{qqq}(x_1,\vec k_1; x_2,\vec k_2; x_3,\vec k_3)\,
\nn\\
& &
\int\limits_x^{\mathrm{min}(x_1,1-x_2)}\frac{\dd x_g}{x_g} \int\dd^2 k_g
\frac{z_1\vec p_1-\vec k_g}
     {\left(z_1\vec p_1-\vec k_g \right)^2}
\cdot
\frac{z_2\vec p_2 - (1-z_2)(\vec k_g -\vec q_1-\vec q_2)}
  {\left(z_2\vec p_2 - (1-z_2)(\vec k_g -\vec q_1-\vec q_2)\right)^2}
  \nn\\
  &  & \sqrt{\frac{x_1}{x_1-x_g}}\sqrt{\frac{x_2}{x_2+x_g}}
  \left(1-\frac{z_1+z_2}{2}+\frac{z_1 z_2}{6}  \right) \nn\\
  & &
  \Psi_{qqq}^*(x_1-x_g,\vec k_1+x_1\vec q - \vec k_g + x_g\vec K;
x_2+x_g,\vec k_2 -(1-x_2)\vec q + \vec k_g - x_g\vec K;
x_3,\vec k_3 +x_3\vec q)~,
     \label{eq:rho-rho_gl-gl-b}
\eea
with $z_1 = x_g / x_1$ and $z_2 = x_g/(x_2+x_g)$.  Here, the symmetry
factor is 6 which includes a factor of 2 for interchanging the gluon
emission and absorption vertices between quarks 1 and 2.  Note that
this expression is invariant under translations in 2d transverse
momentum space corresponding to a constant shift of both $\vec P$ and
$\vec K$; this is evident upon shifting the integration variable $\vec
k_g \to \vec k_g + x_g \vec K$.

\subsubsection{Coupling to one quark and the gluon, $\langle
  \rho^a_\mathrm{qu}(\vec q_1)\, \rho^b_\mathrm{gl}(\vec q_2)\rangle$}

In this section we compute $\langle \rho^a_\mathrm{qu}(\vec q_1)\,
\rho^b_\mathrm{gl}(\vec q_2)\rangle$. We can then obtain $\langle
\rho^a_\mathrm{gl}(\vec q_1)\, \rho^b_\mathrm{qu}(\vec q_2)\rangle$
simply by exchanging $a\leftrightarrow b$, $\vec q_1 \leftrightarrow
\vec q_2$ since the two charge operators commute.
\begin{figure}[ht]
\centering
\begin{minipage}[b]{\linewidth}
\includegraphics[width=0.28\linewidth]{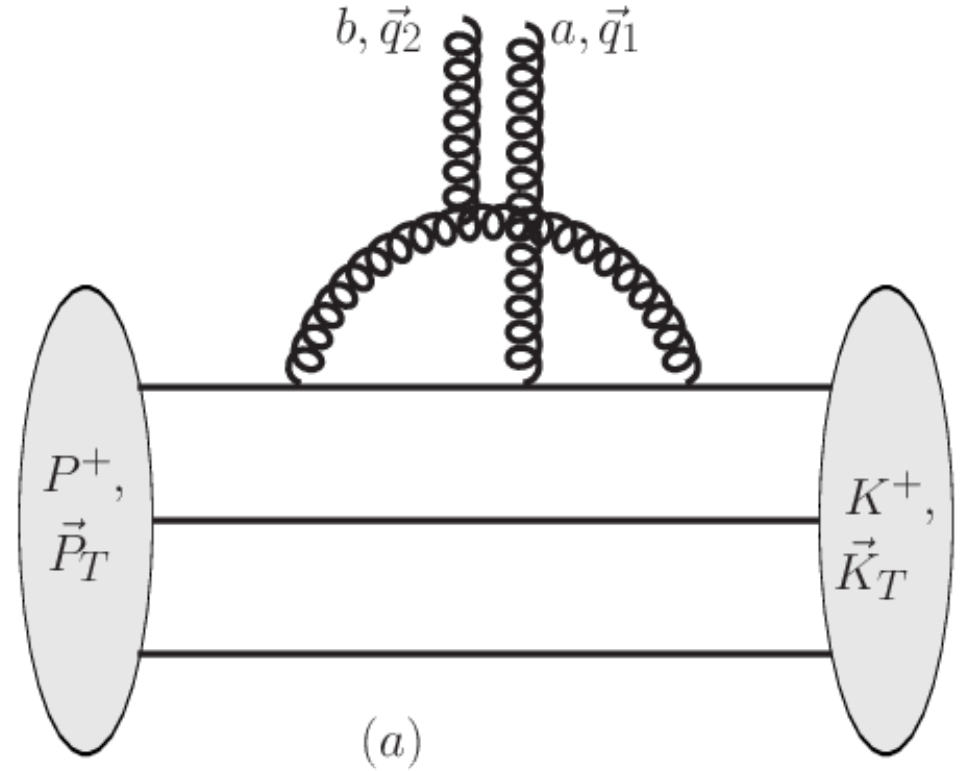}
\hspace*{.4cm}
\includegraphics[width=0.28\linewidth]{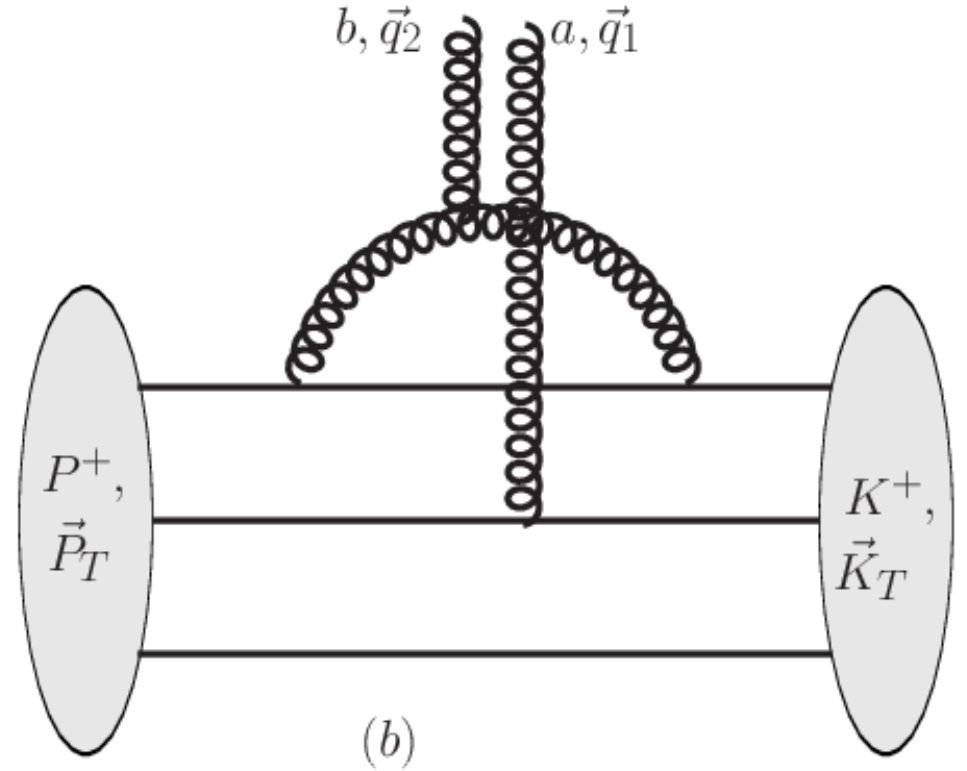}
\end{minipage}
\begin{minipage}[b]{\linewidth}
\includegraphics[width=0.28\linewidth]{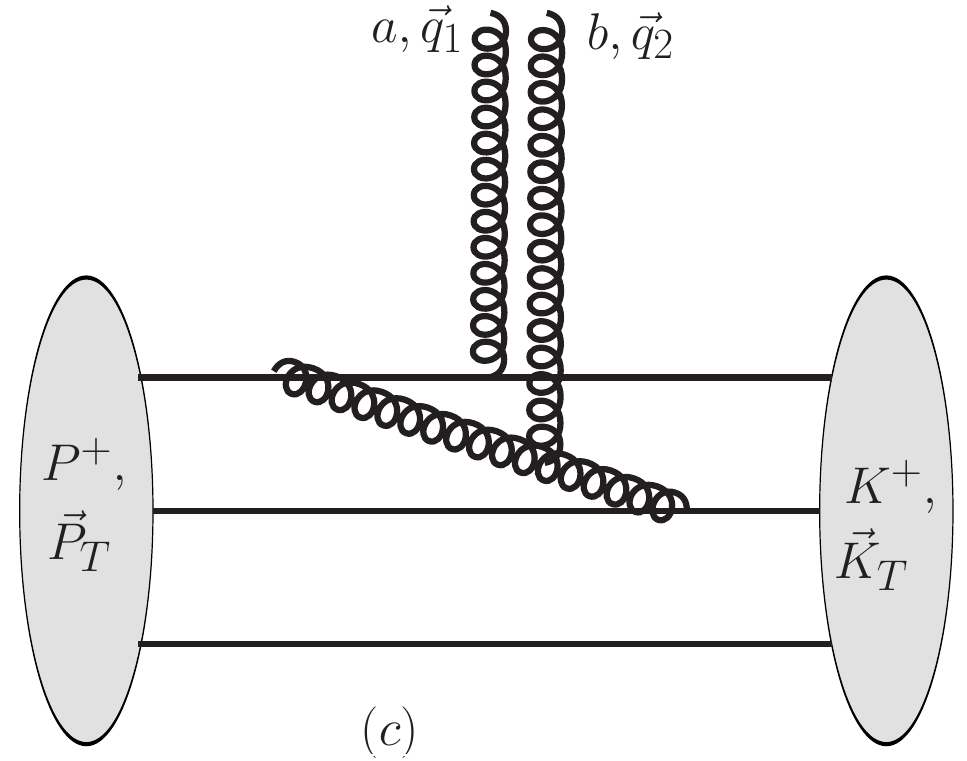}
\hspace*{.4cm}
\includegraphics[width=0.28\linewidth]{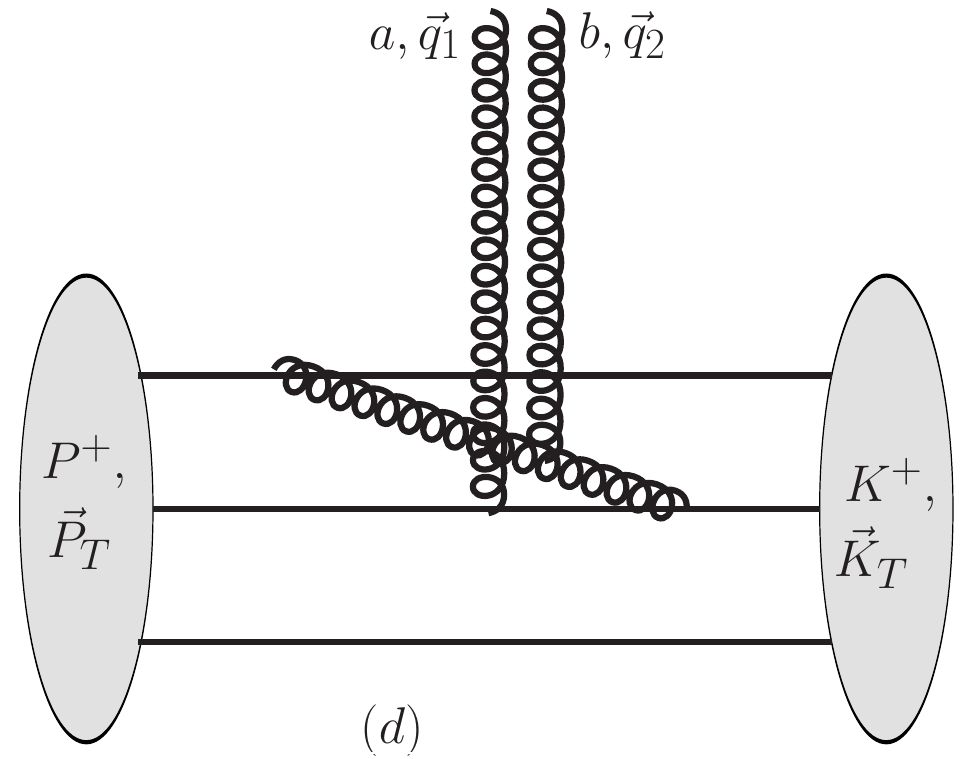}
\hspace*{.4cm}
\includegraphics[width=0.28\linewidth]{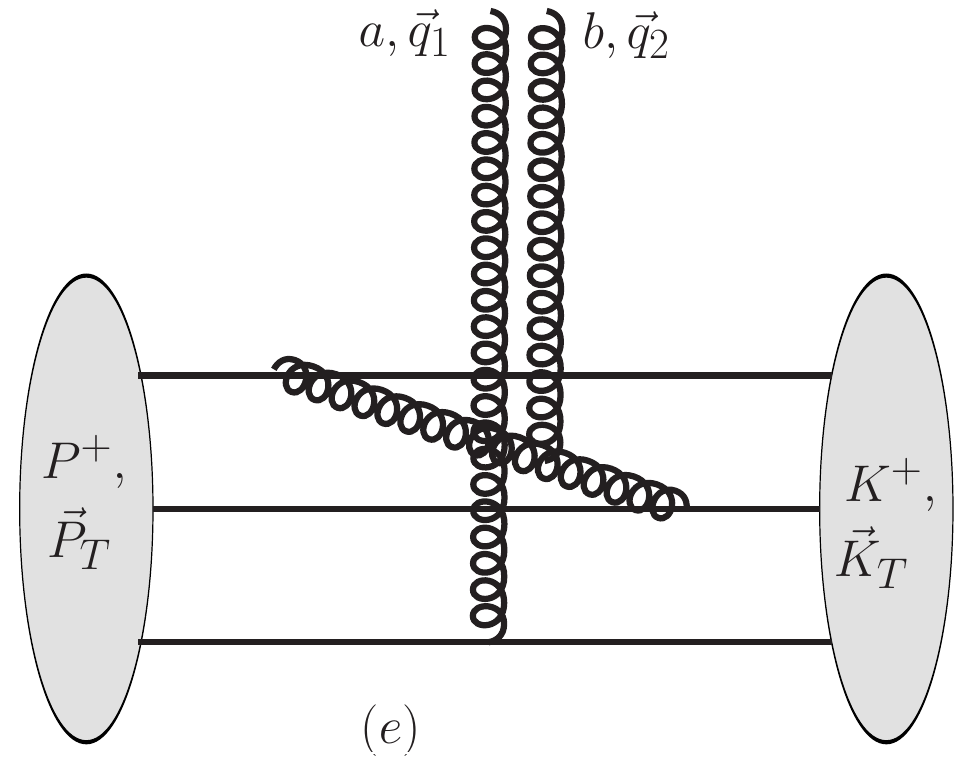}
\caption{Diagrams for $\langle \rho^a_\mathrm{qu}(\vec q_1)\,
  \rho^b_\mathrm{gl}(\vec q_2)\rangle$.}
\label{fig:rho-rho_qu-gl}
\end{minipage}
\end{figure}

To prepare this calculation, we first list the matrix elements of
$\rho^a_\mathrm{qu}(\vec q)$ and $\rho^a_\mathrm{gl}(\vec q)$ between one quark
and one gluon states, respectively:
\be
\begin{split}
  \left< k, h', i' \vert\, \rho^a_\mathrm{qu}(\vec q)\,
  \vert p, h, i \right> &= g \, (t^a)_{i' i} \, 16\pi^3 \, \delta_{h' h}\,
  p^+ \delta(p^+ - k^+)\, \delta(\vec p -\vec k - \vec q)~, \\
    \left< k, \rho, b \vert\, \rho^a_\mathrm{gl}(\vec q)\,
    \vert p, \sigma, c \right> &= g (T^a)_{bc} \, (2\pi)^{D-1} \,
    \delta_{\rho\sigma}\,
  2p^+ \delta(p^+ - k^+)\, \delta(\vec p -\vec k - \vec q)~.
\end{split}
\ee

We then obtain
\bea
\mathrm{fig.}\, \ref{fig:rho-rho_qu-gl}\mathrm{a} &=&
-\frac{g^4}{3\cdot 16\pi^3} \,\tr T^a T^b
\int \left[\dd x_i\right]
  \int \left[\dd^2 k_i\right]\,
  \Psi_{qqq}(x_1,\vec k_1; x_2,\vec k_2; x_3,\vec k_3)\,
  \nn\\
     & &     \Psi_{qqq}^*(x_1,\vec k_1-(1-x_1)\vec q;
x_2,\vec k_2 +x_2\vec q;
x_3,\vec k_3 +x_3\vec q) \nn\\  
& &
\frac{(2\pi)^{D-1}}{2p_1^+} \int \frac{\dk_g}{2(p_1^+ - k_g^+)}
\langle S| \hat\psi_{q\to qg}(p_1; p_1-k_g, k_g)\,
\hat\psi^*_{q\to qg}(p_1- q; p_1-k_g- q_1,
k_g-q_2) |S\rangle~.
\label{eq:rho_qu-gl-a}
\eea
The symmetry factor is 3. Similarly,
\\

\bea
\mathrm{fig.}\, \ref{fig:rho-rho_qu-gl}\mathrm{b} &=&
-\frac{g^4}{6\cdot 16\pi^3} \,\tr T^a T^b
\int \left[\dd x_i\right]
  \int \left[\dd^2 k_i\right] \,
  \Psi_{qqq}(x_1,\vec k_1; x_2,\vec k_2; x_3,\vec k_3)\,
\nn\\
& &     \Psi_{qqq}^*(x_1,\vec k_1+x_1\vec q-\vec q_2;
x_2,\vec k_2 +x_2\vec q-\vec q_1;
x_3,\vec k_3 +x_3\vec q)  \nn\\  
& &
\frac{(2\pi)^{D-1}}{2p_1^+} \int \frac{\dk_g}{2(p_1^+ - k_g^+)}
\langle S| \hat\psi_{q\to qg}(p_1; p_1-k_g, k_g)\,
\hat\psi^*_{q\to qg}(p_1-q_2; p_1-k_g,
k_g-q_2) |S\rangle~.
     \label{eq:rho_qu-gl-b}
\eea
The symmetry factor is 6.\\

The remaining diagrams are finite.
\bea
\mathrm{fig.}\, \ref{fig:rho-rho_qu-gl}\mathrm{c} &=&
\frac{g^4}{12\cdot 16\pi^3} \,\tr T^a T^b
\int \left[\dd x_i\right]
  \int \left[\dd^2 k_i\right] \,
  \Psi_{qqq}(x_1,\vec k_1; x_2,\vec k_2; x_3,\vec k_3)\,
\nn\\
& &
\int\limits_x^{\mathrm{min}(x_1,1-x_2)}\frac{\dd x_g}{x_g}
\left(1-\frac{z_1+z_2}{2}+\frac{z_1 z_2}{6}\right)
\sqrt{\frac{x_1}{x_1-x_g}}\sqrt{\frac{x_2}{x_2+x_g}}
\nn\\
& & \int\dd^2 k_g
\frac{z_1\vec p_1-\vec k_g}
     {\left(z_1\vec p_1-\vec k_g \right)^2}
\cdot
\frac{z_2\vec p_2 - (1-z_2)(\vec k_g -\vec q_2)}
     {\left( z_2\vec p_2 - (1-z_2)(\vec k_g -\vec q_2) \right)^2}
\nn\\
& &  \Psi_{qqq}^*(x_1-x_g,\vec k_1+x_1\vec q-\vec
q_1-\vec k_g + x_g\vec K;
x_2+x_g,\vec k_2 +x_2\vec q-\vec q_2+\vec k_g - x_g\vec K;
x_3,\vec k_3 +x_3\vec q)~,
     \label{eq:rho_qu-gl-c}
\eea
with the same $z_1$ and $z_2$ as above. The symmetry factor is 6.\\

\bea
\mathrm{fig.}\, \ref{fig:rho-rho_qu-gl}\mathrm{d} &=&
\frac{g^4}{12\cdot 16\pi^3} \,\tr T^a T^b
\int \left[\dd x_i\right]
  \int \left[\dd^2 k_i\right] \,
  \Psi_{qqq}(x_1,\vec k_1; x_2,\vec k_2; x_3,\vec k_3)\,
\nn\\
& &
\int\limits_x^{\mathrm{min}(x_1,1-x_2)}\frac{\dd x_g}{x_g}
\left(1-\frac{z_1+z_2}{2}+\frac{z_1 z_2}{6}\right)
\sqrt{\frac{x_1}{x_1-x_g}}\sqrt{\frac{x_2}{x_2+x_g}}
\nn\\
& & \int\dd^2 k_g
\frac{z_1\vec p_1-\vec k_g}
     {\left(z_1\vec p_1-\vec k_g \right)^2}
\cdot
\frac{z_2(\vec p_2-\vec q_1)  - (1-z_2)(\vec k_g -\vec q_2)}
     {\left( z_2(\vec p_2-\vec q_1)  - (1-z_2)(\vec k_g -\vec q_2)\right)^2}
\nn\\
& &  \Psi_{qqq}^*(x_1-x_g,\vec k_1+x_1\vec q-\vec k_g + x_g\vec K;
x_2+x_g,\vec k_2 -(1-x_2)\vec q+\vec k_g - x_g\vec K;
x_3,\vec k_3 +x_3\vec q)~,
     \label{eq:rho_qu-gl-d}
\eea
The symmetry factor is 6.\\

\bea
\mathrm{fig.}\, \ref{fig:rho-rho_qu-gl}\mathrm{e} &=& 0
     \label{eq:rho_qu-gl-e}
\eea
\\~~\\
As already mentioned above, eqs.~(\ref{eq:rho_qu-gl-a} -
\ref{eq:rho_qu-gl-d}) should be duplicated with $\vec q_1
\leftrightarrow \vec q_2$ to include the contribution from $\langle
\rho^a_\mathrm{gl}(\vec q_1)\, \rho^b_\mathrm{qu}(\vec q_2)\rangle$.

\subsubsection{Coupling to quarks, $\langle \rho^a_\mathrm{qu}(\vec q_1)\,
\rho^b_\mathrm{qu}(\vec q_2)\rangle$}

Now consider the diagrams where both external probes couple
to quarks. The matrix element of $\rho^a_\mathrm{qu}(\vec q_1)\,
\rho^b_\mathrm{qu}(\vec q_2)$ in a one quark state is
\be
\langle p',j'\vert \rho^a_\mathrm{qu}(\vec q_1)\,
\rho^b_\mathrm{qu}(\vec q_2)\vert p,j\rangle =
g^2 (t^a t^b)_{j' j} \, 16\pi^3\, p^+\, \delta(p^{\prime +} - p^+)\,
\delta(\vec p - \vec p' -\vec q)~.
\ee

The first contribution where the color charge operators
are sandwiched between the bare 3-quark Fock states~\cite{Dumitru:2018vpr}
\footnote{In eq.~(\ref{eq:rho-rho-3q}), to sum only color charges with
  light-cone momentum fractions beyond a lower cutoff $x$, one would
  restrict the integrations over the active quarks to $x_i>x$. That
  is, the first integrand would be multiplied by
  $\Theta(x_1-x)$, the other two terms by $\Theta(x_1-x)\, \Theta(x_2-x)$,
  respectively. However, we are interested primarily in color charge
  correlations at $x$ much less than the typical valence quark
  light-cone momentum fraction, so these restrictions may be dropped.} is
\bea
\langle \rho^a_\mathrm{qu}(\vec q_1)\, \rho^b_\mathrm{qu}(\vec q_2)\rangle
&=& \frac{g^2}{6} \,
\tr t^at^b\, \int\left[\dd x_i\right]\int\left[\dd^2 k_i\right]
\,
\Psi_\mathrm{qqq}(x_1,\vec k_1; x_2, \vec k_2; x_3, \vec k_3)\nn\\
& & \left[\Psi_\mathrm{qqq}^*(x_1,\vec k_1-(1-x_1)\vec q;
  x_2, \vec k_2+x_2\vec q; x_3, \vec k_3+x_3\vec q)\right.\nn\\
  & & - \frac{1}{2}
  \Psi_\mathrm{qqq}^*(x_1,\vec k_1+x_1\vec q-\vec q_1;
  x_2, \vec k_2+x_2\vec q-\vec q_2; x_3, \vec k_3+x_3\vec q) \nn\\
  & & - \left. \frac{1}{2}
  \Psi_\mathrm{qqq}^*(x_1,\vec k_1+x_1\vec q-\vec q_2;
  x_2, \vec k_2+x_2\vec q-\vec q_1; x_3, \vec k_3+x_3\vec q)
  \right]~.  \label{eq:rho-rho-3q}
\eea
The symmetry factor is 3. The first term (``handbag diagram'') is
proportional to the quark GPD $\int \dd x \,H_q(x,-\vec q^2)$ at
vanishing skewness; c.f.\ appendix~B in ref.~\cite{Burkardt:2002hr}.
The second and third terms (``cat's ears diagrams'') are two-body
diagrams where the gluon probes attach to different quarks in the
proton. They ensure that the color charge correlator satisfies a Ward
identity and vanishes when either $\vec q_1\to0$ or $\vec
q_2\to0$. Also, these contributions are ``higher twist'' suppressed
when both $q_i$ {\em as well as} $|\vec q_1 - \vec q_2|$ are much
greater than the typical transverse momentum of quarks in the 3-quark
Fock state of the proton; on the other hand, the two-body
contributions dominate when the two probes share a large momentum
transfer, $\vec q_1 \approx \vec q_2$, such as in exclusive
$\gamma^{(*)} p \to J/\Psi\, p$ at large $-t$~\cite{Dumitru:2019qec}.

\begin{figure}[htb]
\centerline{
  \includegraphics[width=0.3\linewidth]{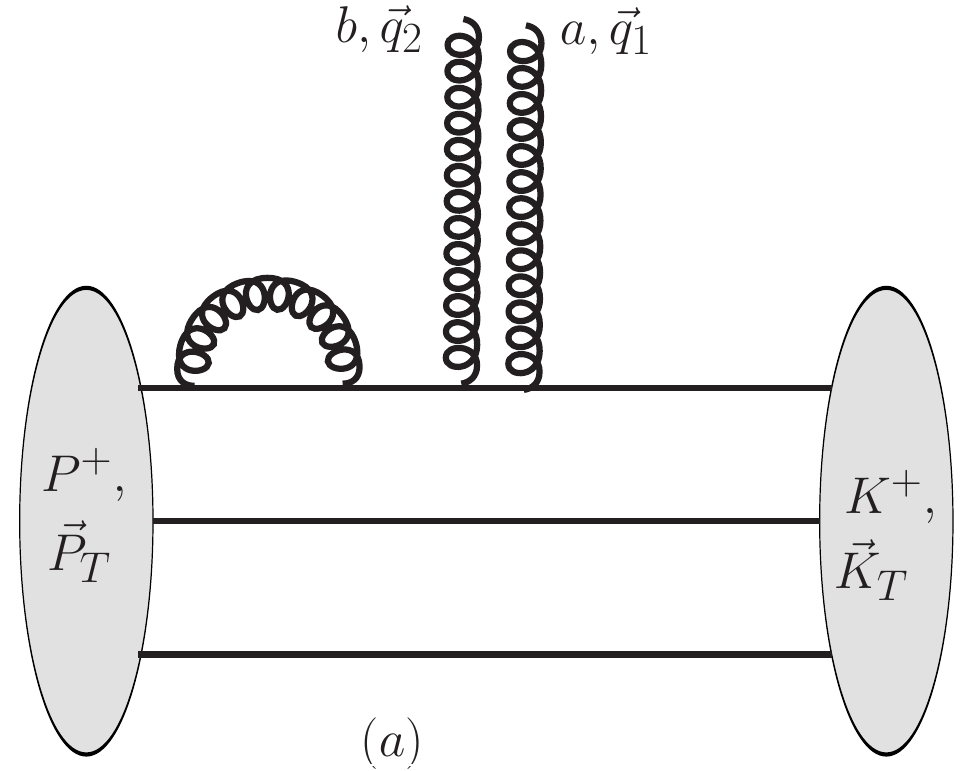}
    \hspace*{.4cm}
    \includegraphics[width=0.3\linewidth]{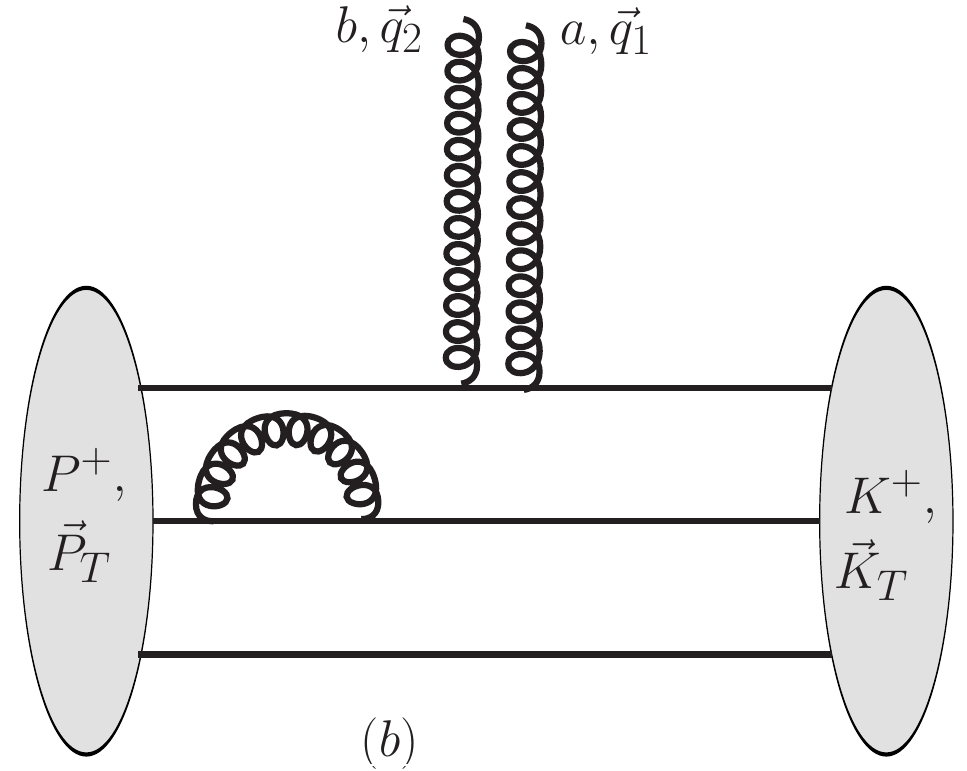}
    \hspace*{.4cm}
  \includegraphics[width=0.3\linewidth]{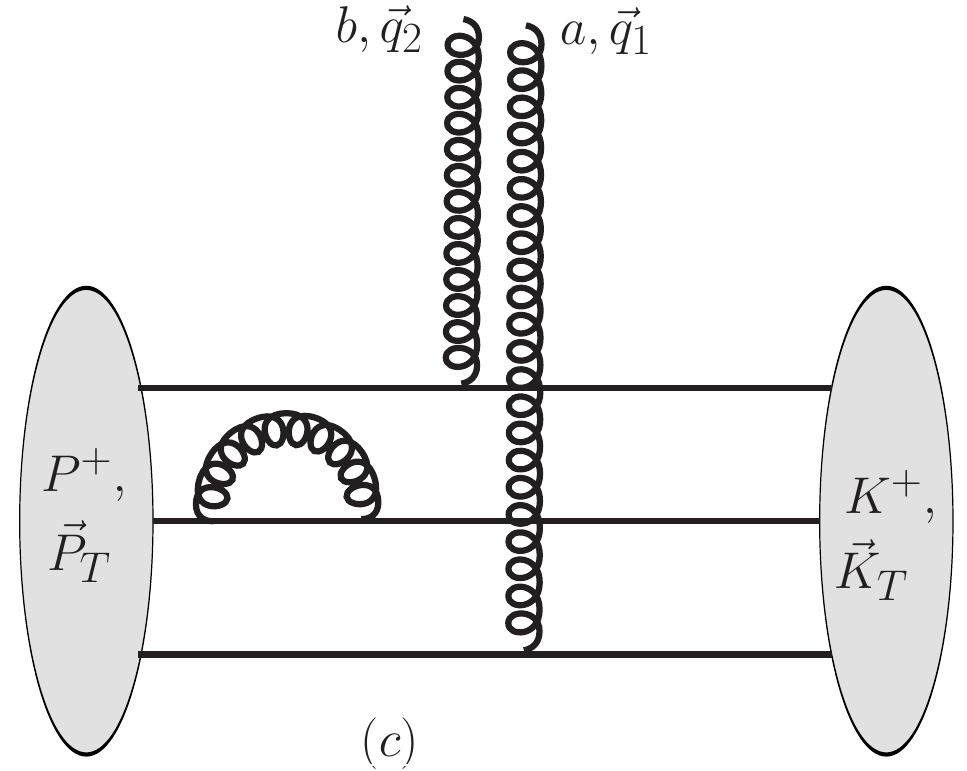}
  }
\caption{Some of the diagrams for the ${\cal O}(g^2)$ correction to
  $\langle \rho^a_\mathrm{qu}(\vec q_1)\, \rho^b_\mathrm{qu}(\vec
  q_2)\rangle$ in the three-quark Fock state.  The cut is located at
  the insertion of the two color charge operators.}
\label{fig:rho-rho_qu_Z}
\end{figure}
To account for the quark wave function renormalization factor, we
multiply the integrand in the previous equation by $Z_q(x_1)\,
Z_q(x_2)\, Z_q(x_3) = 1 - C_q(x_1) - C_q(x_2) - C_q(x_3)$. This
renormalization factor is given in eq.~(\ref{eq:eq:Zqatg2epsform}),
where $\alpha \to x/x_i$. Some of the corresponding diagrams are shown
in fig.~\ref{fig:rho-rho_qu_Z}.\\

\begin{figure}[htb]
\centerline{
  \includegraphics[width=0.25\linewidth]{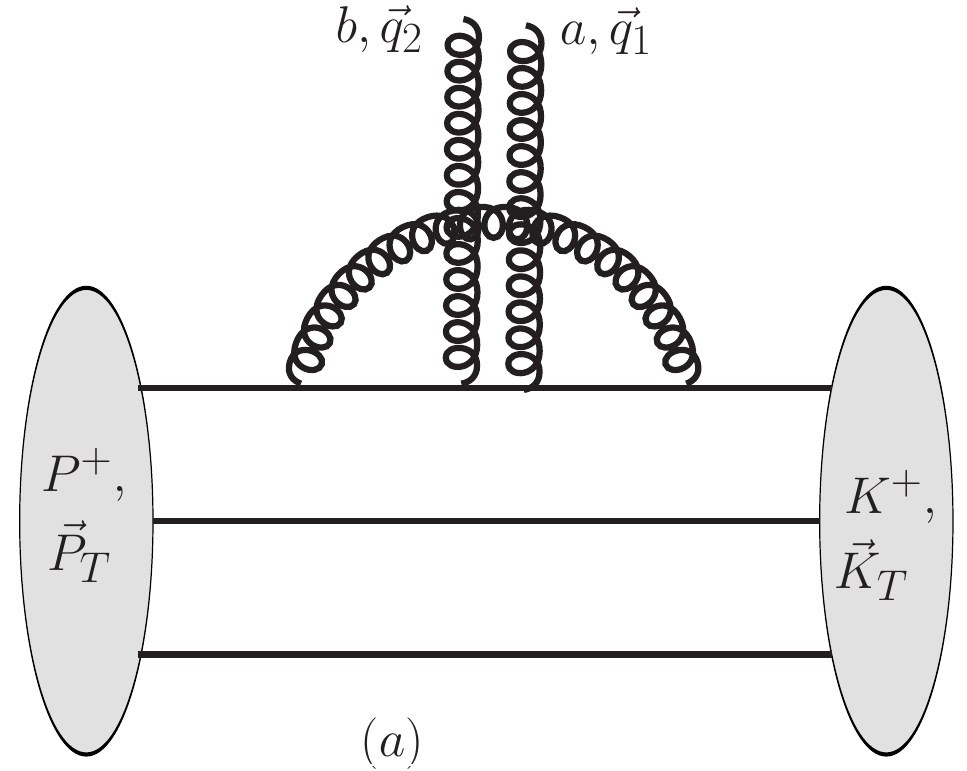}
    \hspace*{.4cm}
    \includegraphics[width=0.25\linewidth]{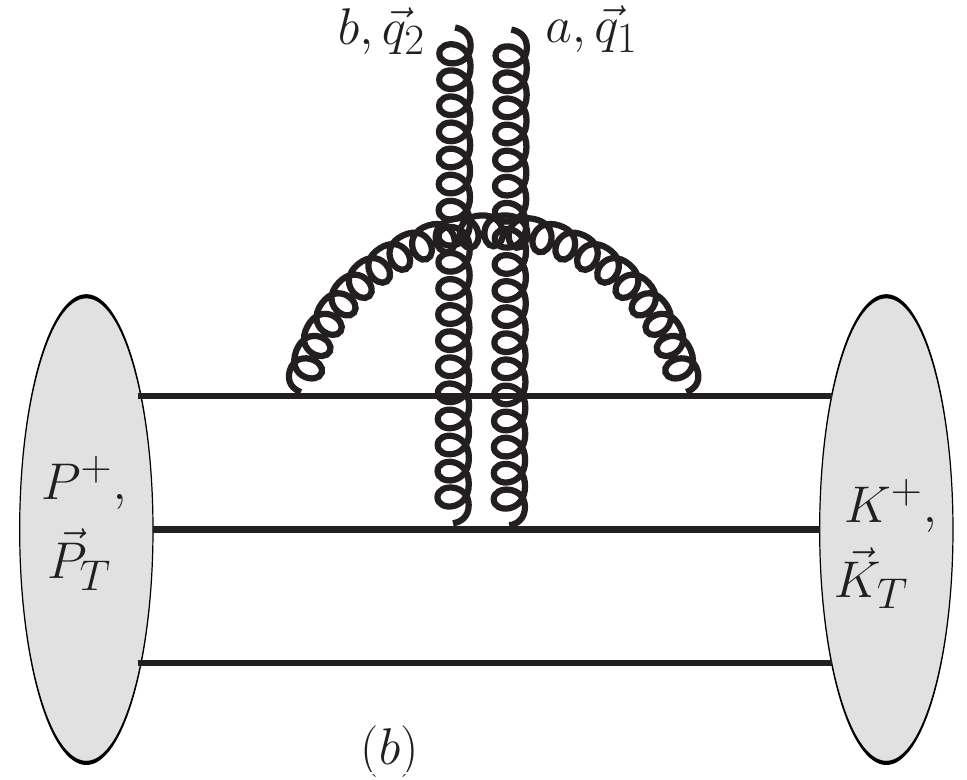}
    \hspace*{.4cm}
  \includegraphics[width=0.25\linewidth]{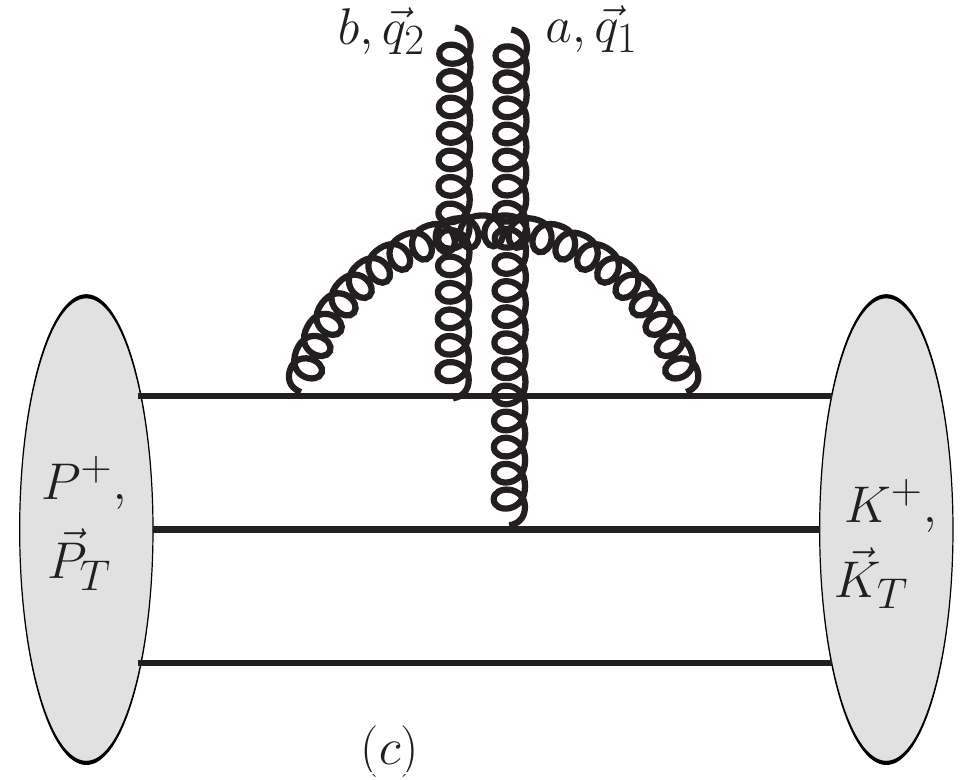}
    \hspace*{.4cm}
  \includegraphics[width=0.25\linewidth]{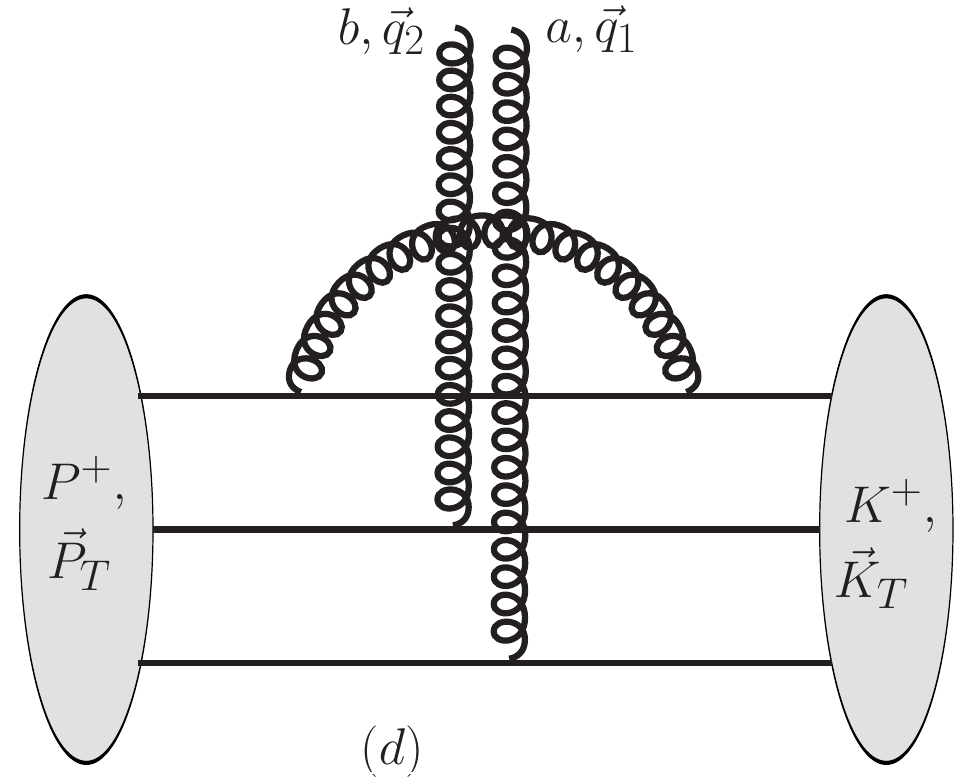}
  }
\caption{First set of diagrams for $\langle \rho^a_\mathrm{qu}(\vec q_1)\,
  \rho^b_\mathrm{qu}(\vec q_2)\rangle$. The cut is located at the
  insertion of the two color charge operators.}
\label{fig:rho-rho_qu}
\end{figure}
We now turn to the diagrams where $\rho^a_\mathrm{qu}(\vec q_1)\,
\rho^b_\mathrm{qu}(\vec q_2)$ is sandwiched between $|qqqg\rangle$
Fock states or where two quarks exchange a gluon on either side of the
operator insertion.
\bea
\mathrm{fig.}\, \ref{fig:rho-rho_qu}\mathrm{a} &=&
\frac{4g^4C_F}{3\cdot 16\pi^3} \,\tr t^a t^b
\int \left[\dd x_i\right]
  \int \left[\dd^2 k_i\right]\,
  \Psi_{qqq}(x_1,\vec k_1; x_2,\vec k_2; x_3,\vec k_3)\nn\\
  & &
\Psi_{qqq}^*(x_1,\vec k_1-(1-x_1)\vec q;
x_2,\vec k_2 +x_2\vec q; x_3,\vec k_3 +x_3\vec q)
\nn\\
& &
\frac{(2\pi)^{D-1}}{2p_1^+} \int \frac{\dk_g}{2(p_1^+ - k_g^+)}
\langle S| \hat\psi_{q\to qg}(p_1; p_1- k_g, k_g)\,
\hat\psi^*_{q\to qg}(p_1-  q; p_1- k_g- q, k_g) |S\rangle ~.
     \label{eq:rho_qu-qu-a}
\eea
The symmetry factor is 3.

\bea
\mathrm{fig.}\, \ref{fig:rho-rho_qu}\mathrm{b} &=&
\frac{4g^4C_F}{3\cdot 16\pi^3} \,\tr t^a t^b
\int \left[\dd x_i\right]
  \int \left[\dd^2 k_i\right]\,
  \Psi_{qqq}(x_1,\vec k_1; x_2,\vec k_2; x_3,\vec k_3)\,
  \nn\\
 & &   \Psi_{qqq}^*(x_1,\vec k_1+x_1\vec q;
x_2,\vec k_2 -(1-x_2)\vec q; x_3,\vec k_3 +x_3\vec q)
\nn\\ & &
\frac{(2\pi)^{D-1}}{2p_1^+} \int \frac{\dk_g}{2(p_1^+ - k_g^+)}
\langle S| \hat\psi_{q\to qg}(p_1; p_1-k_g, k_g)\,
\hat\psi^*_{q\to qg}(p_1; p_1-k_g, k_g) |S\rangle ~.
     \label{eq:rho_qu-qu-b}
\eea
The symmetry factor is 6.

\bea
\mathrm{fig.}\, \ref{fig:rho-rho_qu}\mathrm{c} &=&
\frac{2g^4}{3\cdot 6\cdot 16\pi^3} \,\tr t^a t^b
\int \left[\dd x_i\right]
  \int \left[\dd^2 k_i\right]\,
  \Psi_{qqq}(x_1,\vec k_1; x_2,\vec k_2; x_3,\vec k_3)\,
  \nn\\
& & \Psi_{qqq}^*(x_1,\vec k_1+x_1\vec q-\vec q_2;
x_2,\vec k_2 +x_2\vec q-\vec q_1;
x_3,\vec k_3 +x_3\vec q) \nn\\
& &
\frac{(2\pi)^{D-1}}{2p_1^+} \int \frac{\dk_g}{2(p_1^+ - k_g^+)}
\langle S| \hat\psi_{q\to qg}(p_1; p_1-k_g, k_g)\,
\hat\psi^*_{q\to qg}(p_1- q_2; p_1-k_g- q_2, k_g)
|S\rangle ~.
\label{eq:rho_qu-qu-c}
\eea
The symmetry factor is 6.

The diagram where the two probe gluons connect oppositely to quarks 1
and 2 is
\bea
\mathrm{fig.}\, \ref{fig:rho-rho_qu}\mathrm{c'} &=&
\mathrm{fig.}\, \ref{fig:rho-rho_qu}\mathrm{c}\, (\vec q_1
\leftrightarrow \vec q_2)
     \label{eq:rho_qu-qu-c'}
\eea
Again, the symmetry factor is 6.

Next,
\bea
\mathrm{fig.}\, \ref{fig:rho-rho_qu}\mathrm{d} &=&
-\frac{2g^4C_F}{3\cdot 16\pi^3} \,\tr t^a t^b
\int \left[\dd x_i\right]
  \int \left[\dd^2 k_i\right]\,
  \Psi_{qqq}(x_1,\vec k_1; x_2,\vec k_2; x_3,\vec k_3)\,
  \nn\\
& &  \Psi_{qqq}^*(x_1,\vec k_1+x_1\vec q;
x_2,\vec k_2 +x_2\vec q-\vec q_2;
x_3,\vec k_3 +x_3\vec q-\vec q_1) \nn\\
& &
\frac{(2\pi)^{D-1}}{2p_1^+} \int \frac{\dk_g}{2(p_1^+ - k_g^+)}
\langle S| \hat\psi_{q\to qg}(p_1; p_1-k_g, k_g)\,
\hat\psi^*_{q\to qg}(p_1; p_1-k_g, k_g)
|S\rangle ~.
\label{eq:rho_qu-qu-d}
\eea
Because of the symmetry of this diagram under $a \leftrightarrow b$
and $\vec q_1 \leftrightarrow \vec q_2$ its symmetry factor is 6.
\\~~\\

\begin{figure}[htb]
  \centering
  \begin{minipage}[b]{\linewidth}
    \includegraphics[width=0.28\linewidth]{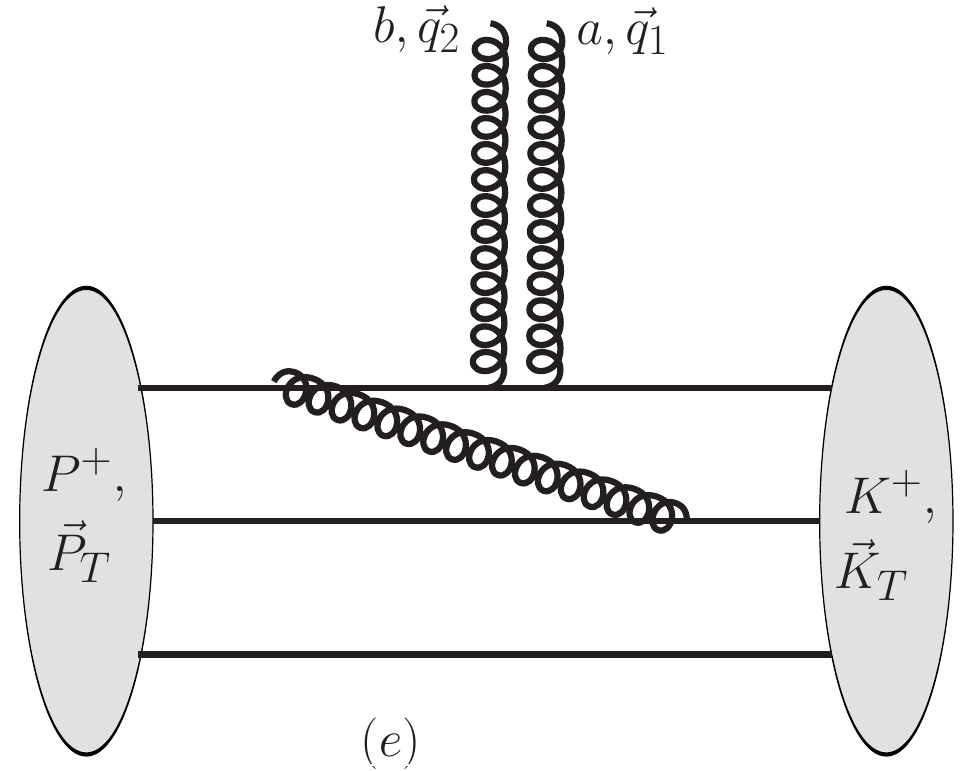}
    \hspace*{.4cm}
    \includegraphics[width=0.28\linewidth]{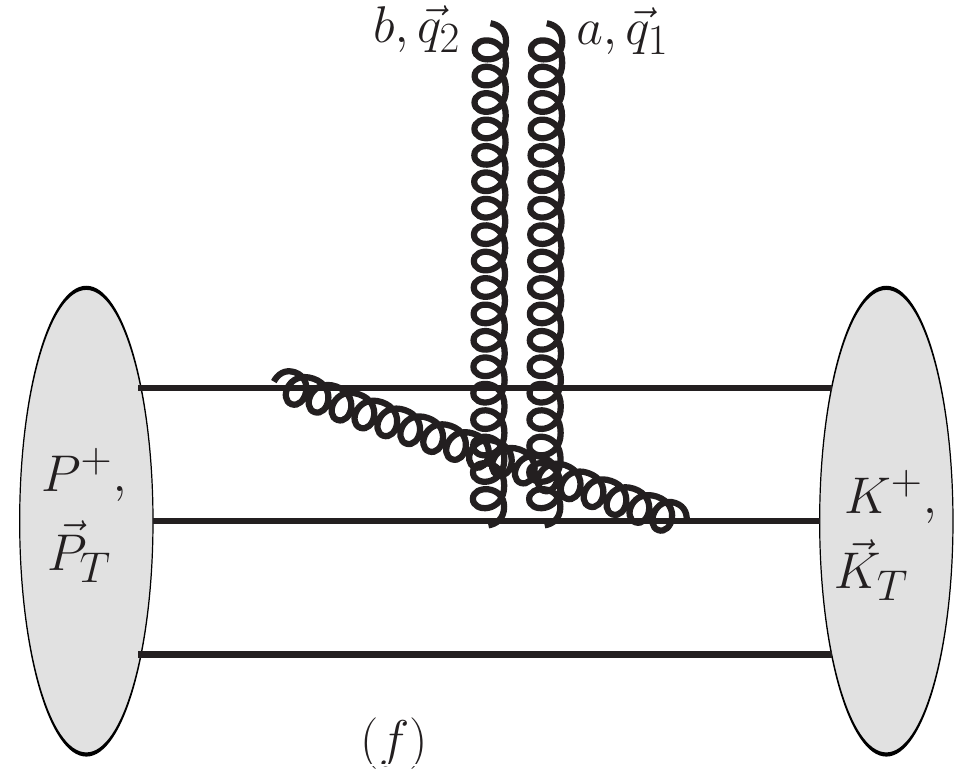}
    \hspace*{.4cm}
    \includegraphics[width=0.28\linewidth]{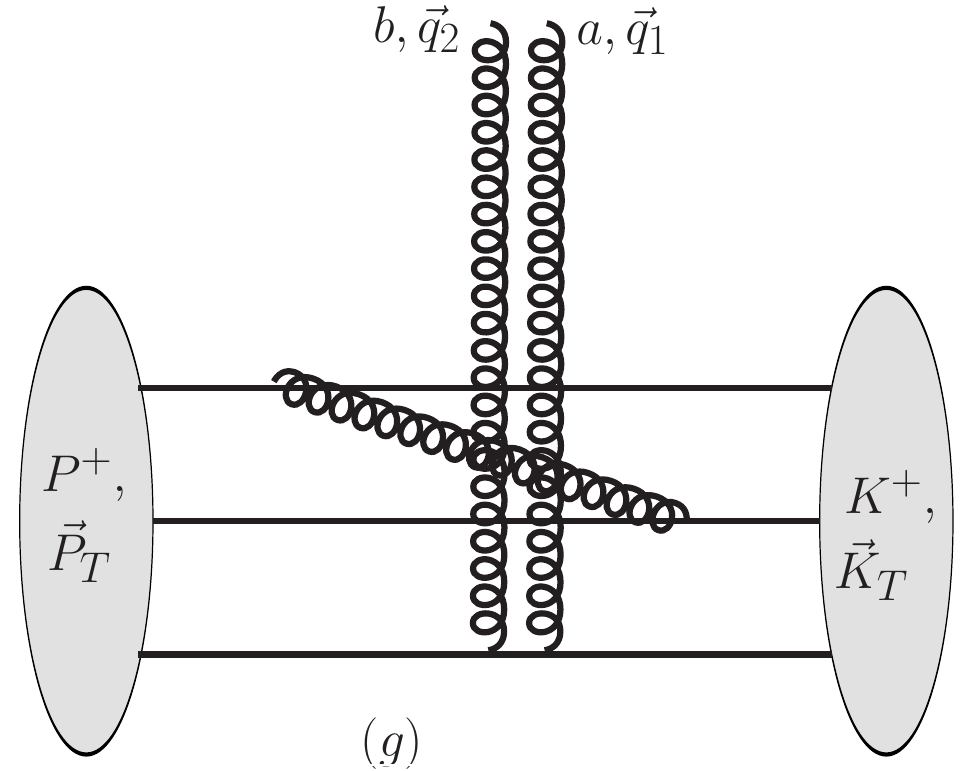}
  \end{minipage}
  \begin{minipage}[b]{\linewidth}
    \includegraphics[width=0.28\linewidth]{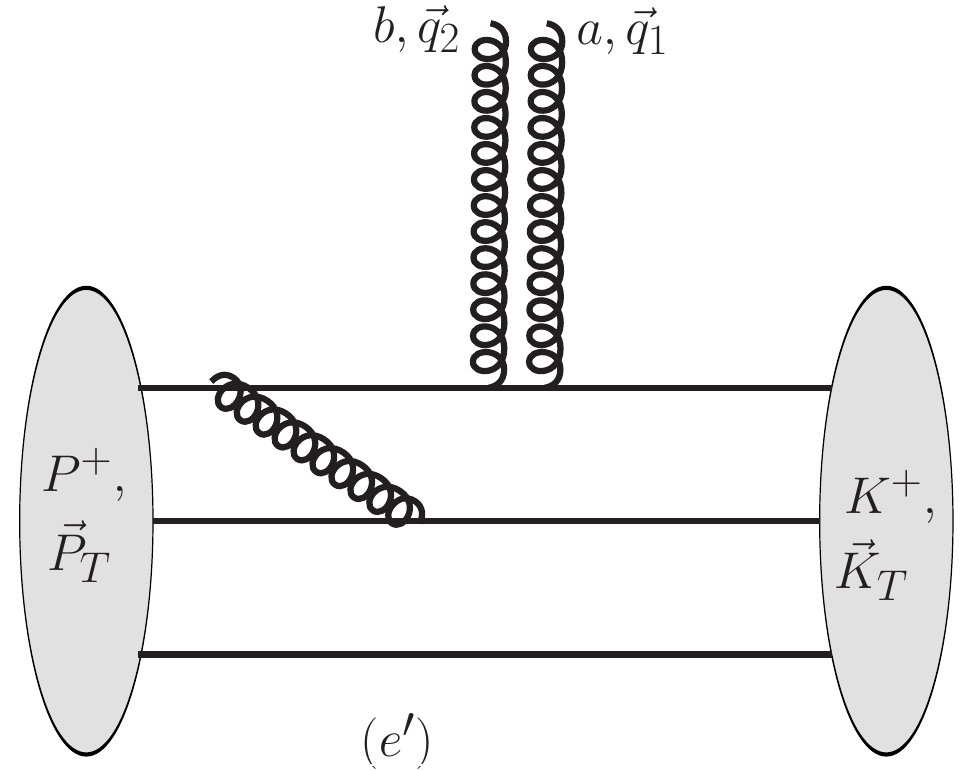}
    \hspace*{.4cm}
    \includegraphics[width=0.28\linewidth]{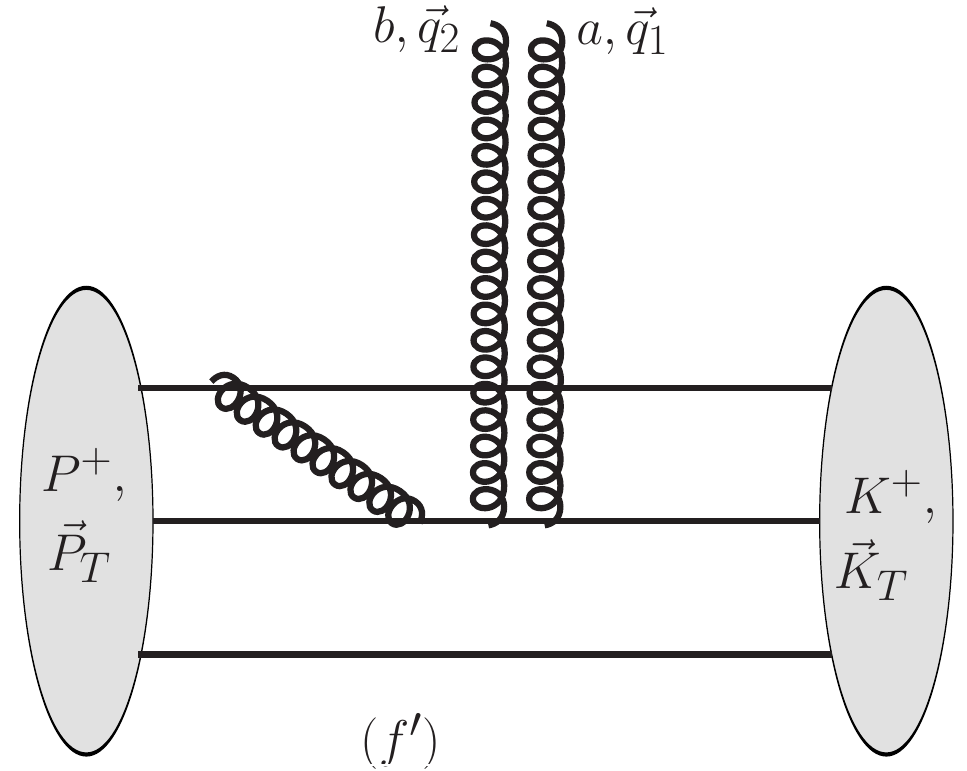}
    \hspace*{.4cm}
    \includegraphics[width=0.28\linewidth]{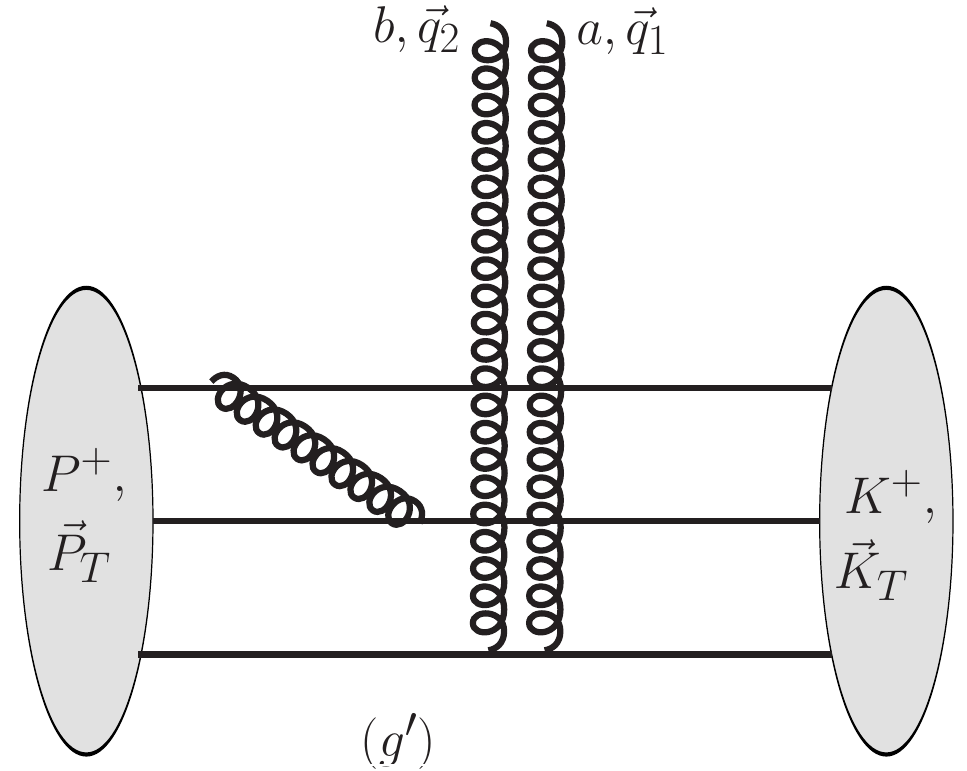}
    \hspace*{.4cm}
  \end{minipage}
\caption{Second set of diagrams for $\langle \rho^a_\mathrm{qu}(\vec q_1)\,
  \rho^b_\mathrm{qu}(\vec q_2)\rangle$. The cut is located at the
  insertion of the two color charge operators.}
\label{fig:rho-rho_qu'}
\end{figure}
The remaining diagrams are finite because
the transverse momentum of the gluon shifts the arguments of
$\Psi^*_\mathrm{qqq}$.
\bea
\mathrm{fig.}\, \ref{fig:rho-rho_qu'}\mathrm{e} &=&
- \frac{g^4}{3\cdot 16\pi^3} \, C_F\, \tr t^at^b
 \int \left[\dd x_i\right]
  \int \left[\dd^2 k_i\right]\,
  \Psi_{qqq}(x_1,\vec k_1; x_2,\vec k_2; x_3,\vec k_3)\,
\nn\\
& &
\int\limits_x^{\mathrm{min}(x_1,1-x_2)}\frac{\dd x_g}{x_g}
\left(1-\frac{z_1+z_2}{2}+\frac{z_1 z_2}{6}\right)
\sqrt{\frac{x_1}{x_1-x_g}}\sqrt{\frac{x_2}{x_2+x_g}}
\nn\\
& &
\int\dd^2 k_g
\frac{z_1\vec p_1-\vec k_g}
     {\left(z_1\vec p_1-\vec k_g \right)^2}
\cdot
\frac{z_2\vec p_2-(1-z_2)\vec k_g}
     {\left(z_2\vec p_2-(1-z_2)\vec k_g \right)^2} \nn\\
     & &
\Psi_{qqq}^*(x_1-x_g,\vec k_1-(1-x_1)\vec q-\vec k_g + x_g\vec K;
x_2+x_g,\vec k_2 +x_2\vec q + \vec k_g - x_g\vec K;
x_3,\vec k_3 + x_3\vec q)~,
     \label{eq:rho-rho_qu-e}
\eea
with a symmetry factor of 6 (because the gluon may also be absorbed
across the insertion by quark 3). The diagram where the gluon is
emitted by quark 2 and absorbed by quark 1 is equal to that from
fig.~\ref{fig:rho-rho_qu'}f and will be included in its symmetry
factor.
\bea
\mathrm{fig.}\, \ref{fig:rho-rho_qu'}\mathrm{f} &=&
- \frac{g^4}{3\cdot 16\pi^3} \, C_F\,\tr t^at^b
 \int \left[\dd x_i\right]
  \int \left[\dd^2 k_i\right]\,
  \Psi_{qqq}(x_1,\vec k_1; x_2,\vec k_2; x_3,\vec k_3)\,
\nn\\
& &
\int\limits_x^{\mathrm{min}(x_1,1-x_2)}\frac{\dd x_g}{x_g}
\left(1-\frac{z_1+z_2}{2}+\frac{z_1 z_2}{6}\right)
\sqrt{\frac{x_1}{x_1-x_g}}\sqrt{\frac{x_2}{x_2+x_g}}  \nn\\
& &
\int\dd^2 k_g
\frac{z_1\vec p_1-\vec k_g}
     {\left(z_1\vec p_1-\vec k_g \right)^2}
\cdot
\frac{z_2(\vec p_2-\vec q)-(1-z_2)\vec k_g}
     {\left(z_2(\vec p_2-\vec q)-(1-z_2)\vec k_g \right)^2} \nn\\
     & &
     \Psi_{qqq}^*(x_1-x_g,\vec k_1+x_1\vec q-\vec k_g + x_g\vec K;
     x_2+x_g,\vec k_2 -(1-x_2)\vec q + \vec k_g - x_g\vec K;
     x_3,\vec k_3 + x_3\vec q)~,
     \label{eq:rho-rho_qu-f}
\eea
again with a symmetry factor of 6.

The diagram where both probes attach to the third quark gives
\bea
\mathrm{fig.}\, \ref{fig:rho-rho_qu'}\mathrm{g} &=&
- \frac{g^4}{3\cdot 16\pi^3} \, C_F\,\tr t^at^b
 \int \left[\dd x_i\right]
  \int \left[\dd^2 k_i\right]\,
  \Psi_{qqq}(x_1,\vec k_1; x_2,\vec k_2; x_3,\vec k_3)\,
\nn\\
& &
\int\limits_x^{\mathrm{min}(x_1,1-x_2)}\frac{\dd x_g}{x_g}
\left(1-\frac{z_1+z_2}{2}+\frac{z_1 z_2}{6}\right)
\sqrt{\frac{x_1}{x_1-x_g}}\sqrt{\frac{x_2}{x_2+x_g}}
\nn\\
& &
\int\dd^2 k_g
\frac{z_1\vec p_1-\vec k_g}
     {\left(z_1\vec p_1-\vec k_g \right)^2}
\cdot
\frac{z_2\vec p_2-(1-z_2)\vec k_g}
     {\left(z_2\vec p_2-(1-z_2)\vec k_g \right)^2} \nn\\
     & &
     \Psi_{qqq}^*(x_1-x_g,\vec k_1+x_1\vec q-\vec k_g + x_g\vec K;
     x_2+x_g,\vec k_2 +x_2\vec q + \vec k_g - x_g\vec K;
     x_3,\vec k_3 -(1-x_3)\vec q)~.
     \label{eq:rho-rho_qu-g}
\eea
Here the symmetry factor is 6 to include the contribution where the gluon
emission/absorption vertices are swapped.
\\

Now we list the diagrams where two quarks exchange a gluon on one side
of the $\rho^a(\vec q_1)\, \rho^b(\vec q_2)$ insertion.
\bea
\mathrm{fig.}\, \ref{fig:rho-rho_qu'}\mathrm{e'} &=&
\frac{g^4}{3\cdot 16\pi^3} \, C_F\,\tr t^at^b
 \int \left[\dd x_i\right]
  \int \left[\dd^2 k_i\right]\,
  \Psi_{qqq}(x_1,\vec k_1; x_2,\vec k_2; x_3,\vec k_3)\,
\nn\\
& &
\int\limits_x^{\mathrm{min}(x_1,1-x_2)}\frac{\dd x_g}{x_g}
\left(1-\frac{z_1+z_2}{2}+\frac{z_1 z_2}{6}\right)
\sqrt{\frac{x_1}{x_1-x_g}}\sqrt{\frac{x_2}{x_2+x_g}}
\int\dd^2 k_g
\frac{z_1\vec p_1-\vec k_g}
     {\left(z_1\vec p_1-\vec k_g \right)^2}
\cdot
\frac{z_2\vec p_2-(1-z_2)\vec k_g}
     {\left(z_2\vec p_2-(1-z_2)\vec k_g \right)^2} \nn\\
     & &
     \Psi_{qqq}^*(x_1-x_g,\vec k_1-(1-x_1)\vec q-\vec k_g+x_g\vec K;
     x_2+x_g,\vec k_2 +x_2\vec q + \vec k_g-x_g\vec K;
     x_3,\vec k_3 +x_3\vec q)~.
     \label{eq:rho-rho_qu-e'}
\eea
The symmetry factor is $\frac{1}{2}\times6=3$; the factor of
.$\frac{1}{2}$ arises because we sum over all permutations of the
gluon exchange within $|P\rangle$.

There is also a diagram (not shown) where quark 2 emits and quark 1
absorbs the gluon on the other side of the insertion:
\bea
\mathrm{fig.}\, \ref{fig:rho-rho_qu'}\mathrm{e''} &=&
\frac{g^4}{3\cdot 16\pi^3} \, C_F\,\tr t^at^b
 \int \left[\dd x_i\right]
  \int \left[\dd^2 k_i\right]\,
  \Psi_{qqq}(x_1,\vec k_1; x_2,\vec k_2; x_3,\vec k_3)\,
\nn\\
& &
\int\limits_x^{\mathrm{min}(x_1,1-x_2)}\frac{\dd x_g}{x_g}
\left(1-\frac{z_1+z_2}{2}+\frac{z_1 z_2}{6}\right)
\sqrt{\frac{x_1}{x_1-x_g}}\sqrt{\frac{x_2}{x_2+x_g}} \nn\\
& & \int\dd^2 k_g
\frac{z_1(\vec p_1-\vec q)-\vec k_g}
     {\left(z_1(\vec p_1-\vec q)-\vec k_g \right)^2}
\cdot
\frac{z_2\vec p_2-(1-z_2)\vec k_g}
     {\left(z_2\vec p_2-(1-z_2)\vec k_g \right)^2} \nn\\
     & &
     \Psi_{qqq}^*(x_1-x_g,\vec k_1-(1-x_1)\vec q-\vec k_g+x_g\vec K;
     x_2+x_g,\vec k_2 +x_2\vec q + \vec k_g-x_g\vec K;
     x_3,\vec k_3 +x_3\vec q)~.
     \label{eq:rho-rho_qu-e''}
\eea
Note that here $z_1 = x_g / x_1$ and $z_2 = x_g/(x_2+x_g)$, as before.
The symmetry factor for this diagram is also 3. Once again, the
diagram with swapped emission and absorption vertices is equal to
diagram fig.~\ref{fig:rho-rho_qu'}f'' and will be included in its
symmetry factor.

\bea
\mathrm{fig.}\, \ref{fig:rho-rho_qu'}\mathrm{f'} &=&
\frac{g^4}{3\cdot 16\pi^3} \, C_F\,\tr t^at^b
 \int \left[\dd x_i\right]
  \int \left[\dd^2 k_i\right]\,
  \Psi_{qqq}(x_1,\vec k_1; x_2,\vec k_2; x_3,\vec k_3)\,
\nn\\
& &
\int\limits_x^{\mathrm{min}(x_1,1-x_2)}\frac{\dd x_g}{x_g}
\left(1-\frac{z_1+z_2}{2}+\frac{z_1 z_2}{6}\right)
\sqrt{\frac{x_1}{x_1-x_g}}\sqrt{\frac{x_2}{x_2+x_g}}
\int\dd^2 k_g
\frac{z_1\vec p_1-\vec k_g}
     {\left(z_1\vec p_1-\vec k_g \right)^2}
\cdot
\frac{z_2\vec p_2-(1-z_2)\vec k_g}
     {\left(z_2\vec p_2-(1-z_2)\vec k_g \right)^2} \nn\\
     & &
     \Psi_{qqq}^*(x_1-x_g,\vec k_1+x_1\vec q-\vec k_g+x_g\vec K;
     x_2+x_g,\vec k_2 -(1-x_2)\vec q + \vec k_g-x_g\vec K;
     x_3,\vec k_3 +x_3\vec q)~.
     \label{eq:rho-rho_qu-f'}
\eea
The symmetry factor is 3.

Again, there is a diagram (not shown) where quark 2 emits and quark 1
absorbs the gluon on the other side of the insertion:
\bea
\mathrm{fig.}\, \ref{fig:rho-rho_qu'}\mathrm{f''} &=&
\frac{g^4}{3\cdot 16\pi^3} \, C_F\,\tr t^at^b
 \int \left[\dd x_i\right]
  \int \left[\dd^2 k_i\right]\,
  \Psi_{qqq}(x_1,\vec k_1; x_2,\vec k_2; x_3,\vec k_3)\,
\nn\\
& &
\int\limits_x^{\mathrm{min}(x_1,1-x_2)}\frac{\dd x_g}{x_g}
\left(1-\frac{z_1+z_2}{2}+\frac{z_1 z_2}{6}\right)
\sqrt{\frac{x_1}{x_1-x_g}}\sqrt{\frac{x_2}{x_2+x_g}} \nn\\
& & \int\dd^2 k_g
\frac{z_1\vec p_1-\vec k_g}
     {\left(z_1\vec p_1-\vec k_g \right)^2}
\cdot
\frac{z_2(\vec p_2-\vec q)-(1-z_2)\vec k_g}
     {\left(z_2(\vec p_2-\vec q)-(1-z_2)\vec k_g \right)^2} \nn\\
     & &
     \Psi_{qqq}^*(x_1-x_g,\vec k_1+x_1\vec q-\vec k_g+x_g\vec K;
     x_2+x_g,\vec k_2 -(1-x_2)\vec q + \vec k_g-x_g\vec K;
     x_3,\vec k_3 +x_3\vec q)~.
     \label{eq:rho-rho_qu-f''}
\eea
The symmetry factor is 3.

\bea
\mathrm{fig.}\, \ref{fig:rho-rho_qu'}\mathrm{g'} &=&
\frac{g^4}{3\cdot 16\pi^3} \, C_F\,\tr t^at^b
 \int \left[\dd x_i\right]
  \int \left[\dd^2 k_i\right]\,
  \Psi_{qqq}(x_1,\vec k_1; x_2,\vec k_2; x_3,\vec k_3)\,
\nn\\
& &
\int\limits_x^{\mathrm{min}(x_1,1-x_2)}\frac{\dd x_g}{x_g}
\left(1-\frac{z_1+z_2}{2}+\frac{z_1 z_2}{6}\right)
\sqrt{\frac{x_1}{x_1-x_g}}\sqrt{\frac{x_2}{x_2+x_g}}
\int\dd^2 k_g
\frac{z_1\vec p_1-\vec k_g}
     {\left(z_1\vec p_1-\vec k_g \right)^2}
\cdot
\frac{z_2\vec p_2-(1-z_2)\vec k_g}
     {\left(z_2\vec p_2-(1-z_2)\vec k_g \right)^2} \nn\\
     & &
     \Psi_{qqq}^*(x_1-x_g,\vec k_1+x_1\vec q-\vec k_g+x_g\vec K;
     x_2+x_g,\vec k_2 +x_2\vec q + \vec k_g-x_g\vec K;
     x_3,\vec k_3 -(1-x_3)\vec q)~.
     \label{eq:rho-rho_qu-g'}
\eea
The symmetry factor is 3.

\bea
\mathrm{fig.}\, \ref{fig:rho-rho_qu'}\mathrm{g''} &=&
\frac{g^4}{3\cdot 16\pi^3} \, C_F\,\tr t^at^b
 \int \left[\dd x_i\right]
  \int \left[\dd^2 k_i\right]\,
  \Psi_{qqq}(x_1,\vec k_1; x_2,\vec k_2; x_3,\vec k_3)\,
\nn\\
& &
\int\limits_x^{\mathrm{min}(x_1,1-x_2)}\frac{\dd x_g}{x_g}
\left(1-\frac{z_1+z_2}{2}+\frac{z_1 z_2}{6}\right)
\sqrt{\frac{x_1}{x_1-x_g}}\sqrt{\frac{x_2}{x_2+x_g}} \nn\\
& & \int\dd^2 k_g
\frac{z_1\vec p_1-\vec k_g}
     {\left(z_1\vec p_1-\vec k_g \right)^2}
\cdot
\frac{z_2\vec p_2-(1-z_2)\vec k_g}
     {\left(z_2\vec p_2-(1-z_2)\vec k_g \right)^2} \nn\\
     & &
     \Psi_{qqq}^*(x_1-x_g,\vec k_1+x_1\vec q-\vec k_g+x_g\vec K;
     x_2+x_g,\vec k_2 +x_2\vec q + \vec k_g-x_g\vec K;
     x_3,\vec k_3 -(1-x_3)\vec q)~.
     \label{eq:rho-rho_qu-g''}
\eea
The symmetry factor is 3.
\\

\begin{figure}[ht]
\centering
\begin{minipage}[b]{\linewidth}
\includegraphics[width=0.28\linewidth]{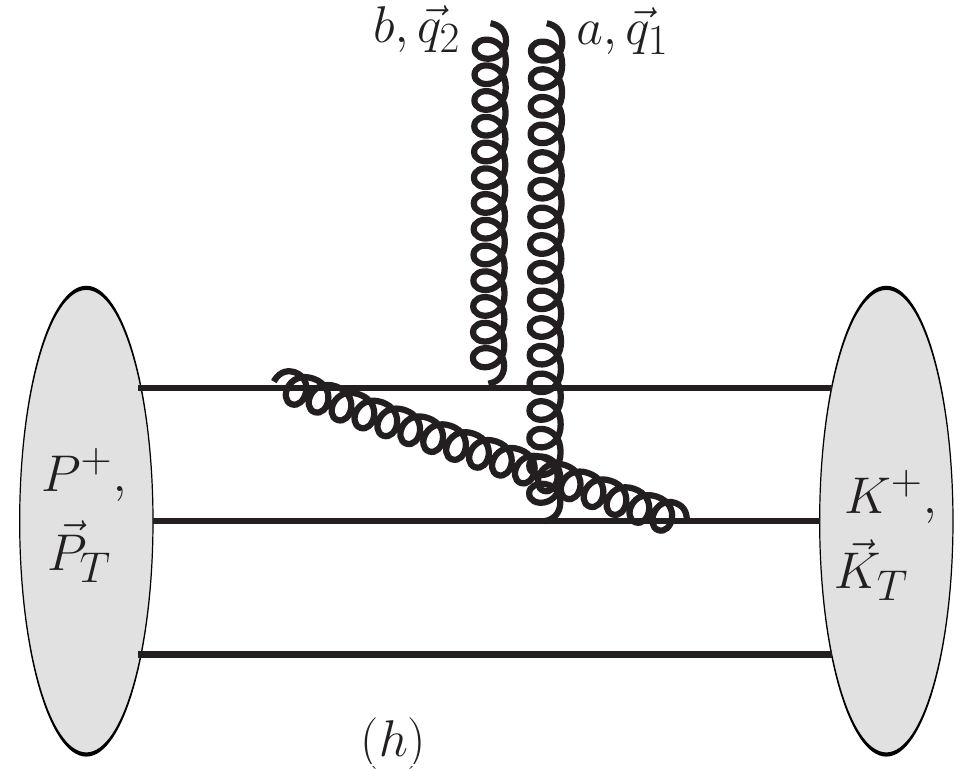}
\hspace*{.4cm}
\includegraphics[width=0.28\linewidth]{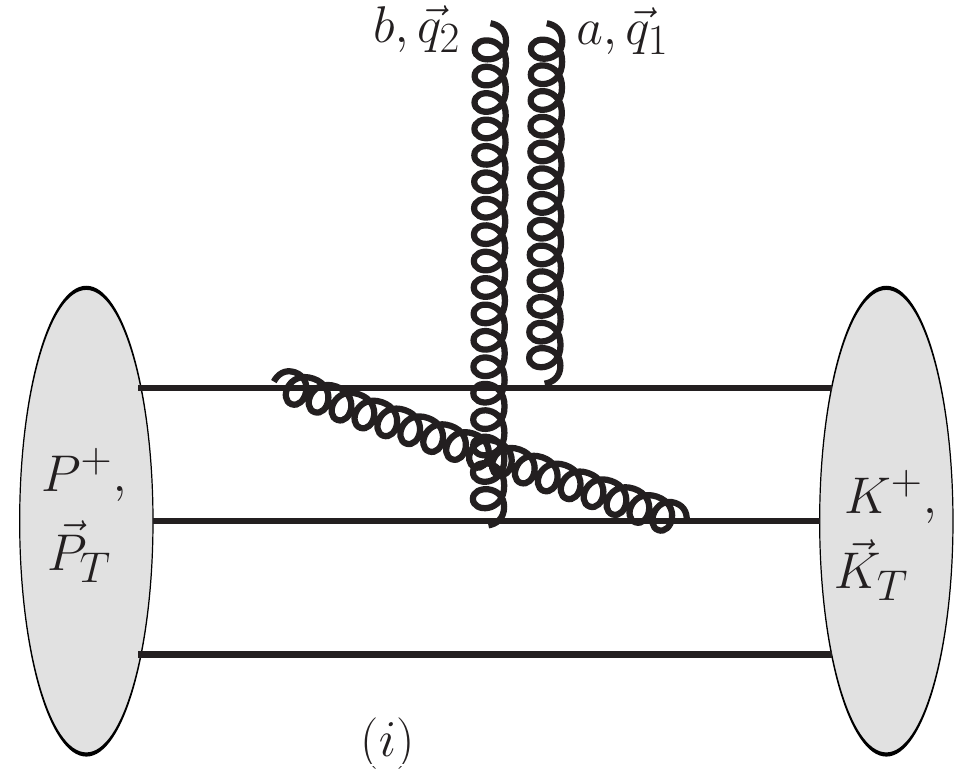}
\hspace*{.4cm}
\includegraphics[width=0.28\linewidth]{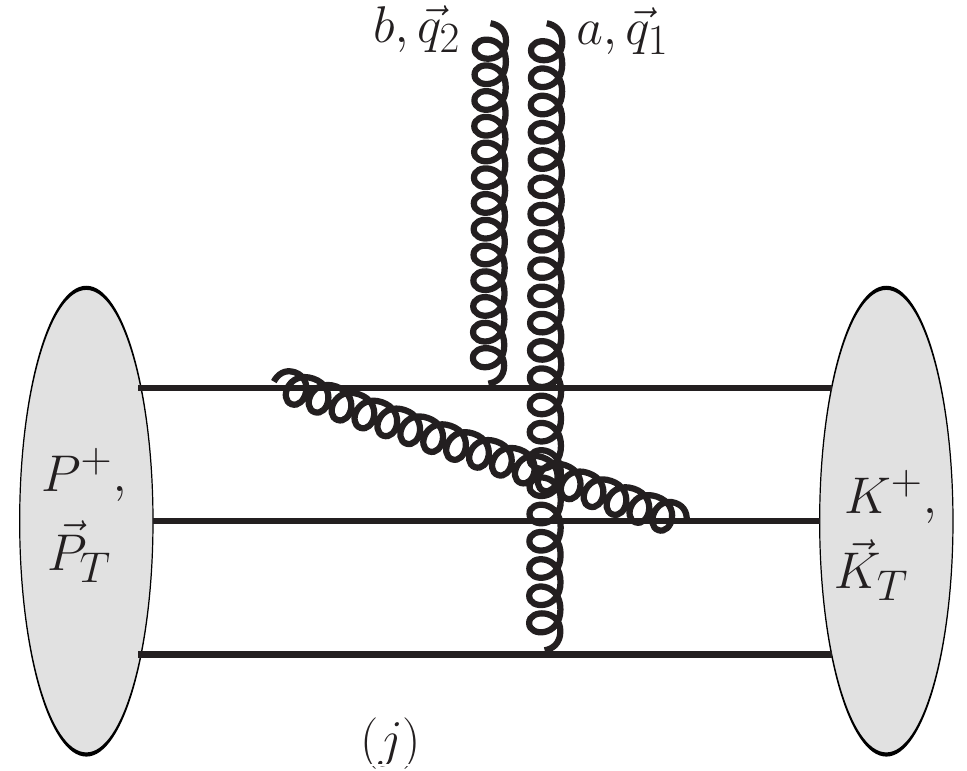}
\end{minipage}
\vspace{1cm}
\begin{minipage}[b]{\linewidth}
    \includegraphics[width=0.28\linewidth]{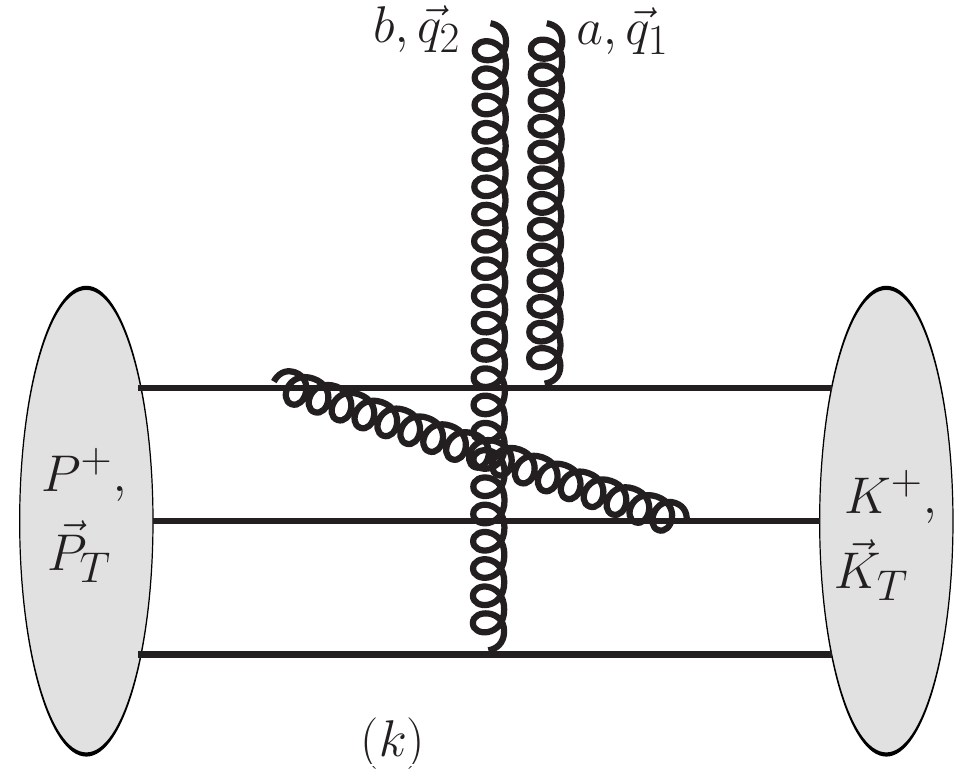}
    \hspace*{.4cm}
    \includegraphics[width=0.28\linewidth]{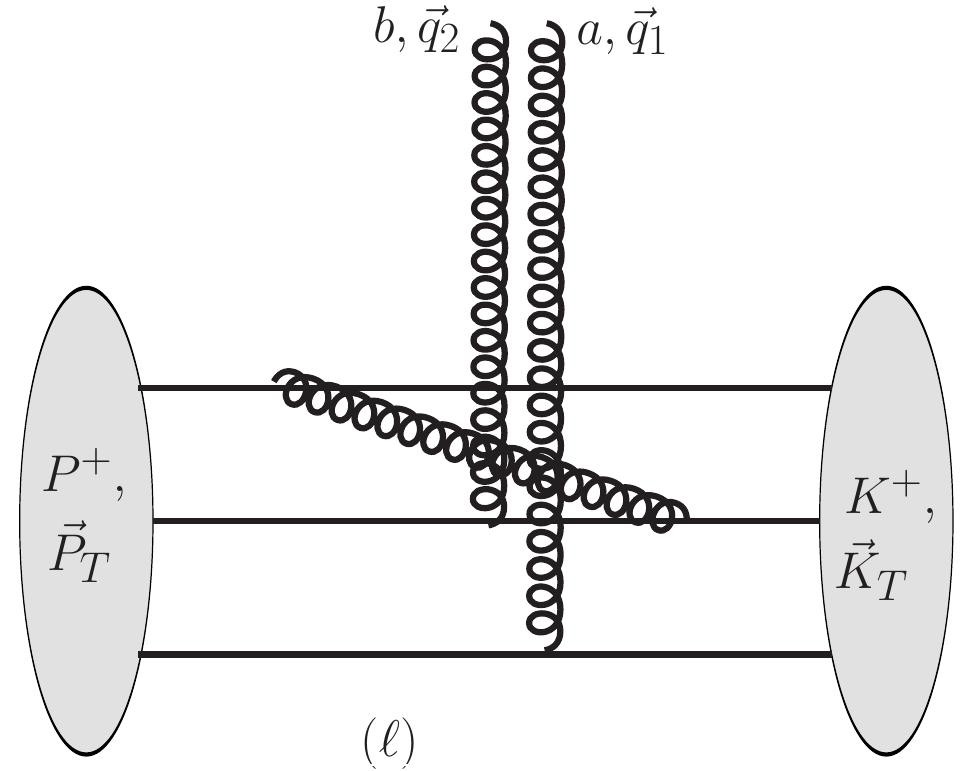}
    \hspace*{.4cm}
    \includegraphics[width=0.28\linewidth]{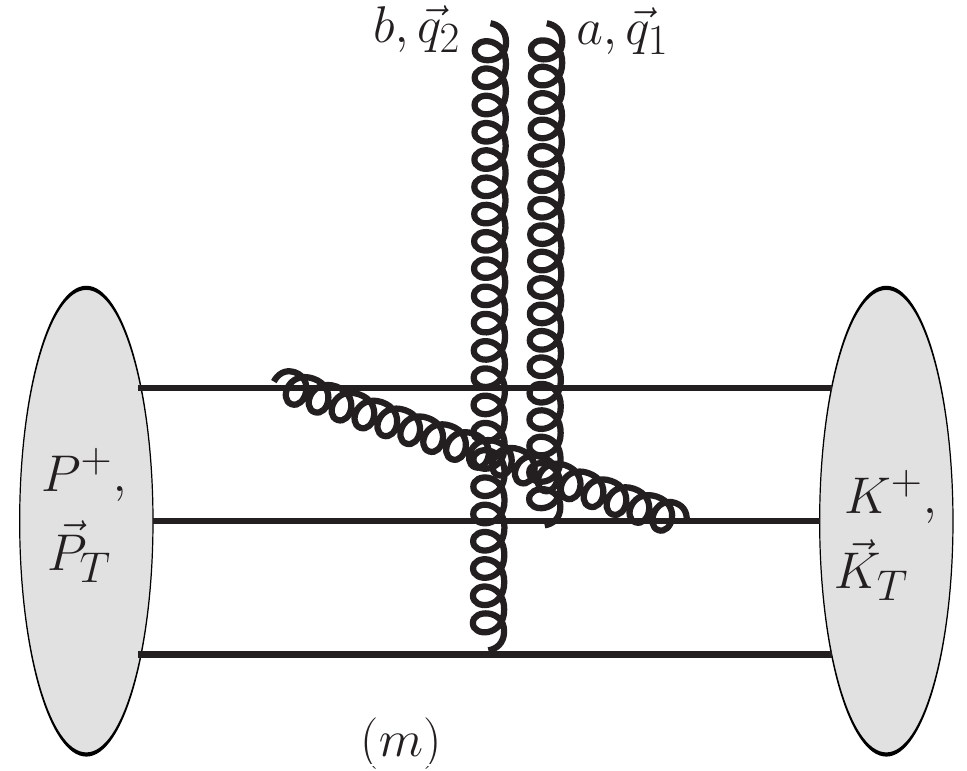}
\caption{Third set of diagrams for $\langle \rho^a_\mathrm{qu}(\vec q_1)\,
  \rho^b_\mathrm{qu}(\vec q_2)\rangle$.}
\label{fig:rho-rho_qu''}
\end{minipage}
\end{figure}
The third set of diagrams is shown in fig.~\ref{fig:rho-rho_qu''}; all their
symmetry factors are 6.
\bea
\mathrm{fig.}\, \ref{fig:rho-rho_qu''}\mathrm{h} &=&
\frac{g^4}{3\cdot 16\pi^3} \, (\frac{1}{2}-C_F)\,\tr t^at^b
 \int \left[\dd x_i\right]
  \int \left[\dd^2 k_i\right]\,
  \Psi_{qqq}(x_1,\vec k_1; x_2,\vec k_2; x_3,\vec k_3)\,
\nn\\
& &
\int\limits_x^{\mathrm{min}(x_1,1-x_2)}\frac{\dd x_g}{x_g}
\left(1-\frac{z_1+z_2}{2}+\frac{z_1 z_2}{6}\right)
\sqrt\frac{x_1}{x_1-x_g} \sqrt\frac{x_2}{x_2+x_g}
\int\dd^2 k_g
\frac{z_1\vec p_1-\vec k_g}
     {\left(z_1\vec p_1-\vec k_g \right)^2}
\cdot
\frac{z_2(\vec p_2-\vec q_1)-(1-z_2)\vec k_g}
     {\left( z_2(\vec p_2-\vec q_1)-(1-z_2)\vec k_g\right)^2} \nn\\
     & &
     \Psi_{qqq}^*(x_1-x_g,\vec k_1+x_1\vec q-\vec q_2 -\vec k_g+x_g\vec K;
     x_2+x_g,\vec k_2 +x_2\vec q -\vec q_1 + \vec k_g-x_g\vec K;
     x_3,\vec k_3 +x_3\vec q)~.
     \label{eq:rho-rho_qu-h}
\eea
\bea
\mathrm{fig.}\, \ref{fig:rho-rho_qu''}\mathrm{i} &=&
\mathrm{fig.}\,
\ref{fig:rho-rho_qu''}\mathrm{h}\, (\vec q_1 \leftrightarrow \vec q_2)~.
     \label{eq:rho-rho_qu-i}
\eea
\bea
\mathrm{fig.}\, \ref{fig:rho-rho_qu''}\mathrm{j} &=&
\frac{g^4}{3\cdot 16\pi^3} \, (C_F-\frac{1}{2}-\frac{1}{6})\,\tr t^at^b
 \int \left[\dd x_i\right]
  \int \left[\dd^2 k_i\right]\,
  \Psi_{qqq}(x_1,\vec k_1; x_2,\vec k_2; x_3,\vec k_3)\,
\nn\\
& &
\int\limits_x^{\mathrm{min}(x_1,1-x_2)}\frac{\dd x_g}{x_g}
\left(1-\frac{z_1+z_2}{2}+\frac{z_1 z_2}{6}\right)
\sqrt\frac{x_1}{x_1-x_g} \sqrt\frac{x_2}{x_2+x_g}
\int\dd^2 k_g
\frac{z_1\vec p_1-\vec k_g}
     {\left(z_1\vec p_1-\vec k_g \right)^2}
\cdot
\frac{z_2\vec p_2-(1-z_2)\vec k_g}
     {\left(z_2\vec p_2-(1-z_2)\vec k_g \right)^2} \nn\\
     & &
     \Psi_{qqq}^*(x_1-x_g,\vec k_1+x_1\vec q-\vec q_2 - \vec k_g+x_g\vec K;
     x_2+x_g,\vec k_2 +x_2\vec q + \vec k_g-x_g\vec K;
     x_3,\vec k_3 +x_3\vec q-\vec q_1)~.
     \label{eq:rho-rho_qu-j}
\eea
\bea
\mathrm{fig.}\, \ref{fig:rho-rho_qu''}\mathrm{k} &=&
\mathrm{fig.}\,
\ref{fig:rho-rho_qu''}\mathrm{j}\, (\vec q_1 \leftrightarrow \vec q_2)~.
     \label{eq:rho-rho_qu-k}
\eea
\bea
\mathrm{fig.}\, \ref{fig:rho-rho_qu''}\ell &=&
\frac{g^4}{3\cdot 16\pi^3} \, (C_F-\frac{1}{2}-\frac{1}{6})\,\tr t^at^b
 \int \left[\dd x_i\right]
  \int \left[\dd^2 k_i\right]\,
  \Psi_{qqq}(x_1,\vec k_1; x_2,\vec k_2; x_3,\vec k_3)\,
\nn\\
& &
\int\limits_x^{\mathrm{min}(x_1,1-x_2)}\frac{\dd x_g}{x_g}
\left(1-\frac{z_1+z_2}{2}+\frac{z_1 z_2}{6}\right)
\sqrt\frac{x_1}{x_1-x_g} \sqrt\frac{x_2}{x_2+x_g}
\int\dd^2 k_g
\frac{z_1\vec p_1-\vec k_g}
     {\left(z_1\vec p_1-\vec k_g \right)^2}
\cdot
\frac{z_2(\vec p_2-\vec q_2)-(1-z_2)\vec k_g}
     {\left(z_2(\vec p_2-\vec q_2)-(1-z_2)\vec k_g
       \right)^2} \nn\\
     & &
     \Psi_{qqq}^*(x_1-x_g,\vec k_1+x_1\vec q - \vec k_g+x_g\vec K;
     x_2+x_g,\vec k_2 +x_2\vec q -\vec q_2 + \vec k_g-x_g\vec K;
     x_3,\vec k_3 +x_3\vec q-\vec q_1)~.
     \label{eq:rho-rho_qu-l}
\eea
\bea
\mathrm{fig.}\, \ref{fig:rho-rho_qu''}\mathrm{m} &=&
\mathrm{fig.}\,
\ref{fig:rho-rho_qu''}\ell\, (\vec q_1 \leftrightarrow \vec q_2)~.
     \label{eq:rho-rho_qu-m}
\eea
\\

\begin{figure}[ht]
\centering
\begin{minipage}[b]{\linewidth}
\includegraphics[width=0.28\linewidth]{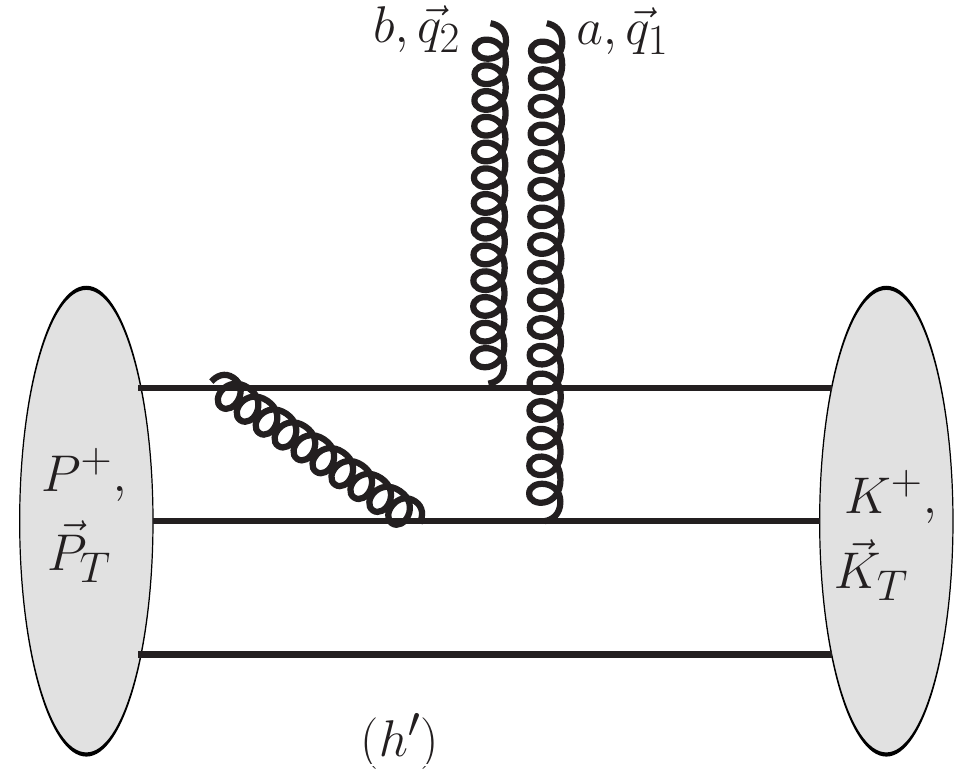}
\hspace*{.4cm}
\includegraphics[width=0.28\linewidth]{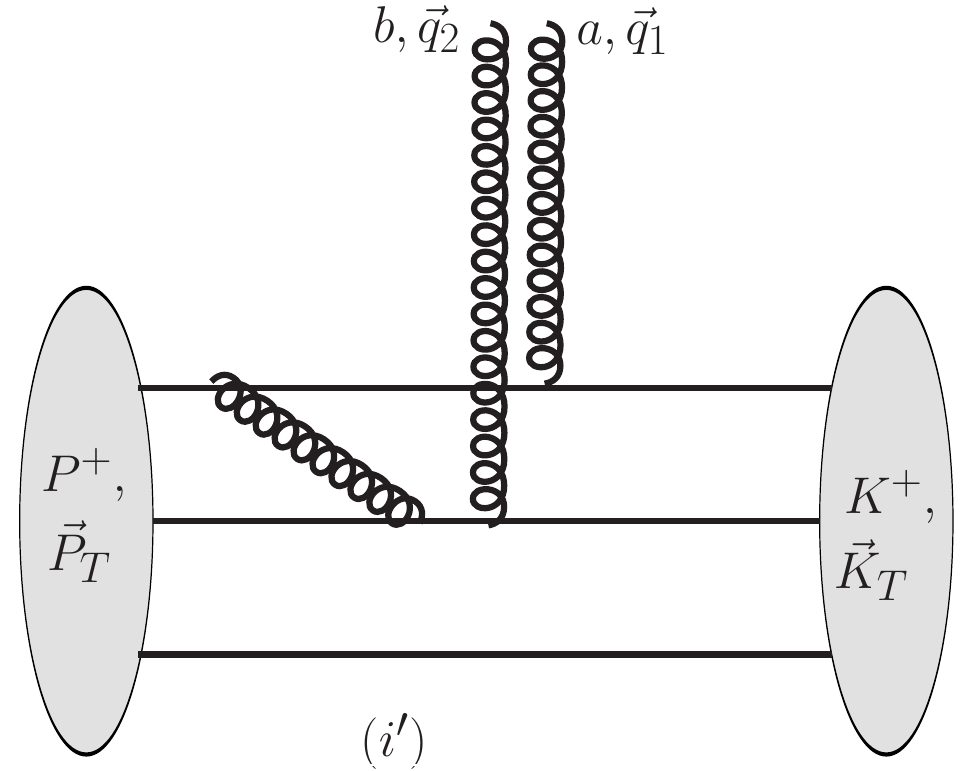}
\hspace*{.4cm}
\includegraphics[width=0.28\linewidth]{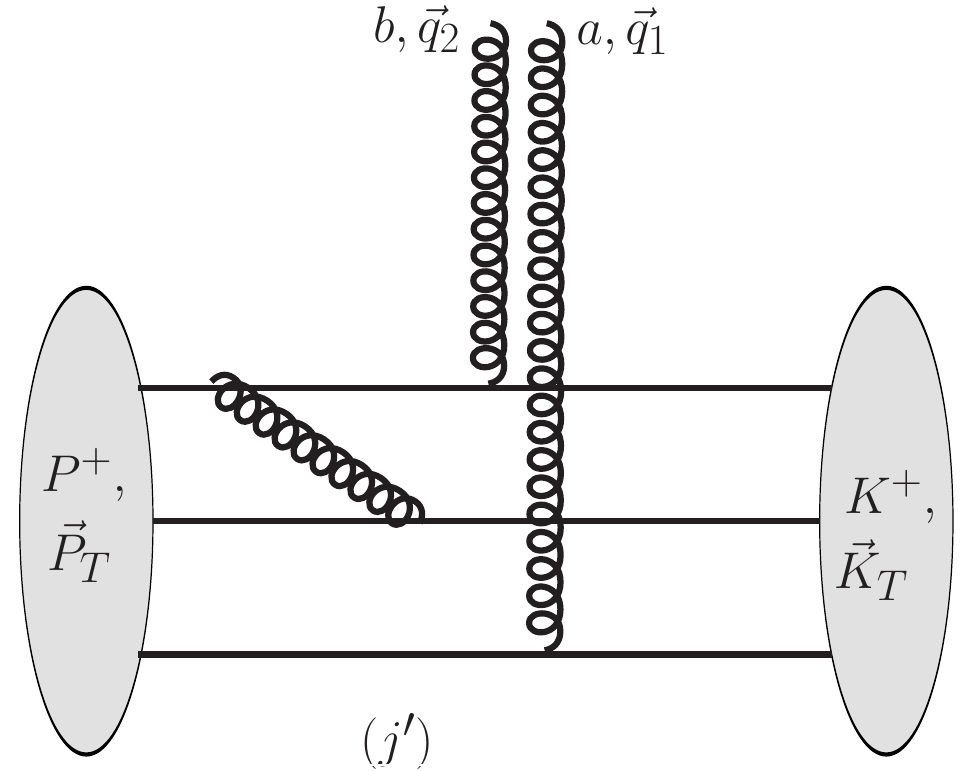}
\end{minipage}
\vspace{1cm}
\begin{minipage}[b]{\linewidth}
  \includegraphics[width=0.28\linewidth]{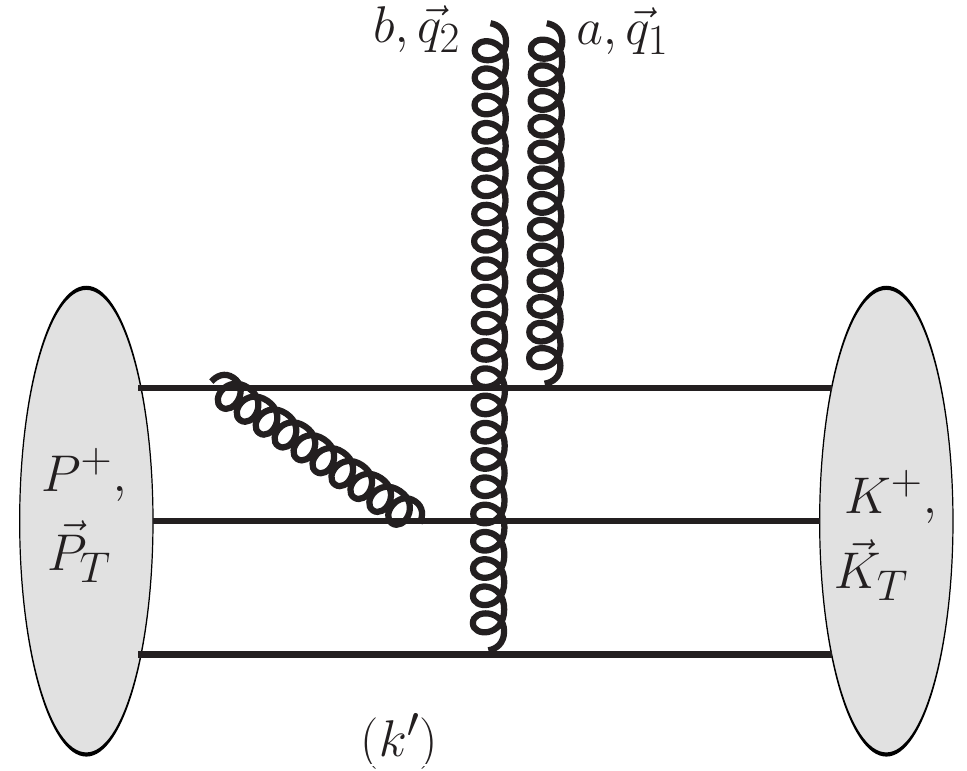}
  \hspace*{.4cm}
  \includegraphics[width=0.28\linewidth]{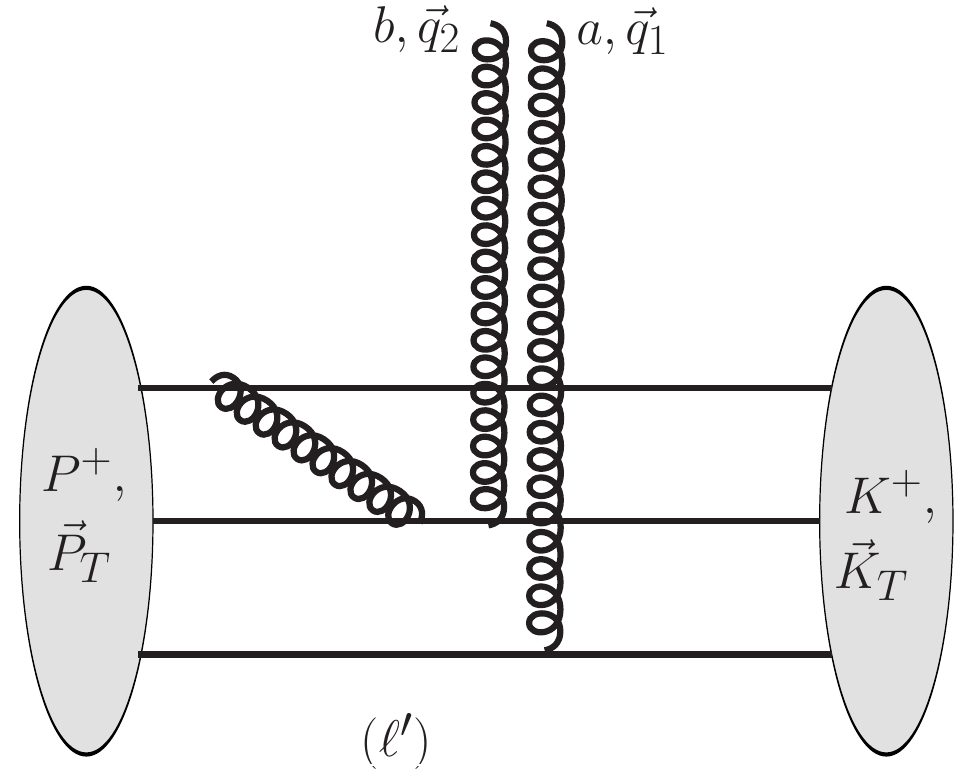}
  \hspace*{.4cm}
  \includegraphics[width=0.28\linewidth]{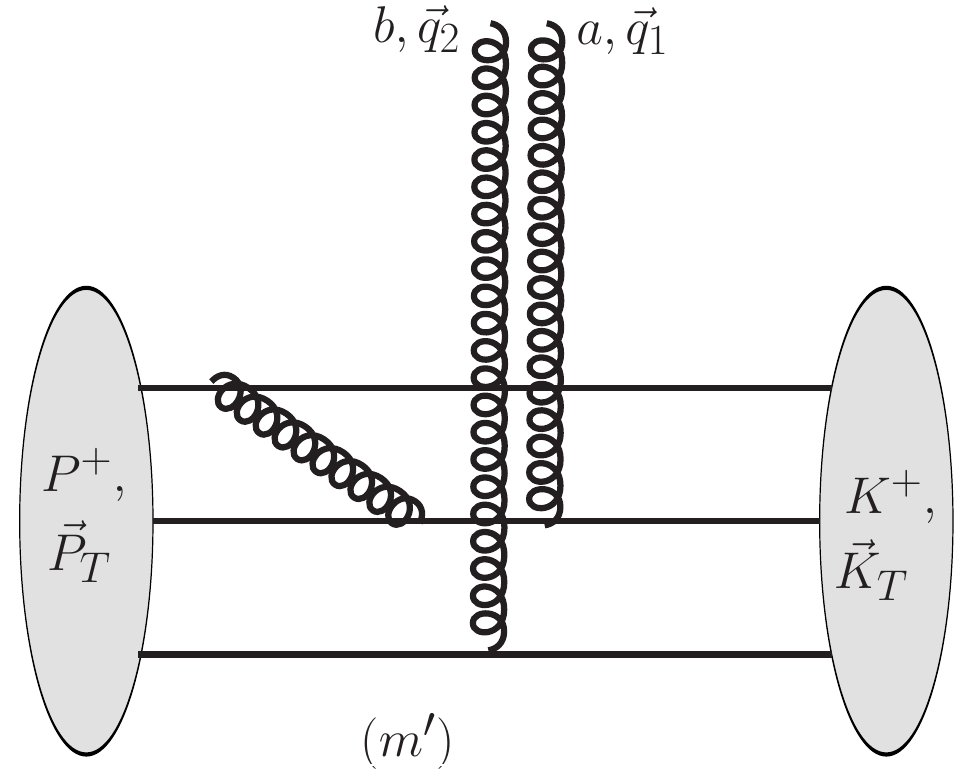}
\caption{Fourth set of diagrams for $\langle \rho^a_\mathrm{qu}(\vec q_1)\,
  \rho^b_\mathrm{qu}(\vec q_2)\rangle$.}
\label{fig:rho-rho_qu'''}
\end{minipage}
\end{figure}
The final set of (finite) diagrams is shown in
fig.~\ref{fig:rho-rho_qu'''}.  Here, in $|P\rangle$, quark 1 emits a
gluon which is absorbed by quark 2.  We will quote with double primes
the diagram (not shown) where in $\langle K|$ the exchange occurs from
quark 2 to quark 1. All these diagrams have a symmetry factor of
$\frac{1}{2} \times 6=3$. As before, in the following expressions
$z_1=x_g/x_1$ and $z_2 = x_g/(x_2+x_g)$.

\bea
\mathrm{fig.}\, \ref{fig:rho-rho_qu'''}\mathrm{h'} &=&
- \frac{2g^4}{9\cdot 16\pi^3} \, \tr t^at^b
 \int \left[\dd x_i\right]
  \int \left[\dd^2 k_i\right]\,
  \Psi_{qqq}(x_1,\vec k_1; x_2,\vec k_2; x_3,\vec k_3)\,
\nn\\
& &
\int\limits_x^{\mathrm{min}(x_1,1-x_2)}\frac{\dd x_g}{x_g}
\left(1-\frac{z_1+z_2}{2}+\frac{z_1 z_2}{6}\right)
\sqrt{\frac{x_1}{x_1-x_g}}\sqrt{\frac{x_2}{x_2+x_g}}
\int\dd^2 k_g
\frac{z_1\vec p_1-\vec k_g}
     {\left(z_1\vec p_1-\vec k_g \right)^2}
\cdot
\frac{z_2\vec p_2-(1-z_2)\vec k_g}
     {\left(z_2\vec p_2-(1-z_2)\vec k_g \right)^2} \nn\\
     & &
     \Psi_{qqq}^*(x_1-x_g,\vec k_1+x_1\vec q-\vec k_g-\vec q_2+x_g\vec K;
     x_2+x_g,\vec k_2 +x_2\vec q + \vec k_g - \vec q_1-x_g\vec K;
     x_3,\vec k_3 +x_3\vec q)~.
     \label{eq:rho-rho_qu-h'}
\eea
\bea
\mathrm{fig.}\, \ref{fig:rho-rho_qu'''}\mathrm{h''} &=&
- \frac{2g^4}{9\cdot 16\pi^3} \, \tr t^at^b
 \int \left[\dd x_i\right]
  \int \left[\dd^2 k_i\right]\,
  \Psi_{qqq}(x_1,\vec k_1; x_2,\vec k_2; x_3,\vec k_3)\,
\nn\\
& &
\int\limits_x^{\mathrm{min}(x_1,1-x_2)}\frac{\dd x_g}{x_g}
\left(1-\frac{z_1+z_2}{2}+\frac{z_1 z_2}{6}\right)
\sqrt{\frac{x_1}{x_1-x_g}}\sqrt{\frac{x_2}{x_2+x_g}}
\int\dd^2 k_g
\frac{z_1(\vec p_1-\vec q_2)-\vec k_g}
     {\left(z_1(\vec p_1-\vec q_2)-\vec k_g \right)^2}
\cdot
\frac{z_2(\vec p_2-\vec q_1)-(1-z_2)\vec k_g}
     {\left(z_2(\vec p_2-\vec q_1)-(1-z_2)\vec k_g \right)^2} \nn\\
     & &
     \Psi_{qqq}^*(x_1-x_g,\vec k_1+x_1\vec q-\vec k_g-\vec q_2+x_g\vec K;
     x_2+x_g,\vec k_2 +x_2\vec q + \vec k_g - \vec q_1-x_g\vec K;
     x_3,\vec k_3 +x_3\vec q)~.
     \label{eq:rho-rho_qu-h''}
\eea
\bea
\mathrm{fig.}\, \ref{fig:rho-rho_qu'''}\mathrm{i'} &=&
\mathrm{fig.}\, \ref{fig:rho-rho_qu'''}\mathrm{h'}
(\vec q_1 \leftrightarrow \vec q_2)~.
     \label{eq:rho-rho_qu-i'}
\eea
\bea
\mathrm{fig.}\, \ref{fig:rho-rho_qu'''}\mathrm{i''} &=&
\mathrm{fig.}\, \ref{fig:rho-rho_qu'''}\mathrm{h''}
(\vec q_1 \leftrightarrow \vec q_2)~.
     \label{eq:rho-rho_qu-i''}
\eea
\bea
\mathrm{fig.}\, \ref{fig:rho-rho_qu'''}\mathrm{j'} &=&
- \frac{g^4}{3\cdot 16\pi^3} \, (C_F-\frac{2}{3})\, \tr t^at^b
 \int \left[\dd x_i\right]
  \int \left[\dd^2 k_i\right]\,
  \Psi_{qqq}(x_1,\vec k_1; x_2,\vec k_2; x_3,\vec k_3)\,
\nn\\
& &
\int\limits_x^{\mathrm{min}(x_1,1-x_2)}\frac{\dd x_g}{x_g}
\left(1-\frac{z_1+z_2}{2}+\frac{z_1 z_2}{6}\right)
\sqrt{\frac{x_1}{x_1-x_g}}\sqrt{\frac{x_2}{x_2+x_g}}
\int\dd^2 k_g
\frac{z_1\vec p_1-\vec k_g}
     {\left(z_1\vec p_1-\vec k_g \right)^2}
\cdot
\frac{z_2\vec p_2-(1-z_2)\vec k_g}
     {\left(z_2\vec p_2-(1-z_2)\vec k_g \right)^2} \nn\\
     & &
     \Psi_{qqq}^*(x_1-x_g,\vec k_1+x_1\vec q-\vec k_g-\vec q_2+x_g\vec K;
     x_2+x_g,\vec k_2 +x_2\vec q + \vec k_g-x_g\vec K;
     x_3,\vec k_3 +x_3\vec q-\vec q_1)~.
     \label{eq:rho-rho_qu-j'}
\eea
\bea
\mathrm{fig.}\, \ref{fig:rho-rho_qu'''}\mathrm{j''} &=&
- \frac{g^4}{3\cdot 16\pi^3} \, (C_F-\frac{2}{3})\, \tr t^at^b
 \int \left[\dd x_i\right]
  \int \left[\dd^2 k_i\right]\,
  \Psi_{qqq}(x_1,\vec k_1; x_2,\vec k_2; x_3,\vec k_3)\,
\nn\\
& &
\int\limits_x^{\mathrm{min}(x_1,1-x_2)}\frac{\dd x_g}{x_g}
\left(1-\frac{z_1+z_2}{2}+\frac{z_1 z_2}{6}\right)
\sqrt{\frac{x_1}{x_1-x_g}}\sqrt{\frac{x_2}{x_2+x_g}}
\int\dd^2 k_g
\frac{z_1(\vec p_1-\vec q_2)-\vec k_g}
     {\left(z_1(\vec p_1-\vec q_2)-\vec k_g \right)^2}
\cdot
\frac{z_2\vec p_2-(1-z_2)\vec k_g}
     {\left(z_2\vec p_2-(1-z_2)\vec k_g \right)^2} \nn\\
     & &
     \Psi_{qqq}^*(x_1-x_g,\vec k_1+x_1\vec q-\vec k_g-\vec q_2+x_g\vec K;
     x_2+x_g,\vec k_2 +x_2\vec q + \vec k_g-x_g\vec K;
     x_3,\vec k_3 +x_3\vec q - \vec q_1)~.
     \label{eq:rho-rho_qu-j''}
\eea
\bea
\mathrm{fig.}\, \ref{fig:rho-rho_qu'''}\mathrm{k'} &=&
\mathrm{fig.}\, \ref{fig:rho-rho_qu'''}\mathrm{j'}
(\vec q_1 \leftrightarrow \vec q_2)~.
     \label{eq:rho-rho_qu-k'}
\eea
\bea
\mathrm{fig.}\, \ref{fig:rho-rho_qu'''}\mathrm{k''} &=&
\mathrm{fig.}\, \ref{fig:rho-rho_qu'''}\mathrm{j''}
(\vec q_1 \leftrightarrow \vec q_2)~.
     \label{eq:rho-rho_qu-k''}
\eea
\bea
\mathrm{fig.}\, \ref{fig:rho-rho_qu'''}\mathrm{\ell'} &=&
- \frac{g^4}{3\cdot 16\pi^3} \, (C_F-\frac{2}{3})\, \tr t^at^b
 \int \left[\dd x_i\right]
  \int \left[\dd^2 k_i\right]\,
  \Psi_{qqq}(x_1,\vec k_1; x_2,\vec k_2; x_3,\vec k_3)\,
\nn\\
& &
\int\limits_x^{\mathrm{min}(x_1,1-x_2)}\frac{\dd x_g}{x_g}
\left(1-\frac{z_1+z_2}{2}+\frac{z_1 z_2}{6}\right)
\sqrt{\frac{x_1}{x_1-x_g}}\sqrt{\frac{x_2}{x_2+x_g}}
\int\dd^2 k_g
\frac{z_1\vec p_1-\vec k_g}
     {\left(z_1\vec p_1-\vec k_g \right)^2}
\cdot
\frac{z_2\vec p_2-(1-z_2)\vec k_g}
     {\left(z_2\vec p_2-(1-z_2)\vec k_g \right)^2} \nn\\
     & &
     \Psi_{qqq}^*(x_1-x_g,\vec k_1+x_1\vec q-\vec k_g+x_g\vec K;
     x_2+x_g,\vec k_2 +x_2\vec q + \vec k_g -\vec q_2-x_g\vec K;
     x_3,\vec k_3 +x_3\vec q-\vec q_1)~.
     \label{eq:rho-rho_qu-l'}
\eea
\bea
\mathrm{fig.}\, \ref{fig:rho-rho_qu'''}\mathrm{\ell''} &=&
- \frac{g^4}{3\cdot 16\pi^3} \, (C_F-\frac{2}{3})\, \tr t^at^b
 \int \left[\dd x_i\right]
  \int \left[\dd^2 k_i\right]\,
  \Psi_{qqq}(x_1,\vec k_1; x_2,\vec k_2; x_3,\vec k_3)\,
\nn\\
& &
\int\limits_x^{\mathrm{min}(x_1,1-x_2)}\frac{\dd x_g}{x_g}
\left(1-\frac{z_1+z_2}{2}+\frac{z_1 z_2}{6}\right)
\sqrt{\frac{x_1}{x_1-x_g}}\sqrt{\frac{x_2}{x_2+x_g}}
\int\dd^2 k_g
\frac{z_1\vec p_1-\vec k_g}
     {\left(z_1\vec p_1-\vec k_g \right)^2}
\cdot
\frac{z_2(\vec p_2-\vec q_2)-(1-z_2)\vec k_g}
     {\left(z_2(\vec p_2-\vec q_2)-(1-z_2)\vec k_g \right)^2} \nn\\
     & &
     \Psi_{qqq}^*(x_1-x_g,\vec k_1+x_1\vec q-\vec k_g+x_g\vec K;
     x_2+x_g,\vec k_2 +x_2\vec q + \vec k_g - \vec q_2-x_g\vec K;
     x_3,\vec k_3 +x_3\vec q - \vec q_1)~.
     \label{eq:rho-rho_qu-l''}
\eea
\bea
\mathrm{fig.}\, \ref{fig:rho-rho_qu'''}\mathrm{m'} &=&
\mathrm{fig.}\, \ref{fig:rho-rho_qu'''}\mathrm{\ell'}
(\vec q_1 \leftrightarrow \vec q_2)~.
     \label{eq:rho-rho_qu-m'}
\eea
\bea
\mathrm{fig.}\, \ref{fig:rho-rho_qu'''}\mathrm{m''} &=&
\mathrm{fig.}\, \ref{fig:rho-rho_qu'''}\mathrm{\ell''}
(\vec q_1 \leftrightarrow \vec q_2)~.
     \label{eq:rho-rho_qu-m''}
\eea
%

\subsection{Cancellation of UV divergences}
\label{sec:UVcancel}

In this section we collect all UV divergent diagrams to verify that
the divergent contributions to $\langle\rho^a(\vec q_1)\, \rho^b(\vec
q_2)\rangle$ cancel, to leave just the finite parts. This implies that
at this order $\langle\rho^a(\vec q_1)\, \rho^b(\vec
q_2)\rangle$ is independent of the renormalization scale.\\

We consider the diagrams where quark 1 exchanges a gluon with itself.\\

We begin with the diagrams where the charge operators couple to
either the gluon or quark 1. The UV divergent contribution of all
these diagrams is proportional to\footnote{The arguments of $\Psi_{qqq}^*$
  may differ from eq.~(\ref{eq:UV-diagrams-factor}) but will match
across all cancelling diagrams, of course.}
\bea
& &
2\pi^3 \frac{g^2}{3\cdot 16\pi^3} \,\delta^{ab}
\int \left[\dd x_i\right]
\int \left[\dd^2 k_i\right]\,
C_q(x_1)\,
  \Psi_{qqq}(x_1,\vec k_1; x_2,\vec k_2; x_3,\vec k_3)\,
\nn\\
& &     ~~~~~~
\Psi_{qqq}^*(x_1,\vec k_1-(1-x_1)\vec q;
x_2,\vec k_2 +x_2\vec q;
x_3,\vec k_3 +x_3\vec q) ~.  \label{eq:UV-diagrams-factor}
\eea
We now collect all the prefactors. Fig.~\ref{fig:rho-rho_gl}a comes
with a factor of $2\cdot 3 /C_F$ while fig.~\ref{fig:rho-rho_qu-gl}a,
plus the corresponding diagram for $\vec q_1\leftrightarrow\vec q_2$,
contributes $-2\cdot 3 /C_F$.

From fig.~\ref{fig:rho-rho_qu_Z}a, plus the corresponding diagram
where the gluon emission and reabsorption occurs on the other side of
the insertion, we get $- \frac{1}{6}\cdot \frac{1}{2}\cdot 3 \cdot
16\pi^3 / 2\pi^3 = - 2$. On the other hand,
fig.~\ref{fig:rho-rho_qu}a contributes $4\cdot\frac{1}{2}$.
\\

Next is the contribution where both probes couple to the second quark.
From the diagram analogous to fig.~\ref{fig:rho-rho_qu_Z}a, with both
probes attached to quark 2, we again get $- \frac{1}{6}\cdot
\frac{1}{2}\cdot 3 \cdot 16\pi^3 / 2\pi^3 = -
2$. Fig.~\ref{fig:rho-rho_qu}b contributes $4\cdot\frac{1}{2}=2$.  \\

Now consider the diagrams where the first probe ($a, \vec q_1$) attaches
to quark 2 while the second probe ($b, \vec q_2$) attaches either
to quark 1 or to the gluon in the proton. Fig.~\ref{fig:rho-rho_qu-gl}b
comes with a prefactor of $-\frac{1}{2}\cdot 3 / C_F = -9/8$. The prefactor
of fig.~\ref{fig:rho-rho_qu}c is $\frac{2}{6}\cdot\frac{1}{2}/C_F = 1/8$.
Lastly, from the third term in eq.~(\ref{eq:rho-rho-3q}), which is a
diagram like fig.~\ref{fig:rho-rho_qu_Z}a but with the first probe attached
to quark 2, we get $\frac{1}{6}\cdot \frac{1}{2}\cdot \frac{1}{2}
\cdot 3 \cdot 16\pi^3 / 2\pi^3 = 1$.\\

Finally, we turn to the diagrams where the first probe ($a, \vec q_1$)
attaches to quark 3 while the second probe ($b, \vec q_2$) attaches to
quark 2. The contribution from diagrams of the type of
fig.~\ref{fig:rho-rho_qu_Z} where quark 1 exchanges a gluon with
itself on either side of the insertion (second term in
  eq.~(\ref{eq:rho-rho-3q}) with quarks 1 and 3 interchanged,
  multiplied by $-C_q(x_1)$) we get $\frac{1}{6}\cdot \frac{1}{2}\cdot
\frac{1}{2} \cdot 3 \cdot 16\pi^3 / 2\pi^3 =
1$. Fig.~\ref{fig:rho-rho_qu}d contributes $-2\cdot\frac{1}{2}$.

\subsection{Decoupling of gluon probes with infinite transverse wave length}
\label{sec:WardIdentity}

Here we verify that $\langle \rho^a(\vec q_1)\, \rho^b(\vec q_2)\rangle$
vanishes when either one of the transverse momenta goes to zero; we
consider $\vec q_1 \to 0$, the other case follows by symmetry. In
intuitive terms this reflects the fact that a gluon with infinite wave
length does not couple to a color singlet proton.\\

Since we have already verified in the previous section that all UV
divergences cancel we may now discard the divergent pieces of all
diagrams and focus on their finite parts.\\

The charge correlator at ${\cal O}(g^2)$ given in
eq.~(\ref{eq:rho-rho-3q}) does indeed vanish when $\vec q_1 \to 0$, on
account of the symmetry of the three quark wave function
$\Psi_{\mathrm{qqq}}$ under exchange of any two quarks.  For the rest
of this section we consider the contributions to $\langle \rho^a(\vec
q_1)\, \rho^b(\vec q_2)\rangle$ at ${\cal O}(g^4)$.\\

For the purpose of more compact expressions we will split off the
``pre-factor''
\be \label{eq:Ward_prefactor1}
3 \frac{g^4}{16\pi^3} \, \delta^{ab}\, \int [\dd x_i] \int[\dd^2 k_i]
\Psi_{qqq}(x_1,\vec k_1; x_2,\vec k_2; x_3,\vec k_3)~,
\ee
which includes a symmetry factor of 3, from the following expressions.

We collect first all the terms from the divergent diagrams which
involve $\Psi_{qqq}^*(x_1,\vec k_1-(1-x_1)\vec q_2; x_2,\vec k_2
+x_2\vec q_2; x_3,\vec k_3 +x_3\vec q_2)$:
eq.~(\ref{eq:rho-rho_gl-gl-a}), eq.~(\ref{eq:rho_qu-gl-a}) plus $\vec
q_1 \leftrightarrow \vec q_2$, two times eq.~(\ref{eq:rho_qu-gl-b})
plus $\vec q_1 \leftrightarrow \vec q_2$ (with quarks 1 and 2
interchanged), eq.~(\ref{eq:rho_qu-qu-a}), eq.~(\ref{eq:rho_qu-qu-b}),
eq.~(\ref{eq:rho_qu-qu-c}) plus $\vec q_1 \leftrightarrow \vec q_2$
(with quarks 1 and 2 interchanged), and eq.~(\ref{eq:rho_qu-qu-d})
(with quarks 1 and 2 interchanged):
\be
\begin{split}
  & 2 \cdot 2\pi^3 F(\vec q_2,0,\frac{x}{x_1},m^2) \\  
  - & 2\pi^3 F(\vec q_2,0,\frac{x}{x_1},m^2) \\
  - & 2\pi^3 F(\vec q_2,\vec q_2,\frac{x}{x_1},m^2) \\
  - & 2 \cdot \frac{1}{2}\cdot  2\pi^3 F(\vec q_2,0,\frac{x}{x_1},m^2)\\
  - & 2 \cdot \frac{1}{2}\cdot  2\pi^3 F(0,0,\frac{x}{x_2},m^2)\\
  + & \frac{4}{3}\cdot C_F\cdot \frac{1}{2}\cdot
  2\pi^3 F(\vec q_2,\vec q_2,\frac{x}{x_1},m^2)\\
  + & 2\cdot\frac{4}{3}\cdot C_F\cdot \frac{1}{2}\cdot
    2\pi^3 F(0,0,\frac{x}{x_2},m^2)\\
  + & 2\cdot\frac{2}{18}\cdot \frac{1}{2}\cdot   
    2\pi^3 F(\vec q_2,\vec q_2,\frac{x}{x_1},m^2)\\
  + & 2\cdot\frac{2}{18}\cdot \frac{1}{2}\cdot   
    2\pi^3 F(0,0,\frac{x}{x_2},m^2)\\
  - &2\cdot \frac{2}{3}\cdot C_F\cdot\frac{1}{2}\cdot   
    2\pi^3 F(0,0,\frac{x}{x_2},m^2)\\
  = & ~ 0~.
\end{split}
\ee
\\

The remaining terms which involve $\Psi_{qqq}^*(x_1-x_g, \vec
k_1-(1-x_1)\vec q_2 - \vec k_g+x_g\vec K; x_2+x_g,\vec k_2 +x_2\vec q_2 + \vec
k_g-x_g\vec K; x_3,\vec k_3 +x_3\vec q_2)$, again with the
``pre-factor''~(\ref{eq:Ward_prefactor2}): eq.~(\ref{eq:rho_qu-gl-c})
with $\vec q_1 \leftrightarrow \vec q_2$, eq.~(\ref{eq:rho-rho_qu-e},
\ref{eq:rho-rho_qu-e'}, \ref{eq:rho-rho_qu-h}, \ref{eq:rho-rho_qu-j},
\ref{eq:rho-rho_qu-h'}, \ref{eq:rho-rho_qu-j'}),
\be
\begin{split}
 & 2\cdot \frac{1}{12} \cdot 3   
\cdot
\frac{z_2\vec p_2 - (1-z_2)\vec k_g}
     {\left( z_2\vec p_2 - (1-z_2)\vec k_g \right)^2}\\
- & 2\cdot \frac{1}{3} \cdot C_F \cdot \frac{1}{2}   
\cdot
\frac{z_2\vec p_2 - (1-z_2)\vec k_g}
     {\left( z_2\vec p_2 - (1-z_2)\vec k_g \right)^2}\\
+ & \frac{1}{3} \cdot C_F \cdot \frac{1}{2}   
\cdot
\frac{z_2\vec p_2-(1-z_2)\vec k_g}
     {\left(z_2\vec p_2-(1-z_2)\vec k_g \right)^2}\\
+ & 2\cdot \frac{1}{3} \cdot (\frac{1}{2}-C_F) \cdot \frac{1}{2}   
\cdot
\frac{z_2\vec p_2-(1-z_2)\vec k_g}
     {\left(z_2\vec p_2-(1-z_2)\vec k_g \right)^2}\\
+ & 2\cdot \frac{1}{3} \cdot (C_F-\frac{2}{3}) \cdot \frac{1}{2}   
\cdot
\frac{z_2\vec p_2-(1-z_2)\vec k_g}
     {\left(z_2\vec p_2-(1-z_2)\vec k_g \right)^2}\\
- & \frac{2}{9} \cdot \frac{1}{2}   
\cdot
\frac{z_2\vec p_2-(1-z_2)\vec k_g}
     {\left(z_2\vec p_2-(1-z_2)\vec k_g \right)^2}\\
- & \frac{1}{3} \cdot (C_F-\frac{2}{3}) \cdot \frac{1}{2}   
\cdot
\frac{z_2\vec p_2-(1-z_2)\vec k_g}
     {\left(z_2\vec p_2-(1-z_2) \vec k_g\right)^2} \\
& = 0 ~.
\end{split}
\ee
There are three more contributions which involve
$\Psi_{qqq}^*(x_1-x_g, \vec k_1-(1-x_1)\vec q_2 - \vec k_g+x_g\vec K;
x_2+x_g,\vec k_2 +x_2\vec q_2 + \vec k_g-x_g\vec K; x_3,\vec k_3
+x_3\vec q_2)$ but the structure
\be
\frac{z_1(\vec p_1 - \vec q_2)-\vec k_g}
     {(z_1(\vec p_1 - \vec q_2)-\vec k_g)^2}
     \cdot
\frac{z_2\vec p_2 -(1-z_2)\vec k_g}{(z_2\vec p_2 -(1-z_2)\vec k_g)^2}~.
\ee
These are eq.~(\ref{eq:rho-rho_qu-e''}) with a prefactor of $2/9$, and
eqs.~(\ref{eq:rho-rho_qu-h''}, \ref{eq:rho-rho_qu-j''}) each
with a prefactor of $-1/9$.
\\

Next, we collect terms which involve $\Psi_{qqq}^*(x_1-x_g, \vec
k_1+x_1\vec q_2 - \vec k_g+x_g\vec K; x_2+x_g,\vec k_2 -(1-x_2)\vec q_2 + \vec
k_g-x_g\vec K; x_3,\vec k_3 +x_3\vec q_2)$. Here, we split off the
``pre-factor''
\be
\begin{split}
& 3 \frac{g^4}{16\pi^3} \,\delta^{ab}
\int \left[\dd x_i\right]
  \int \left[\dd^2 k_i\right]\,
  \Psi_{qqq}(x_1,\vec k_1; x_2,\vec k_2; x_3,\vec k_3)\,
  \int\limits_x^{\mathrm{min}(x_1,1-x_2)}\frac{\dd x_g}{x_g} \\
& \int\dd^2 k_g
  \sqrt{\frac{x_1}{x_1-x_g}}\sqrt{\frac{x_2}{x_2+x_g}}
  \left(1-\frac{z_1+z_2}{2}+\frac{z_1 z_2}{6}  \right)
  \frac{z_1\vec p_1-\vec k_g}
     {\left(z_1\vec p_1-\vec k_g \right)^2}
\cdot~,  \label{eq:Ward_prefactor2}
\end{split}
\ee
which includes a factor of 3 out of the symmetry factors of the
diagrams.  These are: eqs.~(\ref{eq:rho-rho_gl-gl-b},
\ref{eq:rho_qu-gl-c}), eq.~(\ref{eq:rho_qu-gl-d}) plus $\vec q_1
\leftrightarrow \vec q_2$, eq.~(\ref{eq:rho-rho_qu-f},
\ref{eq:rho-rho_qu-f'},
\ref{eq:rho-rho_qu-f''}, \ref{eq:rho-rho_qu-i}, \ref{eq:rho-rho_qu-l},
\ref{eq:rho-rho_qu-i'}, \ref{eq:rho-rho_qu-i''},
\ref{eq:rho-rho_qu-l'}, \ref{eq:rho-rho_qu-l''})
\be
\begin{split}
- & 2\cdot \frac{1}{6}\cdot 3   
\cdot
\frac{z_2\vec p_2 - (1-z_2)(\vec k_g -\vec q_2)}
     {\left(z_2\vec p_2 - (1-z_2)(\vec k_g -\vec q_2)\right)^2}\\
+ & 2\cdot \frac{1}{12} \cdot 3   
\cdot
\frac{z_2\vec p_2 - (1-z_2)(\vec k_g -\vec q_2)}
     {\left( z_2\vec p_2 - (1-z_2)(\vec k_g -\vec q_2) \right)^2}\\
+ & 2\cdot \frac{1}{12} \cdot 3   
\cdot
\frac{z_2\vec p_2 - (1-z_2)(\vec k_g -\vec q_2)}
     {\left( z_2\vec p_2 - (1-z_2)(\vec k_g -\vec q_2) \right)^2}\\
+ & 2\cdot \frac{1}{12} \cdot 3   
\cdot
\frac{z_2(\vec p_2-\vec q_2) - (1-z_2)\vec k_g}
     {\left( z_2(\vec p_2-\vec q_2) - (1-z_2)\vec k_g \right)^2}\\
- & 2\cdot \frac{1}{3}\cdot C_F \cdot \frac{1}{2}  
\cdot
     \frac{z_2(\vec p_2-\vec q_2) - (1-z_2)\vec k_g}
     {\left( z_2(\vec p_2-\vec q_2) - (1-z_2)\vec k_g\right)^2}\\
+ & \frac{1}{3}\cdot C_F \cdot \frac{1}{2}  
\cdot
     \frac{z_2\vec p_2 - (1-z_2)\vec k_g}
     {\left( z_2\vec p_2 - (1-z_2)\vec k_g\right)^2}\\
+ & \frac{1}{3}\cdot C_F \cdot \frac{1}{2}  
\cdot
     \frac{z_2(\vec p_2-\vec q_2) - (1-z_2)\vec k_g}
     {\left( z_2(\vec p_2-\vec q_2) - (1-z_2)\vec k_g\right)^2}\\
+ & 2\cdot \frac{1}{3}(\frac{1}{2}- C_F) \cdot \frac{1}{2}  
\cdot
     \frac{z_2(\vec p_2-\vec q_2) - (1-z_2)\vec k_g}
     {\left( z_2(\vec p_2-\vec q_2) - (1-z_2)\vec k_g\right)^2}\\
- & 2\cdot \frac{1}{3}(\frac{2}{3}- C_F) \cdot \frac{1}{2}  
\cdot
     \frac{z_2(\vec p_2-\vec q_2) - (1-z_2)\vec k_g}
     {\left( z_2(\vec p_2-\vec q_2) - (1-z_2)\vec k_g\right)^2}\\
- & \frac{2}{9} \cdot \frac{1}{2}  
\cdot
     \frac{z_2\vec p_2 - (1-z_2)\vec k_g}
     {\left( z_2\vec p_2 - (1-z_2)\vec k_g\right)^2}\\
- & \frac{2}{9} \cdot \frac{1}{2}  
\cdot
     \frac{z_2(\vec p_2 -\vec q_2) - (1-z_2)\vec k_g}
     {\left( z_2(\vec p_2 -\vec q_2) - (1-z_2)\vec k_g\right)^2}\\
- & \frac{1}{3} \cdot (C_F-\frac{2}{3}) \cdot \frac{1}{2}  
\cdot
     \frac{z_2\vec p_2 - (1-z_2)\vec k_g}
     {\left( z_2\vec p_2 - (1-z_2)\vec k_g\right)^2}\\
- & \frac{1}{3} \cdot (C_F-\frac{2}{3}) \cdot \frac{1}{2}  
\cdot
     \frac{z_2(\vec p_2-\vec q_2) - (1-z_2)\vec k_g}
          {\left( z_2(\vec p_2-\vec q_2) - (1-z_2)\vec k_g\right)^2} \\
  & = 0~.
\end{split}
\ee
\\

Lastly, we collect terms which involve $\Psi_{qqq}^*(x_1-x_g, \vec
k_1+x_1\vec q_2 - \vec k_g+x_g\vec K;
x_2+x_g,\vec k_2 +x_2\vec q_2 + \vec k_g-x_g\vec K;
x_3,\vec k_3 -(1-x_3)\vec q_2)$, with the ``pre-factor''
(\ref{eq:Ward_prefactor2}): eq.~(\ref{eq:rho-rho_qu-g},
\ref{eq:rho-rho_qu-g'}, \ref{eq:rho-rho_qu-g''}, \ref{eq:rho-rho_qu-k},
\ref{eq:rho-rho_qu-m}, \ref{eq:rho-rho_qu-k'}, \ref{eq:rho-rho_qu-k''},
\ref{eq:rho-rho_qu-m'}, \ref{eq:rho-rho_qu-m''}),
\bea
&-& 2 \cdot \frac{C_F}{3} \cdot \frac{1}{2} \cdot  
\frac{z_2\vec p_2-(1-z_2)\vec k_g}
     {\left(z_2\vec p_2-(1-z_2)\vec k_g \right)^2} \nn\\
&+& \frac{C_F}{3} \cdot \frac{1}{2} \cdot  
\frac{z_2\vec p_2-(1-z_2)\vec k_g}
     {\left(z_2\vec p_2-(1-z_2)\vec k_g \right)^2} \nn\\
&+& \frac{C_F}{3} \cdot \frac{1}{2} \cdot  
\frac{z_2\vec p_2-(1-z_2)\vec k_g}
     {\left(z_2\vec p_2-(1-z_2)\vec k_g \right)^2} \nn\\
&+& 2\cdot\frac{1}{3} \cdot (C_F-\frac{2}{3})\cdot \frac{1}{2} \cdot  
\frac{z_2\vec p_2-(1-z_2)\vec k_g}
     {\left(z_2\vec p_2-(1-z_2)\vec k_g \right)^2} \nn\\
&+& 2\cdot \frac{1}{3} \cdot (C_F-\frac{2}{3})\cdot \frac{1}{2} \cdot 
\frac{z_2\vec p_2-(1-z_2)\vec k_g}
     {\left(z_2\vec p_2-(1-z_2)\vec k_g \right)^2}\nn\\
&-& \frac{1}{3} \cdot (C_F-\frac{2}{3})\cdot \frac{1}{2} \cdot 
\frac{z_2\vec p_2-(1-z_2)\vec k_g}
     {\left(z_2\vec p_2-(1-z_2)\vec k_g \right)^2} \nn\\
&-& \frac{1}{3} \cdot (C_F-\frac{2}{3})\cdot \frac{1}{2} \cdot
\frac{z_2\vec p_2-(1-z_2)\vec k_g}
     {\left(z_2\vec p_2-(1-z_2)\vec k_g \right)^2} \nn\\
&-& \frac{1}{3} \cdot (C_F-\frac{2}{3})\cdot \frac{1}{2} \cdot 
\frac{z_2\vec p_2-(1-z_2)\vec k_g}
     {\left(z_2\vec p_2-(1-z_2)\vec k_g \right)^2} \nn\\
&-& \frac{1}{3} \cdot (C_F-\frac{2}{3})\cdot \frac{1}{2} \cdot
\frac{z_2\vec p_2-(1-z_2)\vec k_g}
     {\left(z_2\vec p_2-(1-z_2)\vec k_g \right)^2} \nn\\
& = 0~.
\eea
%

\section{Outlook}

Ref.~\cite{Dumitru:2020fdh} showed ``sub-femtometer'' scale color
charge correlations in a proton composed of three quarks. These
correlators were found to display interesting dependence on the impact
parameter and on the relative momentum of the gluon probes, rather than
being simply proportional to the one-body particle density in the proton. 

Here, we have computed the expressions for the $\langle\rho^a(\vec
q_1)\, \rho^b(\vec q_2)\rangle$ correlator in a proton made of three
quarks and a perturbative gluon (which is not required to carry a
small light-cone momentum). These results may be used to obtain a more
realistic picture of color charge correlations in the proton at
moderate $x \gsim 0.01$. Also, they could be used to ``jump-start''
small-$x$ BK evolution, in particular impact parameter dependent
evolution~\cite{GolecBiernat:2003ym, Berger:2010sh, Cepila:2018faq,
  Bendova:2019psy}, towards $x \ll 0.01$ from a better constrained and
perhaps more realistic initial condition.

It would be very interesting to obtain numerical results for
the color charge correlator as a function of impact parameter $\vec b$
and relative transverse position $\vec r$ of the probes. For numerical
estimates one could employ a model for the non-perturbative valence
quark wave function such as the one of
ref.~\cite{Schlumpf:1992vq,Brodsky:1994fz}; it encodes the proper
proton radius and average quark longitudinal and transverse momentum,
as well as momentum correlations among the valence quarks.

Furthermore, the present calculation should be extended to the
correlator of three charge operators, $\langle\rho^a(\vec q_1)\,
\rho^b(\vec q_2)\, \rho^c(\vec q_3)\rangle$; the contribution of the
$|qqq\rangle$ Fock state has been analyzed in
ref.~\cite{Dumitru:2020fdh}.  $C$-odd three-gluon exchange gives an
imaginary contribution to the dipole scattering amplitude [or a real
  part, respectively, if ${\cal T} \to i{\cal T}$ in
  eq.~(\ref{eq:S_dipole_b})]. It is related to various spin dependent
Transverse Momentum Dependent (TMD) distributions such as the (dipole)
gluon Sivers function of a transversely polarized
proton~\cite{Yao:2018vcg}. This amplitude is also relevant for
charge asymmetries in diffractive electroproduction of a $\pi^+\,
\pi^-$ pair~\cite{Hagler:2002nh,Hagler:2002nf} or for exclusive
production of a pseudo-scalar meson~\cite{Dumitru:2019qec,
  Czyzewski:1996bv, Engel:1997cga, Kilian:1997ew}.  Work is in
progress to account for corrections to the three-gluon exchange
amplitude due to a perturbative gluon in the proton, and will be
reported elsewhere.

\appendix

\section{Regularization of the integral over
  $\hat\psi_{q\to qg}\, \hat\psi^*_{q\to qg}$}
\label{sec:app-psi-psi*}

Consider the following integral (see \eq\nr{eq:Int-psi-psi*})
\begin{equation}
\label{eq:integralI}
I = \frac{(2\pi)^{D-1}}{2p_1^{+}} \int \frac{\dk_g}{2(p_1^+ - k_g^+)}\hat\psi_{q \to qg}(p_1
;p_1 - k_g, k_g)\hat\psi^{\ast}_{q \to qg}(p_1 - l, p_1 - k_g - l_1, k_g - l + l_1),
\end{equation} 
where $l^+ = l_1^+ = 0$. The quark and gluon helicities are conserved, and
the gluon is emitted and absorbed by the same quark 1. Therefore,
sandwiching $I$ between helicity wave functions gives $\langle S | I | S\rangle
= I\, \langle S| S\rangle = I$.

The reduced LCwf's are given by
\begin{equation}
\label{eq:numpart1}
\begin{split}	
\hat\psi_{q \to qg}(p_1
;p_1 - k_g, k_g) & = \biggl [\biggl (1 - \frac{z_1}{2} \biggr 
)\bar u_{h_1}(p_1 - k_g)\gamma^+ u_{h_1}(p_1)\delta^{ij}
- \frac{z_1}{4}\bar u_{h_1}(p_1 - k_g)\gamma^+[\gamma^i,\gamma^j] u_{h_1}(p_1) \biggr ] \frac{n^i \epst_{\sigma}^{\ast j}}{\nvec^2}
\end{split}
\end{equation}
and
\begin{equation}
\label{eq:numpart2}
\begin{split}	
\hat\psi^{\ast}_{q \to qg}(p_1 - l, p_1 - k_g - l_1, k_g - l + l_1) & = \biggl [\biggl (1 - \frac{z_2}{2} \biggr 
)\bar u_{h_1}(p_1 - l)\gamma^+ u_{h_1}(p_1-k_g-l_1)\delta^{kl}  \\
& + \frac{z_2}{4}\bar u_{h_1}(p_1 - l)\gamma^+[\gamma^k,\gamma^l] u_{h_1}(p_1-k_g-l_1)  \biggr ] \frac{m^k \epst_{\sigma}^{l}}{\mvec^2},
\end{split}
\end{equation}
where $z_1 = k_g^+/p_1^+$, $\nvec = \kvec_g - z_1\pvec$ and similarly
\begin{equation}
z_2 = \frac{k_g^+ - l^+ + l_1^+}{p_1^+ - l^+} = \frac{k_g^+}{p_1^+} = z_1
\end{equation}
and $\mvec = \nvec + \hvec$ with $\hvec = \lvec_1 - (1-z_1)\lvec$.

In order to simplify the spinor algebra, we first note that the following relation between the complete spinors $u_h(p), \bar{u}_h(p)$ and the
good component of the spinors $u^{G}_h(p), \bar{u}^{G}_h(p)$ is
satisfied\footnote{See the discussion e.g.\ in \
ref.~\cite{Hanninen:2017ddy, Beuf:2016wdz}. Also note that the spinor
structure $\bar{u}_h(p)\gamma^+ [\gamma^i, \gamma^j] u_h(k)$ can be
expressed in terms of the good component of the spinors.}
\begin{equation}
\bar{u}_h(p)\gamma^+ u_h(k) = \bar{u}^{G}_h(p^+)\gamma^+ u^{G}_h(k^+) = \sqrt{2p^+}\sqrt{2k^+}, 
\end{equation}
where the good components depend only on $p^+, k^+$ and helicity. For the good components one has the completeness relation
\begin{equation}
\label{eq:comprel}
\sum_h u^{G}_h(p^+)\bar{u}^{G}_h(p^+)\gamma^+ = 2p^+\Lambda_{+},
\end{equation}
where the projection operator\footnote{Note that $(\gamma^+)^2 = 0$ and $\gamma^+ \Lambda_+ = \gamma^+$.} $\Lambda_+ = (\gamma^-\gamma^+)/2$.

By using \eq\nr{eq:comprel} and noting that $l^+ = l_1^+ = 0$, the product of $\hat\psi_{q \to qg}\hat\psi^{\ast}_{q \to qg}$ simplifies to 
\begin{equation}
\label{eq:psipsi}
\begin{split}	
\hat\psi_{q \to qg}\hat\psi^{\ast}_{q \to qg} & = 2p_1^+ \biggl \{\biggl (1 - \frac{z_1}{2}\biggr )^2 \bar u^{G}_{h_1}(p_1^+ - k_g^+)\gamma^+ u^{G}_{h_1}(p_1^+-k_g^+) \delta^{ik} \\
& -  \biggl (1 - \frac{z_1}{2}\biggr )\frac{z_1}{2} \bar u^{G}_{h_1}(p_1^+ - k_g^+)\gamma^+ [\gamma^i,\gamma^k] u^{G}_{h_1}(p_1^+-k_g^+)  \\
& - \biggl (\frac{z_1}{4}\biggr )^2 \bar u^{G}_{h_1}(p_1^+ - k_g^+)\gamma^+ [\gamma^i,\gamma^j][\gamma^k,\gamma^j] u^{G}_{h_1}(p_1^+-k_g^+) \biggr \} \frac{n^i (n + h)^k}{\nvec^2(\nvec + \hvec)^2}\\
& = 2p_1^+\mathcal{M}^{ik}\frac{n^i (n + h)^k}{\nvec^2(\nvec + \hvec)^2},
\end{split}
\end{equation}
where the product of two commutators yields
\begin{equation}
[\gamma^i,\gamma^j][\gamma^k,\gamma^j] = -4\left ((D-3)\delta^{ik} - \frac{(D-4)}{2}[\gamma^i, \gamma^k] \right )
\end{equation}
and thus the spinor structure in $\mathcal{M}^{ik}$ gives
\begin{equation}
\label{eq:Mij}
\begin{split}	
\mathcal{M}^{ik} & = \biggl (1 - \frac{z_1}{2}\biggr )^2 \bar u^{G}_{h_1}(p_1^+ - k_g^+)\gamma^+ u^{G}_{h_1}(p_1^+-k_g^+) \delta^{ik} \\
& -  \biggl (1 - \frac{z_1}{2}\biggr )\frac{z_1}{2}\bar u^{G}_{h_1}(p_1^+ - k_g^+)\gamma^+ [\gamma^i,\gamma^k] u^{G}_{h_1}(p_1^+-k_g^+)  \\
& + \frac{z_1^2}{4} \left ((D-3) \bar u^{G}_{h_1}(p_1^+ - k_g^+)\gamma^+ u^{G}_{h_1}(p_1^+-k_g^+) \delta^{ik} - \frac{(D-4)}{2}\bar u^{G}_{h_1}(p_1^+ - k_g^+)\gamma^+ [\gamma^i,\gamma^k] u^{G}_{h_1}(p_1^+-k_g^+) \right ).
\end{split}
\end{equation}

To simplify the integral $I$ in \eq\nr{eq:integralI} further, we note that the spinor structure in \eq\nr{eq:Mij} is independent of transverse momenta, i.e. $\mathcal{M}^{ik} = \mathcal{M}^{ik}(p_1^+, k_g^+)$. Therefore we can write 
\begin{equation}
\label{eq:integralIv2}
I = (2\pi)^{D-2} \int_{0}^{p_1^+} \frac{\ud k_g^+}{2k_g^+ 2(p_1^+ - k_g^+)} \mathcal{M}^{ik}(p_1^+, k_g^+)
~(\mu^2)^{2-D/2}\int \frac{\ud^{D-2} n}{(2\pi)^{D-2}} \frac{n^i n^k + n^i h^k}{\nvec^2(\nvec + \hvec)^2},
\end{equation}
where we have changed the integration variable from $\kvec_g$ to
$\nvec$. As discussed in sec.~\ref{sec:Zq}, we regulate the collinear IR divergence by rewriting
\begin{equation}
\frac{n^i n^k + n^i h^k}{\nvec^2(\nvec + \hvec)^2} \to \frac{n^i n^k + n^i h^k}{(\nvec^2 + \Delta)((\nvec + \hvec)^2 + \Delta)},
\end{equation}
where $\Delta = z_1^2m^2$ with $m^2 \geq 0$. Consequently, the transverse integral can be rewritten as
\begin{equation}
\label{eq:trasverseintegral}
(\mu^2)^{2-D/2}\int \frac{\ud^{D-2} n}{(2\pi)^{D-2}} \frac{n^i n^k + n^i h^k}{\nvec^2(\nvec + \hvec)^2} = \frac{1}{4\pi} \biggl \{B^{ik}(\Delta,\Delta,\hvec) + h^k B^{i}(\Delta,\Delta,\hvec) \biggr \}.
\end{equation} 
Here we have introduced the following notation (see e.g.\
ref.~\cite{Beuf:2016wdz}):
\begin{equation}
\begin{split}
A_0(\Delta) & = 4\pi(\mu^2)^{2-D/2}\int \frac{\ud^{D-2} n}{(2\pi)^{D-2}} \frac{1}{\nvec^2 + \Delta}, \\
B_0(\Delta,\Delta,\hvec^2) & = 4\pi(\mu^2)^{2-D/2} \int \frac{\ud^{D-2} n}{(2\pi)^{D-2}} \frac{1}{(\nvec^2 + \Delta)((\nvec + \hvec)^2 + \Delta)}, \\
B^i(\Delta,\Delta,\hvec) & = 4\pi(\mu^2)^{2-D/2}\int \frac{\ud^{D-2} n}{(2\pi)^{D-2}} \frac{n^i}{(\nvec^2 + \Delta)((\nvec + \hvec)^2 + \Delta)}, \\
B^{ik}(\Delta,\Delta,\hvec) & = 4\pi(\mu^2)^{2-D/2}\int \frac{\ud^{D-2} n}{(2\pi)^{D-2}} \frac{n^i n^k}{(\nvec^2 + \Delta)((\nvec + \hvec)^2 + \Delta)}, \\
\end{split}
\end{equation} 
where the scalar integrals $A_0$ and $B_0$ are UV divergent and UV
finite, respectively. Then, due to rotational symmetry in $D-2$
dimensions, the rank-one and rank-two tensor integrals satisfy
\begin{equation}
\begin{split}
B^i(\Delta, \Delta, \hvec)
& = h^i B_1(\Delta, \Delta, \hvec^2),\\
B^{ik}(\Delta, \Delta, \hvec) & = \delta^{ik}B_{21}(\Delta, \Delta, \hvec^2) + h^i h^k B_{22}(\Delta, \Delta, \hvec^2),
\end{split}
\end{equation}
where the coefficient $B_1$ is given by (note that we take the limit $m^2 \to 0$
whenever it appears as a prefactor)
\begin{equation}
B_1(\Delta,\Delta,\hvec^2)  = \frac{1}{2\hvec^2}\biggl [\underbrace{A_0(\Delta) - A_0(\Delta)}_{\text{UV part cancel out}} + (2\Delta - \hvec^2) B_0(\Delta,\Delta,\hvec^2) \biggr ] \to  -\frac{1}{2}B_0(\Delta,\Delta,\hvec^2)
\end{equation}
and the coefficients $B_{21}$ and $B_{22}$ satisfy
\begin{equation}
(D-2)B_{21}(\Delta, \Delta, \hvec^2) + \hvec^2 B_{22}(\Delta, \Delta, \hvec^2) = A_0(\Delta) - \Delta B_0(\Delta, \Delta, \hvec^2) \to  A_0(\Delta).
\end{equation}

It is then easy to show that the antisymmetric part $\propto
[\gamma^i, \gamma^k]$ in $\mathcal{M}^{ik}$ contracted with
$\delta^{ik}$ or $h^i h^k$ vanishes. Therefore the remaning part
in \eq\nr{eq:Mij} can be easily simplified to
\begin{equation}
\label{eq:Mijv2}
\begin{split}
\mathcal{M}^{ij}  & = 2(p_1^+ - k_g^+)\biggl \{\biggl (1 - \frac{z_1}{2}\biggr )^2 + (D-3)\frac{z_1^2}{4} \biggr \} \delta^{ik}\\
& =  2(p_1^+ - k_g^+) \frac{1}{2}\biggl \{ 1 + (1 - z_1)^2 + (D-4)\frac{z_1^2}{2}\biggr \} \delta^{ik}.
\end{split}
\end{equation}
All in all, the integral $I$ in \eq\nr{eq:integralIv2} can be cast into the following form
\begin{equation}
\label{eq:integralIv3}
I = \frac{(2\pi)^{D-2}}{16\pi} \int_{\alpha_1}^{1} \frac{\ud z_1}{z_1} \biggl \{ 1 + (1 - z_1)^2 + (D-4)\frac{z_1^2}{2}\biggr \}\biggl [A_0(\Delta) - \frac{\hvec^2}{2}B_0(\Delta,\Delta,\hvec^2) \biggr ],
\end{equation}
where we have changed the integration variable from $k_g^+$ to $z_1$
and regulated the soft IR divergence $z_1 \to 0$ by a cutoff $\alpha_1
= x/x_1$. Furthemore, by using the Feynman parametrization, one can
easily show that the coefficient $B_0$ can be written
as~\cite{Beuf:2016wdz}
\begin{equation}
\begin{split}
B_0(\Delta,\Delta,\hvec^2)  = \int_{0}^{1} \frac{\ud x}{x(1-x)\hvec^2 + \Delta}  
 + \mathcal{O}(D-4).
\end{split}
\end{equation}

Finally, noting that the quark renormalization factor $C_q$, computed in sec.~\ref{sec:Zq}, can be written as
\begin{equation}
\frac{C_q(p_1^+)}{g^2 \cf} = \frac{1}{8\pi^2} \int_{\alpha_1}^{1} \frac{\ud z_1}{z_1} \biggl \{ 1 + (1 - z_1)^2 + (D-4)\frac{z_1^2}{2}\biggr \}A_0(\Delta)
\end{equation}
we obtain for our integral $I$ in \eq\nr{eq:integralI} a very compact final result
\begin{equation}
I = 2\pi^3 \biggl \{\frac{C_q(p_1^+)}{g^2\cf} + F(\lvec, \lvec_1;\alpha_1,m^2) \biggr \}, 
\end{equation}
where we have introduced the UV finite integral
\begin{equation}
F(\lvec, \lvec_1;\alpha_1,m^2) = - \frac{1}{8\pi^2}\int_{\alpha_1}^{1} \frac{\ud z_1}{z_1} \biggl [1 + (1 - z_1)^2 \biggr ]\frac{\hvec^2}{2}B_0(\Delta,\Delta,\hvec^2)
\end{equation}
with $\hvec^2 = \lvec_1^2 + \lvec^2(1-z_1)^2  - 2(\lvec_1\cdot \lvec)(1-z_1)$. 
\\

For example, if $\lvec = \qvec_1 + \qvec_2 \equiv \qvec$ and $\lvec_1 = 0$, then
$\hvec^2 = \qvec^2(1-z_1)^2$ and hence
\begin{equation}
\begin{split}
F(\qvec,0;\alpha_1,m^2) & = -\frac{1}{8\pi^2}\int_{\alpha_1}^{1} \frac{\ud z_1}{z_1} \biggl [1 + (1-z_1)^2 \biggr ] \frac{\qvec^2(1-z_1)^2}{2}\int_{0}^{1} \frac{\ud x}{x(1-x)(1-z_1)^2\qvec^2 + z_1^2m^2}\\
& = -\frac{1}{8\pi^2}\int_{\alpha_1}^{1} \frac{\ud z_1}{z_1} \biggl [1 + (1-z_1)^2 \biggr ] \frac{\qvec^2}{2}\int_{0}^{1} \frac{\ud x}{x(1-x)\qvec^2 + \frac{z_1^2}{(1-z_1)^2}m^2}.\\
\end{split}
\end{equation}
Since $\qvec^2, m^2 \geq 0$, the remaining $x$-integral gives
\begin{equation}
\begin{split}
\int_{0}^{1} \frac{\ud x}{x(1-x)\qvec^2 + \hat m^2} & =
-\frac{2}{\sqrt{\qvec^2(\qvec^2 + 4\hat m^2)}}\log \left (\frac{2\hat m^2}{\qvec^2 + 2\hat m^2 + \sqrt{\qvec^2(\qvec^2 + 4\hat m^2)}} \right )~,
\end{split}
\end{equation}
where $\hat m^2 = z_1^2 m^2/(1-z_1)^2$. This exhibits the ``DGLAP logarithm''~\cite{Gribov:1972ri,
Gribov:1972rt, Altarelli:1977zs, Dokshitzer:1977sg} which emerges from
the collinear divergence.

\section*{Acknowledgements}

A.D.\ acknowledges support by the DOE Office of Nuclear Physics
through Grant No.\ DE-FG02-09ER41620.
R.P.\  is  supported  by  the  European  Research  Council,  grant  no.\
725369 and Academy of Finland grant no.\ 1322507. The figures have been prepared with Jaxodraw~\cite{Binosi:2008ig}.

\bibliography{spires}

\end{document}